\documentclass[useAMS,usenatbib]{pasj01}

\usepackage{tikz}
\usepackage{lscape}
\usepackage{color}
\usepackage{comment}
\usepackage{graphicx}
\usepackage{natbib, aas_macros}
\usepackage{bm}
\usepackage{placeins}

\usepackage{cuted}
\usepackage{flushend}

\usepackage{soul}
\sethlcolor{pink}

\usepackage{lineno}

\newcommand{\simgt}{\lower.5ex\hbox{$\; \buildrel > \over \sim \;$}}
\newcommand{\simlt}{\lower.5ex\hbox{$\; \buildrel < \over \sim \;$}}

\def\btheta{\mbox{\boldmath $\theta$}}

\newcommand{\green}[1]{{\textcolor{green}{#1}}}

\def\h70kpc{\mathrel{h_{70}^{-1}{\rm kpc}}}

\def\Msol{\mathrel{M_\odot}}

\def\h70Msol{\mathrel{h_{70}^{-1}M_\odot}}

\usepackage{version}

\begin{document}

\Received{}
\Accepted{}


\title{ HSC-XXL : Baryon budget of the 136 XXL Groups and Clusters
\thanks{Based on data collected at Subaru Telescope, which is operated
by the National Astronomical Observatory of Japan.}}

\author{Daichi \textsc{Akino}\altaffilmark{1}}
\email{m204964@hiroshima-u.ac.jp}
\altaffiltext{1}{Physics Program, Graduate School of Advanced Science and Engineering, Hiroshima University, 1-3-1 Kagamiyama, Higashi-Hiroshima, Hiroshima 739-8526, Japan}

\author{Dominique \textsc{Eckert}\altaffilmark{2}}
\email{Dominique.Eckert@unige.ch}
\altaffiltext{2}{Department of Astronomy, University of Geneva, ch. d’Ecogia 16, 1290 Versoix, Switzerland}

\author{Nobuhiro \textsc{Okabe}\altaffilmark{1,3,4}}
\email{okabe@hiroshima-u.ac.jp}
\altaffiltext{3}{Hiroshima Astrophysical Science Center, Hiroshima University, 1-3-1 Kagamiyama, Higashi-Hiroshima, Hiroshima 739-8526, Japan}
\altaffiltext{4}{Core Research for Energetic Universe, Hiroshima University, 1-3-1, Kagamiyama, Higashi-Hiroshima, Hiroshima 739-8526, Japan}

\author{Mauro \textsc{Sereno}\altaffilmark{5,6}}
\altaffiltext{5}{INAF - Osservatorio di Astrofisica e Scienza dello Spazio di Bologna,
via Piero Gobetti 93/3, I-40129 Bologna, Italy}
\altaffiltext{6}{INFN, Sezione di Bologna, viale Berti Pichat 6/2, 40127 Bologna, Italy}

\author{Keiichi \textsc{Umetsu}\altaffilmark{7}}
\altaffiltext{7}{Academia Sinica Institute of Astronomy and Astrophysics (ASIAA), No. 1, Section 4, Roosevelt Road, Taipei 10617, Taiwan}

\author{Masamune \textsc{Oguri}\altaffilmark{8,9,10}}
\altaffiltext{8}{Research Center for the Early Universe, University of Tokyo, Tokyo 113-0033, Japan}
\altaffiltext{9}{Department of Physics, University of Tokyo, Tokyo 113-0033, Japan}
\altaffiltext{10}{Kavli Institute for the Physics and Mathematics of the Universe (Kavli IPMU, WPI), University
of Tokyo, Chiba 277-8582, Japan}
\author{Fabio \textsc{Gastaldello}\altaffilmark{11}}
\altaffiltext{11}{INAF - IASF Milano, via Bassini 15, I-20133 Milano, Italy}

\author{I-Non \textsc{Chiu}\altaffilmark{7}}
\author{Stefano \textsc{Ettori}\altaffilmark{5,6}}
\author{August E. \textsc{Evrard}\altaffilmark{12,13}}
\altaffiltext{12}{Department of Physics and Michigan Center for Theoretical Physics, University of Michigan, Ann Arbor, MI 48109, USA}
\altaffiltext{13}{McWilliams Center for Cosmology, Department of Physics, Carnegie Mellon University, Pittsburgh, PA 15213, USA}
\author{Arya \textsc{Farahi}\altaffilmark{12,13}}

\author{Ben \textsc{Maughan}\altaffilmark{14}}
\altaffiltext{14}{H. H. Wills Physics Laboratory, University of Bristol, Tyndall Ave., Bristol BS8 1TL, UK}
\author{Marguerite \textsc{Pierre}\altaffilmark{15}}
 \altaffiltext{15}{AIM, CEA, CNRS, Universit\'e Paris-Saclay, Universit\'e Paris Diderot, Sorbonne Paris Cit\'e, F-91191 Gif-sur-Yvette, France}
 \author{Marina \textsc{Ricci}\altaffilmark{16}}
\altaffiltext{16}{ Laboratoire d’Annecy de Physique des Particules, Universit\'e Savoie
Mont Blanc, CNRS/IN2P3, 74941 Annecy, France}
\author{Ivan \textsc{Valtchanov}\altaffilmark{17}}
\altaffiltext{17}{ Telespazio UK for ESA, European Space Astronomy Centre, Operations Department, E-28691 Villanueva de la Ca\~nada, Spain}

\author{Ian \textsc{McCarthy}\altaffilmark{18}}
\altaffiltext{18}{Astrophysics Research Institute, Liverpool John Moores University, 146 Brownlow Hill, Liverpool L3 5RF, UK}
\author{Sean \textsc{McGee}\altaffilmark{19}}
\altaffiltext{19}{School of Astronomy and Physics, University of Birmingham,Edgbaston,Birmingham,B15,2TT, UK}
\author{Satoshi \textsc{Miyazaki}\altaffilmark{20}}
\altaffiltext{20}{National Astronomical Observatory of Japan, Osawa 2-21-1, Mitaka, Tokyo 181-8588, Japan}
\author{Atsushi J. \textsc{Nishizawa}\altaffilmark{21}}
\altaffiltext{21}{Institute for Advanced Research, Nagoya University Furocho,
Chikusa-ku, Nagoya, 464-8602 Japan}
\author{Masayuki \textsc{Tanaka}\altaffilmark{20}}




\maketitle

\begin{abstract}
We present our determination of the baryon budget for an X-ray-selected XXL sample of 136 galaxy groups and clusters spanning nearly two orders of magnitude in mass ($M_{500}\sim10^{13}-10^{15}\Msol$) and the redshift range $0\simlt z \simlt 1$. Our joint analysis is based on the combination of HSC-SSP weak-lensing mass measurements, XXL X-ray gas mass measurements, and HSC and Sloan Digital Sky Survey multiband photometry.
We carry out a Bayesian analysis of multivariate mass-scaling relations of gas mass, galaxy stellar mass, stellar mass of brightest cluster galaxies (BCGs), and soft-band X-ray luminosity, 
by taking into account the intrinsic covariance between cluster properties, selection effect,  weak-lensing mass calibration,  and observational error covariance matrix. 
The mass-dependent slope of the gas mass--total mass ($M_{500}$) relation is found to be $1.29_{-0.10}^{+0.16}$, which is steeper than the self-similar prediction of unity, whereas the slope of the stellar mass--total mass relation is shallower than unity, $0.85_{-0.09}^{+0.12}$.
The BCG stellar mass weakly depends on cluster mass with a slope of $0.49_{-0.10}^{+0.11}$.
 The baryon, gas mass, and stellar mass fractions as a function of $M_{500}$ agree  with the results from numerical simulations and previous observations.
We successfully constrain the full intrinsic covariance of the baryonic contents. 
The BCG stellar mass shows the larger
intrinsic scatter at a given halo total mass, followed in order by stellar mass and gas mass.
We find a significant positive intrinsic correlation coefficient between total (and satellite) stellar mass and BCG stellar mass and no evidence for intrinsic correlation between gas mass and stellar mass.
  All the baryonic components show no redshift evolution. 
\end{abstract}


\section{Introduction}

Galaxy groups and clusters are self-gravitating objects with total mass between $\sim 10^{13}\Msol$ and $\sim 10^{15}\Msol$. They contain diffuse thin plasma, galaxies and dark matter. 
The diffuse gas, referred to as hot baryon, is observed by X-ray satellites or Sunyaev-Zel'dovich (SZ) effect. 
The cold baryons reside mostly in galaxies, whose stellar component can be observed by optical and/or (near)-infrared telescopes.
Galaxies are mainly classified as central brightest cluster galaxies (BCGs) or satellite galaxies.  
Based on the hierarchical structure formation model, objects form via gravitational collapse of a large volume and 
 thus collect baryons into their halo potentials. Therefore, in the absence of dissipation, the baryon mass fraction is expected to be close to the universal average, $\Omega_{b}/\Omega_m$, measured from cosmic microwave background (CMB) experiments \citep[e.g.][]{1993Natur.366..429W,1997MNRAS.292..289E,2003MNRAS.344L..13E}.  Moreover, the cold and hot baryons affect each other through non-gravitational interactions such as star formation, cluster mergers, and energy feedback process by active galactic nuclei (AGNs) and supernove (SN).

 Recent numerical simulations \citep[e.g.][]{2011MNRAS.413..691Y,2011MNRAS.412.1965M,2013MNRAS.431.1487P,2014MNRAS.440.2290M,2014MNRAS.441.1270L,2015MNRAS.452.1982W,2016MNRAS.457.4063S,2017MNRAS.465.2936M,2017MNRAS.471.1088B,2018MNRAS.478.2618F,2020MNRAS.498.2114H,2020MNRAS.493.1361F} showed that the radiative processes convert gas to stars and significantly affect the evolution of the baryonic components. The details highly depend on AGN models and radiative codes \citep[e.g.][]{2011MNRAS.412.1965M,2014MNRAS.441.1270L,2016MNRAS.459.2973S}.  In general, the star formation rate is more efficient in group scale of $\sim 10^{13}\Msol$ than in massive clusters of $\sim10^{15}\Msol$. 
 Furthermore, AGN feedback in groups is energetic enough to expel hot gas out from their relatively shallow gravitational potentials. It is expected that the baryon contents depend on the halo mass. Therefore, a cluster sample covering as wide mass range as possible provides us with a unique opportunity to understand baryonic physics and its relationship with cluster properties.

  The XXL Survey \citep{2016A&A...592A...1P,2016A&A...592A...2P,2016A&A...592A...3G,2016A&A...592A...4L,2016A&A...592A...6P,2018A&A...620A...5A,2018A&A...620A...7G} is one of the largest observing program undertaken by {\it XMM-Newton}, covering two distinct sky areas for a total of 50 square degrees down to a sensitivity of $6\times10^{-15}\,{\rm erg}\,{\rm cm}^{-2}\,{\rm s}^{-1}$ for point-like sources ([0.5-2] keV band). 
   Nearly four hundreds galaxy clusters and groups have been detected \citep{2018A&A...620A...5A} over a wide range of nearly two orders of magnitude in mass ($10^{13}-10^{15}\Msol$) up to $z\sim 2$.
  The XXL cluster sample is optimal \citep{2018A&A...620A...5A,2016A&A...592A..12E,2020ApJ...890..148U,2020MNRAS.492.4528S,2021MNRAS.503.5624W} for studying the baryon budget of groups and clusters.

   \citet{2016A&A...592A..12E} investigated the gas mass fraction for 100 clusters and the stellar mass fraction for 34 clusters from the XXL first cluster catalog \citep[DR1;][]{2016A&A...592A...2P}.
    Each cluster mass was estimated through their X-ray temperature, calibrated with weak-lensing masses for a subset of 38 clusters covered by the CFHTLS Survey \citep{2016A&A...592A...4L}. They found that the total baryon fraction within $r_{500}$ falls short of $\Omega_b/\Omega_m$ by about a factor of two.
    Here, the subscript $500$ denotes that the mean enclosed density is $500$ times the critical density of the Universe at the cluster redshift.

  The Hyper Suprime-Cam Subaru Strategic Program
\citep[HSC-SSP;][]{HSC1stDR,HSC1styrOverview,Miyazaki18HSC,Komiyama18HSC,Kawanomoto18HSC,Furusawa18HSC,2018PASJ...70S...5B,Haung18HSC,Coupon18HSC,HSC2ndDR,2020arXiv200301511N}
  is an on-going wide-field optical imaging survey composed of three layers of different depths (Wide, Deep and UltraDeep).  The Wide layer is designed to obtain five-band ($grizy$) imaging over $1400$~deg$^2$. The survey footprint significantly overlaps with the northern sky of the XXL Survey. 
The HSC-SSP Survey has excellent imaging quality ($\sim$0.7 arcsec seeing in $i$-band) and reaches a depth of $r\simlt26$~ABmag, enabling the measurement of weak-lensing masses and photometry of the XXL clusters in the overlapped footprint.

\citet{2020ApJ...890..148U} measured weak-lensing masses for the 136 XXL clusters in the HSC-SSP survey footprint using the HSC-SSP shape catalog \citep[see details in][]{HSCWL1styr,Mandelbaum18}. They found that the CFHTLS weak-lensing masses, $M_{500}$, are on average $34\pm20$ percent higher than the HSC-SSP ones. 
\citet{2020MNRAS.492.4528S} studied the multivariate scaling relations of X-ray luminosity, temperature, gas mass, and hydrostatic mass for 118 XXL clusters. 
They measured the gas mass within an overdensity radius, $r_{500}^{\rm Eckert}$, which is computed with an iterative procedure using the
surface brightness profile and the $f_g-M_{500}$ relation from \citet{2016A&A...592A..12E}. However, they used the HSC-SSP weak-lensing masses \citep[$M_{500}^{\rm WL}$;][]{2020ApJ...890..148U} and investgated the scaling relation between the gas mass $M_g(<r_{500}^{\rm Eckert})$ and the weak-lensing mass $M_{500}^{\rm WL}(<r_{500}^{\rm WL})$ defined at different radii.
Furthermore, \citet{2020MNRAS.492.4528S} did not consider the mass of stellar components in the scaling relation analysis. 

Therefore, it is vitally important to measure the gas mass and the stellar mass using the same overdensity radius as the weak-lensing mass and investigate the gas mass, stellar mass, and total baryon mass fractions in a self-consistent manner.

The paper investigates the gas and stellar mass fractions of the 136 XXL clusters, which are consistently measured within the overdensity radii $r_{500}$ determined by weak-lensing masses \citep{2020ApJ...890..148U}. 
We employ Bayesian forward modeling, following \citet{Sereno16} and \cite{2020MNRAS.492.4528S}, to study multivariate scaling relations. 
The error covariance matrix, including error correlation induced by the same apertures of the weak-lenisng masses, is fully propagated into the scaling relation analysis. Our analysis considers both selection effect and weak-lensing mass calibration.
Data analysis is described in Sec.\ref{sec:data}, results are presented in Sec.\ref{sec:result}, discussed in Sec.\ref{sec:dis}, and summarized
 in Sec.\ref{sec:sum}. The paper adopts cosmological parameters of $\Omega_{m,0}=0.28$, $\Omega_{\Lambda,0}=0.72$ and $H_0=70\,{\rm km\,s^{-1}\,Mpc^{-1}}$.

\section{Data Analysis} \label{sec:data}

\subsection{XXL cluster sample} \label{sec:sample}

The parent cluster sample consists of spectroscopically confirmed X-ray-selected systems of class C1 and C2 drawn from the XXL second data release (DR2) catalog \citep[see details;][]{2018A&A...620A...5A}. 
The C1 class has a high purity rate (the fraction of false detections $\sim5\%$) with respect to
spurious detections or blended point sources \citep{2018A&A...620A...5A}.
The C2 class consists of fainter, hence less-well characterized objects and allows up to $50\%$ contamination by misclassified point sources.  On average, the C2 clusters have lower masses than the C1 at fixed redshift.
The C2 clusters used in this study are those clusters that could be spectroscopically confirmed by using currently available galaxy redshifts (from the literature and the XXL spectroscopic surveys). Hence, the C2 selection function is, strictly speaking, currently undefined.
This paper uses the 83 C1 and 53 C2 spectroscopically confirmed clusters (for a total of 136 clusters) found in the region of overlap between the HSC-SSP and XXL surveys ($25$ deg$^2$).  
This is the same sample definition used in \citet{2020ApJ...890..148U}.
\citet{2020ApJ...890..148U} measured weak-lensing masses for the 136 XXL clusters using the HSC-SSP shape catalog \citep{HSCWL1styr,Mandelbaum18}. Only galaxies satisfying the full-color and full-depth criteria from the HSC galaxy catalogue were used for precise shape measurements and photometric redshift estimations.  Background galaxies behind each cluster are securely selected by their photometric redshift probability distribution, following \citet{Medezinski18}. The weighted number
density of background source galaxies is $n_{\rm gal}\simeq22.1\,{\rm arcmin}^{-2}$.
Masses are estimated from posterior probability distributions obtained assuming a Navarro--Frenk--White (hereafter, NFW) density profile \citep{NFW96}. 
The weak-lensing mass range covers from group scales $M_{500}^{\rm WL}\sim 10^{13}\Msol$ to cluster scales $\sim 6\times 10^{14}\Msol$. 
Including the upper bound of the weak-lensing mass uncertainties, the sample reaches $10^{15}\Msol$.  
Since the weak-lensing (WL) mass measurement for a low mass cluster of $\mathcal{O}(10^{13}\Msol)$ is noisy (with a median weak-lensing $S/N$ of $1.1$), \citet{2020ApJ...890..148U} validated the bias and scatter of the weak-lensing mass measurements as a function of true mass through numerical simulations and found a mild underestimation weakly depending on the halo mass.  This study uses $M_{500}^{\rm WL}$ and $r_{500}^{\rm WL}$ measurements and weak-lensing mass calibration from \citet{2020ApJ...890..148U}, which is described in Sec \ref{subsec:wl_calib}. We use the XXL centers as cluster centers \citep{2020ApJ...890..148U}.

To summarize, we use the 136 XXL clusters with 83 C1 and 53 C2 clusters. 
The weak-lensing mass range is $M_{500}^{\rm WL}\sim 10^{13}-6\times10^{14}M_\odot$.
The redshift range covers an interval from $0.031$ to $1.033$. The average and median redshifts for the entire, C1, and C2 samples are $\langle z_c \rangle=0.38,0.34,0.45$ and $z_{c,\rm med}=0.30,0.30,0.43$, respectively.

\subsection{Stellar mass estimation} \label{subsec:Ms_est}

We estimate the stellar masses, $M_*$, of red cluster member galaxies using the photometric data of the HSC-SSP Survey S19A \citep{HSC2ndDR} and the Sloan Digital
Sky Survey \citep[SDSS DR16;][]{2020ApJS..249....3A} as the supplementary photometric data.
We first select red galaxies from the color-magnitude plane as a function of cluster redshift, following \citet{2018PASJ...70S..24N} and \citet{2019PASJ...71...79O}. We use the stellar population synthesis model of \citet{2003MNRAS.344.1000B} to estimate stellar masses from the Wide-layer depth grizy-band photometry at a given  cluster redshift ($z_c$). We adopt a single instantaneous burst at the formation redshift $z=z_f$ \citep{Oguri18}.
We assume $z_f=3$ and the Chabrier IMF \citep{2003PASP..115..763C}. When we use the Salpeter IMF \citep{1955ApJ...121..161S}, the masses are higher by a factor of $\sim1.5$ than those obtained with the Chabrier IMF \citep[e.g.][]{2007A&A...474..443P}. We combine them with spectroscopically identified galaxies selected by a slice of  $|z-z_c|<0.01(1+z_c)$ from public spectroscopic redshifts in the HSC-SSP Survey region \citep{2019ApJS..240...23A,2014ApJS..214...24S,2016ApJS..225...27M,2011ApJ...741....8C,2018A&A...609A..84S}. 
The HSC-SSP photometric data of some bright galaxies, such spectroscopically identified galaxies located in nearby clusters at $z\sim0.1$, are too bright for the 8.2m Subaru telescope to be saturated \citep{HSC1stDR,HSC2ndDR}. Some of them are flagged as saturated. We use complementary SDSS photometry for the missing galaxies. 
Moreover, we visually inspect whether large bright galaxies are missing in the catalog and add them if their visual properties are similar to those of galaxies in the catalog. 
The number of additional galaxies is only about twenty in the whole cluster sample.
We use \texttt{cmodel} magnitudes \citep{2001ASPC..238..269L} from  the HSC-SSP \citep{Haung18HSC,2018PASJ...70S...5B} and SDSS photometric data \citep{2004AJ....128..502A}, which is a linear-combination magnitude derived by the exponential and the de Vaucouleurs fits, and correct them with extinction. 
We confirm that the stellar masses estimated by the HSC and SDSS data agree with each other. The \texttt{cmodel} magnitude is a good total flux indicator to use as a universal magnitude for all types of objects  \citep[e.g.][]{2001ASPC..238..269L,2004AJ....128..502A,Haung18HSC,2018PASJ...70S...5B}.
However, it does not effectively include fluxes from the outer regions of massive galaxies, such as BCGs, where a diffuse intracluster light \citep[ICL; e.g.][]{2018MNRAS.475..648P} is dominant and accounts for some fractions of the stellar mass \citep[e.g.][]{2018MNRAS.475.3348H}. We discuss this component in Sec. \ref{subsubsec:blueICL}.

We then sum up stellar masses of galaxies within the projected, weak-lensing overdensity radii, $r_{500}^{\rm WL}$, of individual clusters \citep{2020ApJ...890..148U} from the XXL centers \citep{2018A&A...620A...5A}. We set the minimum stellar mass to be $10^{10}\Msol$. Since the photometric data around bright stars are masked out, 
we correct the cylindrical, total stellar mass by the area fraction ($F$) which is the ratio of the bright-star-masked area to total area within the overdensity radius.
We here assume that the galaxies are uniformly distributed. We next subtract the stellar components associated with large-scale structure environment surrounding the targeting clusters and refer to them as the background component. The background component is estimated in an annulus between 2 Mpc and 4 Mpc 
to correct the projection effect; $M_{*}^{\rm cyl}=\sum_i M_{*i}(r_i<r_{500}^{\rm WL})F - \sum_i M_{*i}(2\,{\rm Mpc}<r_i<4\,{\rm Mpc})F_b$,  where $i$ denotes the $i$-th galaxy within each region and the background component is also estimated by taking into account bright star mask corrections ($F_b$). 
When we change the background annulus to 3-5 Mpc and 1.5-3.5 Mpc, the stellar masses change only by a few percent.
We convert the cylindrical stellar masses to the the spherical stellar masses by a deprojection using the NFW profile.  
We assume that the stellar mass density profile is described 
by the best-fit NFW mass density profile \citep{2020ApJ...890..148U}. 
In the deprojection method, we separate a central BCG from satellite galaxies, where the central BCG is defined by the largest stellar mass galaxy within 200 kpc from the XXL centers. 
We multiply the cylindrical stellar mass of the satellite galaxies, $M_{*}^{\rm cyl}-M_{\rm BCG}$, by a conversion factor, $D_{\rm dpj}$, which is obtained as the ratio between the spherical NFW mass within the measurement radius and an integration of the projected NFW profile out to the measurement radius. We use the concentration parameters for individual clusters in the computation of the conversion factor. The unweighted average of the concentration parameters is $\langle c_{500}^{\rm WL}\rangle=2.8\pm1.5$. 
When we change the concentration by $\pm 1$, $D_{\rm dpj}$ varies by only $\pm 5$ percent.
We then correct the obtained spherical stellar masses by a stellar mass function to consider the incompleteness of the stellar mass caused by the minimum cut of $10^{10}\Msol$. 
We assume that the stellar mass function follows a Schechter luminosity function \citep{Schechter76} and adopt the stellar mass function of quiescent galaxies from the COSMOS Survey \citep{2013ApJ...777...18M}. We find that the correction factor is $C\simeq1$ and independent of cluster redshifts and we use the single value $C=1.0025$ for all the clusters. In short, the spherical mass estimate is described as
\begin{eqnarray}
      M_*=M_{\rm BCG}+ C D_{\rm dpj}(M_{*}^{\rm cyl}-M_{\rm BCG}) =M_{\rm BCG}+M_{\rm sat}.
\end{eqnarray}
Here, $M_{\rm sat}$ is the spherical stellar mass of the satellite galaxies. We consider both the errors of the stellar mass of individual galaxies and the errors of weak-lensing overdensity radii. However, the errors due to weak-lensing overdensity radii ($\sigma_*^{\rm err,WL}$) account for more than 90 percent of the total error budget ($\sigma_*^{\rm err}$), and thus the other error sources ($((\sigma_*^{\rm err})^2-(\sigma_*^{\rm err,WL})^2)^{1/2}$) are negligible.


\begin{table}[ht]
    \caption{Resulting regression parameters of the scaling relations between the cluster quantities ($M_{500}^{\rm WL}$, $L_X$, $M_*$, $M_{\rm BCG}$, and $M_g$) and the true mass $M_{500}$ for the 136 XXL clusters.
     The normalization, $\alpha$, and the slope, $\beta$, are defined by the linear regressions (eq. \ref{eq:append_alpha_beta}). The intrinsic scatter at a fixed true mass is represented by $\sigma_{\rm int}$. $^\dagger$ : the results using a trivariate Gaussian prior as the WL mass calibration, 
     as described in Sec. \ref{subsec:wl_calib}. The errors denote the $1\sigma$ uncertainty.}
    \begin{center}
    \scalebox{1.1}[1.1]{
    \begin{tabular}{c|rrr}   
              & \multicolumn{1}{c}{$\alpha$}                  &  \multicolumn{1}{c}{$\beta$}                  & \multicolumn{1}{c}{$\sigma_{\rm int}$} \\ \hline
        $M_{500}^{\rm WL}E(z)$     &  $-0.11_{-0.02}^{+0.02}$$^\dagger$              & $1.08^{+0.02}_{-0.02}$$^\dagger$                       & $0.21^{+0.02}_{-0.02}$$^\dagger$              \\
        $L_{X}E(z)^{-1}$             &  $0.29_{-0.13}^{+0.13}$  & $1.38_{-0.18}^{+0.27}$ & $0.73_{-0.14}^{+0.12}$\\
        $M_{*}E(z)$         &   $0.76_{-0.08}^{+0.09}$   &   $0.85_{-0.09}^{+0.12}$   &  $0.52_{-0.06}^{+0.09}$  \\
        $M_{\rm BCG}E(z)$   &  $-0.92_{-0.08}^{+0.08}$   &    $0.49_{-0.10}^{+0.11}$   &    $0.70_{-0.05}^{+0.06}$     \\
        $M_{g}E(z)$     &   $1.95_{-0.08}^{+0.08}$   &    $1.29_{-0.10}^{+0.16}$   &    $0.39_{-0.08}^{+0.08}$     \\
    \end{tabular}
    }
    \end{center}
\label{tab:chabrier_4d}
\end{table}

\begin{table*}[ht]
    \caption{Intrinsic covariance for the 136 XXL clusters ($L_X$, $M_*$ , $M_{\rm BCG}$, and $M_g$). The diagonal elements, the lower triangle elements, and the upper triangle elements represent express the intrinsic scatter ($\sigma_{\ln Y_i}$), and a pair correlation coefficient ($r_{ij}$), and an off-diagonal element of the intrinsic covariance ($r_{i,j}\sigma_{\ln Y_i}\sigma_{\ln Y_j}$), respectively. The errors and the lower bound denote the $1\sigma$ uncertainty and the $1\sigma$ lower limit, respectively.}
    \begin{center}
    \scalebox{1.2}[1.2]{
    \begin{tabular}{c|cccc}
         & $L_{X}E(z)^{-1}$ & $M_*E(z)$ & $M_{\rm BCG}E(z)$ & $M_gE(z)$  \\ \hline
        $L_{X}E(z)^{-1}$  & $0.73_{-0.14}^{+0.12}$ & $0.07_{-0.09}^{+0.14}$ & $0.08_{-0.08}^{+0.10}$ & $0.28_{-0.14}^{+0.12}$ \\
        $M_*E(z)$  &  $0.20_{-0.28}^{+0.23}$ & $0.52_{-0.06}^{+0.09}$ & $0.24_{-0.06}^{+0.07}$ & $0.04_{-0.05}^{+0.10}$ \\
        $M_{\rm BCG}E(z)$  &$0.18_{-0.19}^{+0.14}$ & $0.67_{-0.09}^{+0.06}$ & $0.70_{-0.05}^{+0.06}$ & $0.03_{-0.05}^{+0.06}$ \\
        $M_gE(z)$  & $>0.97$ & $0.24_{-0.33}^{+0.31}$ & $0.16_{-0.22}^{+0.17}$ & $0.39_{-0.08}^{+0.08}$ \\
    \end{tabular}
    }
    \end{center}
\label{tab:Sigma_int_4d}
\end{table*}

\subsection{Gas mass estimation} \label{subsec:Mg_est}

Gas masses within a fixed aperture of 500 kpc for all XXL clusters were published in DR2 \citep{2018A&A...620A...5A}. 
In the present analysis, we consider the gas mass measured within the weak-lensing overdensity radius ($r_{500}^{\rm WL}$).
Here we provide updated gas masses from the \emph{XMM-Newton} survey data for the 136 XXL clusters with available WL masses from HSC \citep{2020ApJ...890..148U}. Following \citet{2018A&A...620A...5A}, we reduce the \emph{XMM-Newton}/EPIC data using XMMSAS v13.5 and extract count images from all the available pointings in the [$0.5-2$] keV band. We use exposure maps for each observation to take the vignetting effect into account. We use a large collection of filter-wheel-closed data to extract models of the particle-induced background, and rescale the filter-wheel-closed data to match the count rates observed in the unexposed corners of each observation.
To include all the available data, we create mosaic images by combining the count images, exposure maps, and background maps of all observations.
To determine the gas masses, we follow the method presented in \citet{2020OJAp....3E..12E} and implemented in the public code \texttt{pyproffit}\footnote{ {https://github.com/domeckert/pyproffit} }. For
each cluster, 
we extract a surface brightness profile by accumulating source counts within concentric annuli around the cluster center as determined from the XXL detection pipeline \textsc{XAmin} \citep{2018A&A...620A...9F}.
We detect X-ray point sources by the XXL pipeline and mask them. The missing area is corrected to compute a surface brightness profile.
The same procedure is applied to the background maps to create a model background count profile.

We model the three-dimensional gas emissivity profile as a combination of a large number of basis functions (King profiles) to allow a wide variety of shapes. 
The emissivity profile is projected along the line of sight and convolved with the instrumental PSF to predict the source brightness in each annulus. The residual sky background is fitted jointly to the data. The total model (source + background) is fitted to the data using the Hamiltonian Monte Carlo code \texttt{PyMC3} \citep{2016ascl.soft10016S}. For more details on the reconstruction method, we refer the reader to \citet{2020OJAp....3E..12E}.

To compute the conversion between emissivity and count rate, we simulate an absorbed APEC model \citep{2001ApJ...556L..91S} using XSPEC. 
Gas temperatures are taken from the spectral analysis of \citet{2016A&A...592A...3G} and the plasma is assumed to have a universal metal abundance of $0.25Z_\odot$ \citep{1989GeCoA..53..197A}. We then relate the norm of the APEC model to the simulated source count rate,
\begin{equation}
K_{\rm APEC} = \frac{10^{-14}}{4\pi(1+z)^2 d_A^2}\int n_e n_H \, dV
\end{equation}
with $n_e$, $n_H$ the number density of electrons and ions, respectively. Finally, the gas mass
within the WL overdensity radii, $M_g\equiv M_{\rm gas, 500}$, is determined by integrating the reconstructed gas density profile
within the weak-lensing $r_{500}$,
\begin{equation}
M_{g} = \int_0^{r_{500}^{\rm WL}} 4\pi r^2 \rho_{\rm gas}(r)\, dr
\label{eq:mgas}
\end{equation}
with $\rho_{\rm gas} = \mu m_p (n_e + n_H)$ the gas mass density, $\mu=0.6$ the mean molecular
weight, and $m_p$ the proton mass. To propagate the uncertainties to the gas mass, for each set of
posterior parameter values we integrate $M_{g}$ using eq. \ref{eq:mgas} and randomize the
value of the overdensity radius according to the posterior $r_{500}^{\rm WL}$ distribution (see Sec \ref{subsubsec:errormatrix}).

\subsection{Multivariate scaling relations} \label{subsubsec:multi_scaling}

We simultaneously estimate the multivariate scaling relations between weak-lensing mass ($M_{500}^{\rm WL}$), X-ray luminosity ($L_X$), stellar mass ($M_*$), BCG mass ($M_{\rm BCG}$), and gas mass ($M_{g}$) by a Bayesian framework considering both selection effect and  regression dilution bias. 
The details are described in Appendix \ref{app:a} and \ref{app:b}. 
For the description of this approach see \citet{Sereno16} and \citet{2020MNRAS.492.4528S}. 
We use the natural logarithm ($\ln$) for observables and their intrinsic scatter. We adopt the Markov Chain Monte Carlo (MCMC) method.  
XXL candidate clusters are classified as C1 or C2 according to total count rate and core size to avoid contamination of point sources \citep{2006MNRAS.372..578P,2016A&A...592A...2P}.
The C1 selection can be assimilated to a surface brightness limit,
while the C2 subsample used in this paper is unconstrained (Sec. \ref{sec:sample}).

We approximately use the X-ray luminosity $L_X$ in the [0.5-2] keV band as a simple selection function instead of the two parameters of the total count rate and core size, following \citet{2020MNRAS.492.4528S}.
The DR2 catalog \citep{2018A&A...620A...5A} includes X-ray flux within 60 arcsec from the XXL centers and X-ray luminosity within $r_{500}$ determined by a mass and temperature scaling relation based on a CFHTLenS shape catalog  \citep{2016A&A...592A...4L}.
However, since the X-ray luminosities of 32 clusters out of 136 clusters are not publicly available, it does not meet the sample size of this paper.
Furthermore, to avoid the systematics inherent in the mass scaling relation and WL mass measurement, we remeasure the total X-ray luminosity in the [0.5-2] keV. We first compute the maximum detection radius ($r_{\rm max}$) of each source as the radius at which the reconstructed surface brightness is 10\% of the locally determined background brightness. We then integrate the
surface brightness within the corresponding circular area to measure the total count rate.
Finally, we use XSPEC to compute the conversion between count rate and luminosity at the
redshift of the source assuming the source spectrum is described by an absorbed APEC
model, and then we obtain the total luminosity $L_X$ which is integrated  within the maximum detection radius $r_{\rm max}$. Here, it is important that the measurement of the X-ray luminosity does not use external information but X-ray data alone. 
The maximum radius $r_{\rm max}$ is positively correlated with $r_{500}^{\rm WL}$ and the scatter is $0.11$ dex. 
When we remove 19 large-offset clusters of which $r_{\rm max}$ and $r_{500}^{\rm WL}$ differ by more than twice as large as $r_{500}^{\rm WL}$ errors, we find that the results do not change significantly. This is caused by no error correlation between $r_{\rm max}$ and $r_{500}^{\rm WL}$.
We approximately use $L_XE(z)^{-1}$ as expected by a self-similar solution because the measurement radii sufficiently covers the X-ray dominated region, where $E(z)=(\Omega_{m,0}(1+z)^3+\Omega_{\Lambda,0})^{1/2}$.
We employ the minimum X-ray luminosity as the threshold in the regression analysis (eq. \ref{eq:likelihood}). 
Since the measurement radius of $L_X$ is independent of the WL overdensity radius, there is no error correlation with other quantities. 
We introduce the natural logarithmic quantities for the observables in Bayesian inference, defined as
\begin{eqnarray}
      x&=&\ln \frac{M_{500}^{\rm WL} E(z)}{10^{14}M_\odot}, \label{eq:define_scaling_main}\\
      \bm{y}&=&\Bigr\{\ln \frac{L_{X}E(z)^{-1}}{10^{43}\,{\rm erg s^{-1}}}, \ln \frac{M_*E(z)}{10^{12}M_\odot},\ln \frac{M_{\rm BCG} E(z)}{10^{12}M_\odot}, \ln \frac{M_g E(z)}{10^{12}M_\odot}\Bigl\}. \nonumber 
\end{eqnarray}
We assume the $E(z)$ dependence expected by the self-similar solution for all the observables.
We refer to baryonic observables as $\bm{y}=\{y_l,y_*,y_{\rm BCG},y_g\}$. We express the true values of their observed quantities as capital letters, $X$ and $\bm{Y}$. The observables and true variables are related by $p(x,\bm{y}|X,\bm{Y})=\mathcal{N}(\{x,\bm{y}\}\big| \{X, \bm{Y}\}, \bm{\Sigma}_{{\rm err}})$, where $\mathcal{N}$ and $\bm{\Sigma}_{\rm err}$ denote a normal distribution and an observational error covariance matrix, respectively (Appendix \ref{app:a}). We aim at measuring the linear regression with respect to the actual quantities of $X$ and $\bm Y$ (Sec. \ref{subsec:linear_sigma}). The observational errors of $x$ and $\bm{y}$ are described by fractional errors $\sigma_x^{\rm err}=\sigma_{\rm WL}^{\rm err }$ and $\sigma_{\bm y}^{\rm err}=\{\sigma_l^{\rm err },\sigma_*^{\rm err },\sigma_{\rm BCG}^{\rm err },\sigma_g^{\rm err }\}$. The diagonal elements in $\bm{\Sigma}_{\rm err}$ are $\{(\sigma_x^{\rm err})^2,(\sigma_{\bm y}^{\rm err})^2\}$. The details of the error covariance matrix are described in Sec. \ref{subsubsec:errormatrix}.

Redshift dependence in eq. (\ref{eq:define_scaling_main}) can be obtained for self-similar evolution. Since the critical density $\rho_{\rm cr}(z)\propto H(z)^2\propto E(z)^2$ and the overdensity radius $r_{500} \propto c/H(z) \propto E(z)^{-1}$, the mass becomes $M_{500}\propto \rho_{\rm cr}(z) r_{500}^3 \propto E(z)^{-1}$. Here, $H(z)$ is the Hubble parameter at given redshift and $c$ is the light velocity. 
Thus, $M_{500}E(z)$ is independent of redshift. 
Assuming that the baryonic mass density is proportional to the critical density, that is, a constant baryon fraction against both mass and redshift, we similarly obtain $M_i E(z)$ where $i=\{*,{\rm BCG},g\}$. Since the soft-band X-ray luminosity is proportional to the square of the electron number density ($n_e$) in the volume, $L_X\propto n_e^2 (c/H(z))^3 \propto E(z)$ where $n_e\propto \rho_{\rm cr}(z)$. Therefore, $L_X E(z)^{-1}$ becomes constant against redshift. We show the result without the $E(z)$ dependence in Appendix \ref{app:d}.

\subsubsection{Observational error covariance matrix}\label{subsubsec:errormatrix}

The relationship between the actual ($X$ and $\bm{Y}$) and observed ($x$ and $\bm{y}$) quantities is expressed by a multivariate Gaussian distribution with an observational error covariance matrix, $\bm{\Sigma}_{\rm err}$. The diagonal elements of the error covariance matrix consists of the variance of the observational errors. 
The error correlation in the off-diagonal elements of the error covariance matrix is expressed by the subscript combinations of $r_{i,j}^{\rm err}\sigma_i^{\rm err}\sigma_{j}^{\rm err}$ between the $i$ and $j$ components, where $r_{i,j}^{\rm err}$ is the error correlation coefficient with $i,j=\{{\rm WL},l,*,{\rm BCG},g\}$. 
The error correlation coefficient, $r_{i,j}^{\rm err}$, describes an error propagation by the same measurement radii of the weak-lensing, stellar, and gas masses because the stellar and gas masses are computed within the spherical overdensity radii ($r_{500}^{\rm WL}$) of the weak-lensing masses \citep{2020ApJ...890..148U}.

We first estimate the error correlation coefficient, $r_{{\rm WL},*}^{\rm err}$, between the weak-lensing mass and the stellar mass.   
Since member galaxies are sparsely distributed, the errors of the weak-lensing masses randomly affect the stellar mass estimations in the individual clusters and it is difficult to independently measure individual error correlations. 
We therefore use the same correlation coefficient, $r_{{\rm WL},*}^{\rm err}$, estimated by the whole sample of clusters. 
We randomly pick up the overdensity radius by drawing values according to the posterior distributions of weak-lensing overdensity radii for the individual clusters. We correspondingly measure stellar masses within the given radii and then evaluate the error correlation coefficient by combining all the clusters. The number of realizations is 1000 for each cluster.
We find $r_{\rm{WL},*}^{\rm err}=0.873$. The error correlation between $M_{\rm BCG}$ and $M_*$ is negligible because the error of $M_*$ is mainly due to the weak-lensing overdensity radii.

Since the gas mass density is smoothly distributed, we easily obtain the correlation coefficient of measurement error $r_{\rm{WL},g}^{\rm err}$ between the weak-lensing mass and the gas mass for individual clusters. We employ the same method as the estimation of $r_{\rm{WL},*}^{\rm err}$. The average correlation coefficient for the whole sample is $\langle r_{{\rm WL},g}^{\rm err} \rangle=0.908$.

The measurement errors for the stellar mass and the gas mass are also correlated through the same overdensity radii. It is, however, difficult to measure $r_{*,g}^{\rm err}$ for individual clusters because the sparse distribution of the member galaxies makes it difficult to estimate individual error correlation coefficients.
We therefore estimate $r_{*,g}^{\rm err}$ by a trigonometric formula \citep[e.g.][]{Rousseuw94} which is derived from the definition that the determinant of covariance matrix must be positive. The expected range of $r_{*,g}^{\rm err}$ is $r_{{\rm WL},*}^{\rm err}r_{{\rm WL},g}^{\rm err}-\sqrt{1-(r_{{\rm WL},*}^{\rm err})^2}\sqrt{1-(r_{{\rm WL},g}^{\rm err})^2}\le r_{*,g}\le r_{\rm{WL},*}^{\rm err}r_{{\rm WL},g}^{\rm err}+\sqrt{1-(r_{\rm{WL},*}^{\rm err})^2}\sqrt{1-(r_{{\rm WL},g}^{\rm err})^2}$.
When we uniformly and randomly pick up values within the ranges, the average result approximates mean value, $r_{*,g}^{\rm err}=r_{\rm{WL},*}^{\rm err}r_{\rm{WL},g}^{\rm err}$.
We therefore adopt the mean value, $r_{\rm{WL},*}^{\rm err}r_{{\rm WL},g}^{\rm err}$, for individual clusters to derive the regression parameters, and then 
 incorporate the results with the lower and upper bounds of the trigonometric formula into the parameter errors. The average value for whole sample is $\langle r_{{\rm WL},*}^{\rm err}r_{{\rm WL},g}^{\rm err} \rangle=0.793$.
The lower and upper ranges of $r_{*,g}^{\rm err}$ increase the measurement error of the intrinsic correlation coefficient between $M_*$ and $M_g$ by $\sim100\%$, while the other parameters are insensitive. If we set $r_{*,g}=0$, an acceptance ratio of the MCMC chain becomes close to zero and the parameters cannot be constrained because it does not satisfy with the condition of the error correlation matrix.

\subsubsection{Linear regression and intrinsic covariance}\label{subsec:linear_sigma}

The linear regressions between the actual quantities ($X$ and $\bm{Y}$) and a true mass ($Z$) are described by  
\begin{eqnarray}
      X_Z&=& \alpha_X + \beta_X Z,  \label{eq:X_z} \\
  {\bm Y}_Z&=&{\bm \alpha} + {\bm \beta} Z, \label{eq:Y_Z} 
\end{eqnarray}
where $Z=\ln (M_{500} E(z)/10^{14}\Msol)$ is the logarithmic value of the true mass, $\alpha_X$ and $\bm{\alpha}$ are the normalizations, and $\beta_X$ and $\bm{\beta}$ are the slopes of the mass dependence. 
We consider the intrinsic covariance matrix, $\bm{\Sigma}_{\rm int}$ \citep{Okabe10c}, which describes  the statistical properties of cluster baryonic components. 
The diagonal elements of the intrinsic covariance are specified by fractional scatter $\sigma_X^2=\sigma_{\rm WL}^2$ and  $\sigma_{\bm{Y}}^2=\{\sigma_{l}^2,\sigma_{*}^2,\sigma_{\rm BCG}^2,\sigma_g^2\}$. 
The intrinsic correlation coefficient in the off-diagonal elements is expressed as the correlation $r_{i,j}$ between the $i$ and $j$ components, where $i,j=\{l,*,{\rm BCG},g\}$. Since it is difficult to constrain the intrinsic correlation coefficient associated with the weak-lensing mass, we fix $r_{{\rm WL},i}=0$.  We use flat prior for the parameters.

We also consider a single Gaussian distribution of $p(Z)$ as a parent population of $Z$ in the Bayesian analysis in order to correct both regression dilution effect and selection effect (Appendix \ref{app:a}). The parent population $p(Z)$ is  simultaneously determined by the scaling relation between the total X-ray luminosity and the mass, where the X-ray luminosity is approximately the tracer of the cluster finder. It also can be determined by weak-lensing masses with the mass calibration, which is discussed in Sec \ref{subsubsec:Pz}.
Due to the cosmological dimming of X-ray emission, we expect that more massive clusters can be found at higher redshift \citep{2020MNRAS.492.4528S}. We therefore introduce a redshift dependence of the parent population, $p(Z,z)={\mathcal N}(\mu_Z(z),\sigma_Z(z))$, of which the mean and standard deviation are described by
\begin{eqnarray}
     \mu_Z(z)&=&\mu_{Z,0}+\gamma_{\mu_Z}\ln E(z), \\ 
     \sigma_Z(z)&=&\sigma_{Z,0}E(z)^{\gamma_{\sigma_Z}},
\end{eqnarray}
where $\gamma_{\mu_Z}$ and $\gamma_{\sigma_Z}$ are the redshift dependence of the mean and standard deviation, respectively. The parameters of $p(Z,z)$ are non-informative, hyper-parameters and simultaneously derived by the Bayesian analysis. Thus, the result of multivariate scaling relations is independent of the cluster number counts and of the cosmological parameters.
We note that $\sigma_Z(z)$ is important to accurately determine the slopes by considering the regression dilution effect (Appendix \ref{app:a} and \ref{app:b}).
Although we tried to fit with double Gaussian distributions of the $Z$ distribution, we could not constrain the parameters of the second Gaussian component. Thus, the single $p(Z,z)$ is sufficient for this analysis.
Other possibilities including no-redshift evolution of $p(Z)$ will be discussed in Sec. \ref{subsubsec:Pz}.

\subsubsection{Weak-lensing mass calibration}\label{subsec:wl_calib}

The parameters, $\alpha_{\rm WL}$, $\beta_{\rm WL}$, and $\sigma_{\rm WL}$, describe our weak-lensing mass calibration.
Weak-lensing mass estimates for individual clusters are scattered from their true values because of their non-spherical halo shape, substructures, and surrounding large-scale structure \citep[e.g.][]{Hoekstra03,Becker11,Oguri11b,Okabe16,2020A&ARv..28....7U}. Moreover, even when averaged over many clusters, their ensemble mass estimates can be biased, if the true mass profiles deviate from the assumed profile \citep[e.g.][]{2020ApJ...890..148U}. \citet{2020ApJ...890..148U} validated their weak-lensing mass estimates for cluster and group scales using both cosmological numerical simulations \citep{2017MNRAS.465.2936M,2018MNRAS.476.2999M} and analytical NFW models, and found that the weak lensing mass bias weakly depends on true masses. 
In our multivariate regression analysis, we consider the bias and the scatter between the weak-lensing mass, $M_{500}^{\rm WL}$, and the true mass, $M_{500}$.

\citet{2020ApJ...890..148U} and \citet{2020MNRAS.492.4528S} only accounted for the
$\pm 5\%$ calibration uncertainty due primarily to observational
systematics in their observable--mass scaling relations. In the mass
forecasting for the $M_{500}-T_{X}$ relation of \citet{2020ApJ...890..148U}, they applied an additional constant mass-modeling bias correction of $-11\%$ evaluated at
the mean mass scale of the XXL sample.

\begin{figure}
    \includegraphics[width=\hsize]{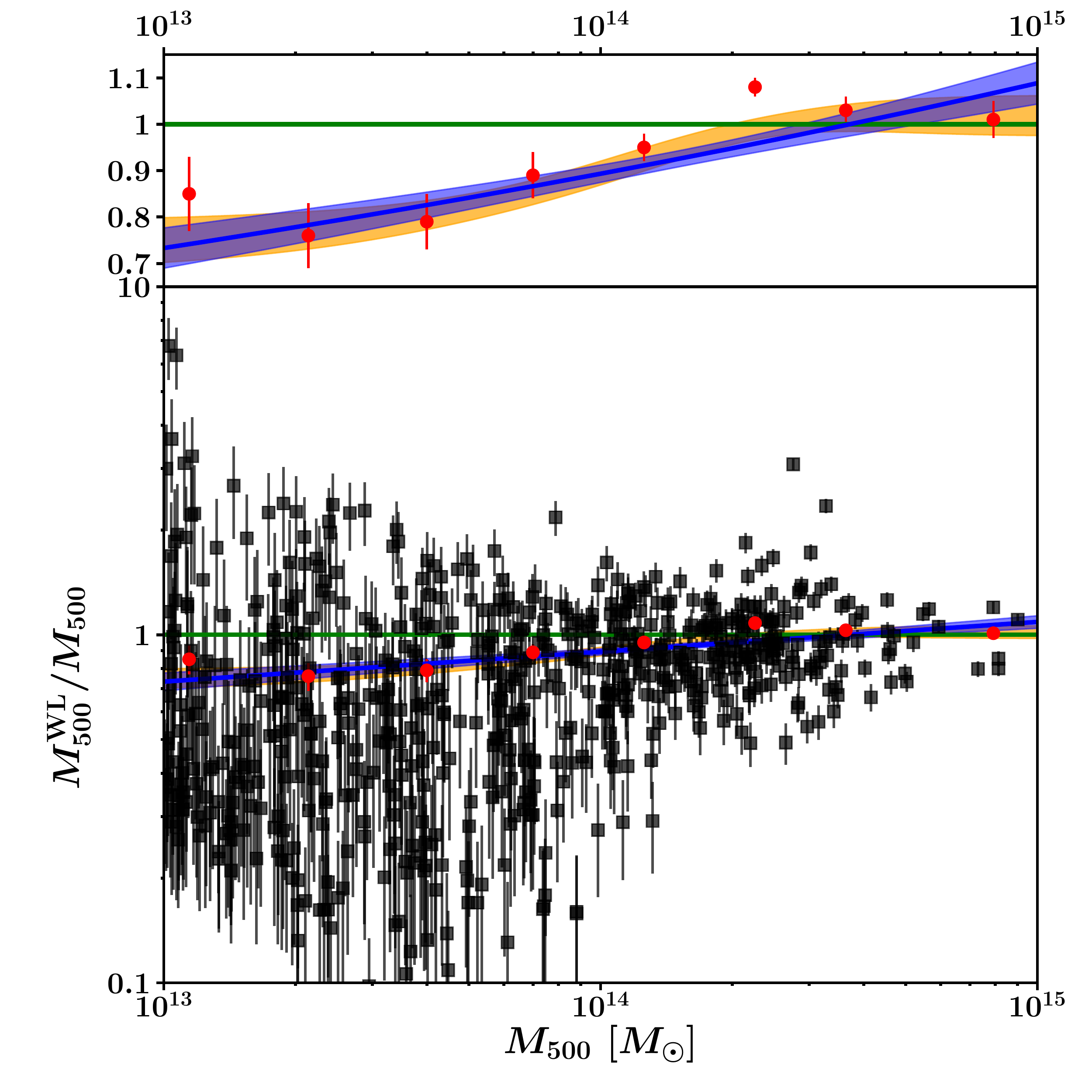}
    \caption{Weak-lensing mass calibration. 
    Red circles denote the the average mass bias at each mass bin from \citet{2020ApJ...890..148U}.
    The solid blue line and regions denote the estimate and $1\sigma$ uncertainty for the power-law relation ($\ln M_{500}^{\rm WL}=\alpha_{\rm WL} + \beta_{\rm WL} \ln M_{500})$, respectively. The orange region is the $1\sigma$ uncertainty for the function of $M_{500}^{\rm WL}/M_{500}=a \tanh M_{500}/b+c$. The green horizontal line expresses no bias.
    The black squares are the results of simulated clusters \citep{2020ApJ...890..148U}. The top and bottom panels display data and result on a linear scale and a logarithmic scale, respectively. The $\tanh$ form better describes the data. 
    }
    \label{fig:masscalib}
\end{figure}

We characterize the mass dependence of the NFW weak-lensing
mass estimates, or the $M_\mathrm{500}^{\rm WL}$--$M_{500}$
relation, using the results of \citet{2020ApJ...890..148U} based on synthetic
weak-lensing observations of 639 cluster halos in the dark-matter-only
run of BAHAMAS simulations \citep{2017MNRAS.465.2936M}. 
As shown in Figure \ref{fig:masscalib}, the mass bias increases with true mass in the regime of low masses and it is nearly constant in the high mass range ($M_\mathrm{500} \simgt 10^{14}M_\odot/h$). This can be approximated with a $\tanh$ functional form. We fit the data with a functional form of $M_{500}^{\rm WL}/M_{500}=a \tanh M_{500} /b +c$ and the intrinsic scatter of $\sigma_{\rm WL}$ . We find $a=0.29\pm0.06$, $b=1.57_{-0.48}^{+0.97}$, $c=0.73_{-0.06}^{+0.06}$, and $\sigma_{\rm WL}=0.21\pm0.02$ ( orange region in Figure \ref{fig:masscalib}).

However, the mathematical formulation in the regression analysis (Appendix \ref{app:a}) requires a power-law relation between the true mass and weak-lensing mass, or a linear relation between their logarithmic quantities. 
We here assume eq. (\ref{eq:X_z}) and find $\alpha_{\rm WL}=\ln (0.89\pm0.02)$, $\beta_{\rm WL}=1.09\pm0.02$ and $\sigma_{\rm WL}=0.21\pm0.02$
for the mean relationship between the weak-lensing mass and the true mass, as represented by the blue line in Figure \ref{fig:masscalib}. The result agrees with that of the $\tanh$ function within the $1\sigma$ uncertainty at $M_{500}E(z) \simlt 1.5\times10^{15}\Msol$.

We use a trivariate Gaussian distribution of ${\mathcal N}_{\rm 3D}(\alpha_{\rm WL},\beta_{\rm WL}, \ln \sigma_{\rm WL})$ as a prior for the weak-lensing mass calibration. 
The covariance matrix in ${\mathcal N}_{\rm 3D}$ employs the error covariance matrix of the linear regression in the power-law mass calibration. Therefore, all the mass calibration uncertainties  are propagated into the results. We discuss a case of the $\tanh$ function in Sec. \ref{subsubsec:dis_WLcalib}.

\section{Results}\label{sec:result}

\subsection{Normalization and slopes of scaling relations}\label{subsec:sc_result}

The Bayesian framework straightforwardly derives the normalization, slopes, and intrinsic covariance of the multivariate scaling relations. The number of parameters is $25$ including 4 hyper-parameters. 
We use biweight estimates of marginalized posterior distributions as the parameter estimates.

\begin{figure} 
    \includegraphics[width=\hsize]{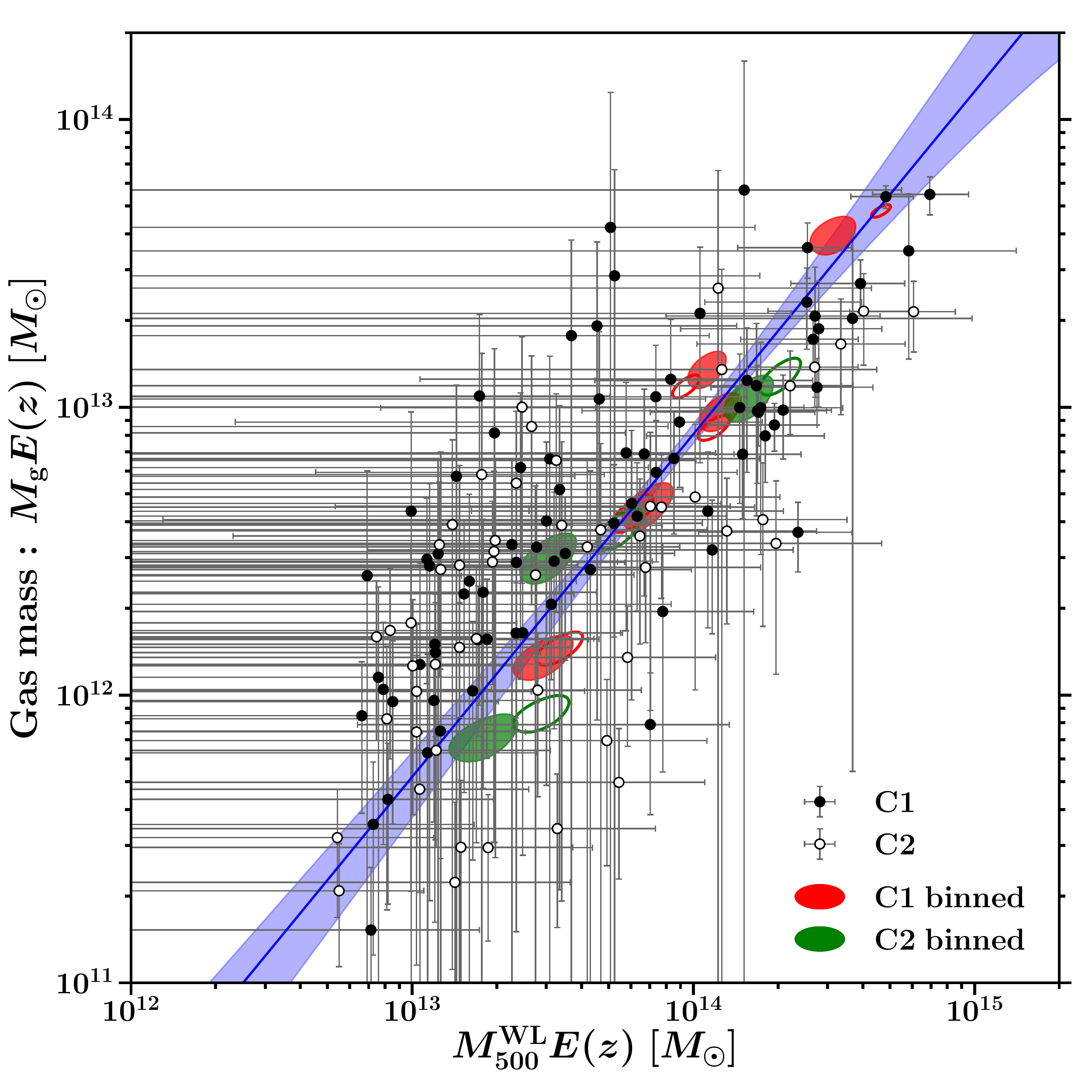}  
    \caption{The gas mass and weak-lensing mass relation. The solid blue line and region denote the estimate and $1\sigma$ uncertainty for the scaling-relation baseline, respectively. The black and white circles are the C1 and C2 subsamples, respectively. The red and green ellipses represent the 68 percent confidence levels for the stacked quantities of the subsamples of the C1 and C2 clusters which are sorted by the X-ray luminosity, respectively. Filled and open ellipses are those weighted with the covariance matrix $\bm{\Sigma}=\bm{\Sigma}_{\rm err}+\bm{\Sigma}_{\rm int}$ and $\bm{\Sigma}_{\rm err}$, respectively. The stacked quantifies follow the resulting baseline well. }  
    \label{fig:mg-mwl}
\end{figure}

\begin{figure} 
\includegraphics[width=\hsize]{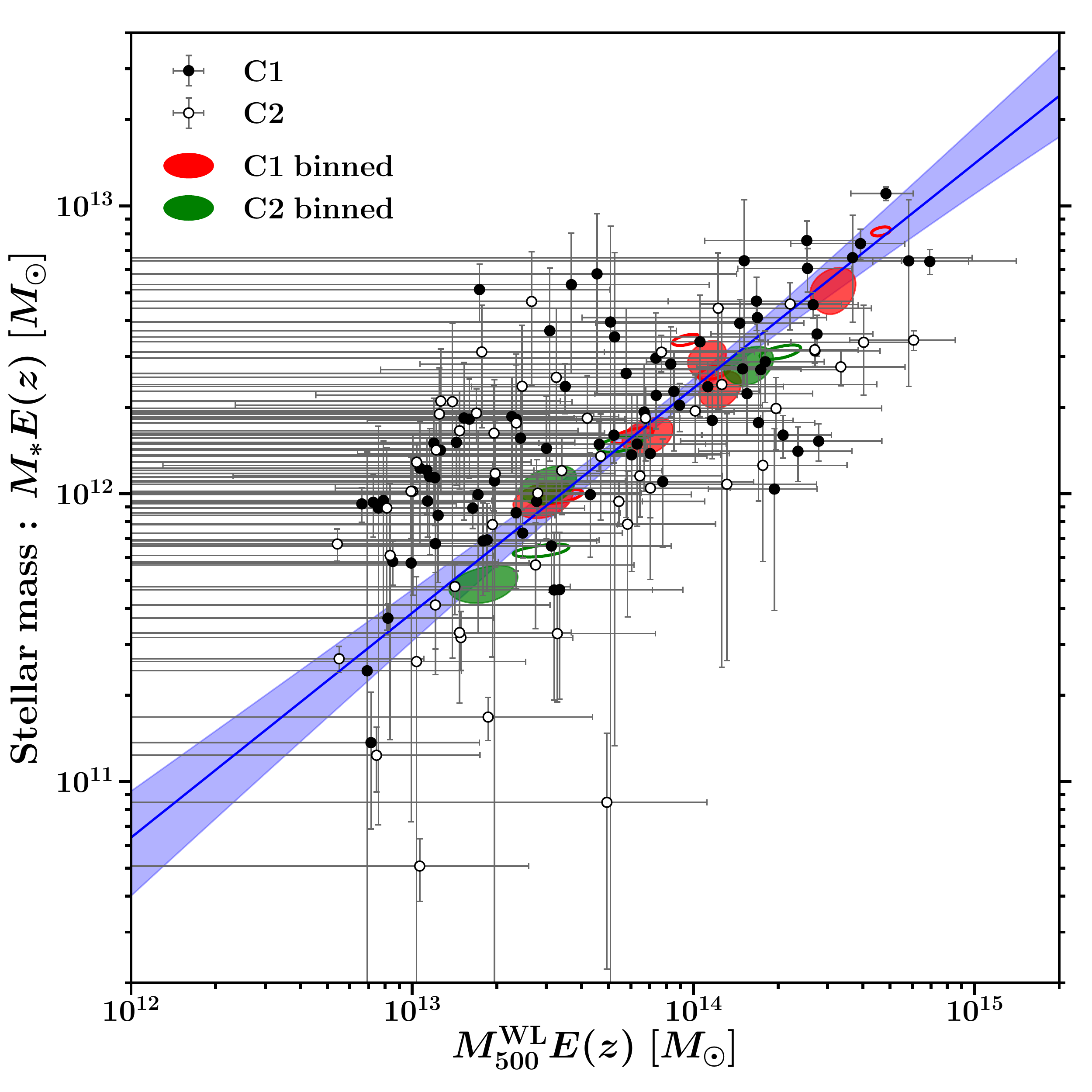}
    \caption{The stellar mass and weak-lensing mass relation. The solid blue line and region denote the estimate and $1\sigma$ uncertainty for the scaling-relation baseline, respectively.  
    The black and white circles and the red and green ellipses have the same meaning as Figure \ref{fig:mg-mwl}.}
    \label{fig:ms-mwl}
\end{figure}

\begin{figure} 
    \includegraphics[width=\hsize]{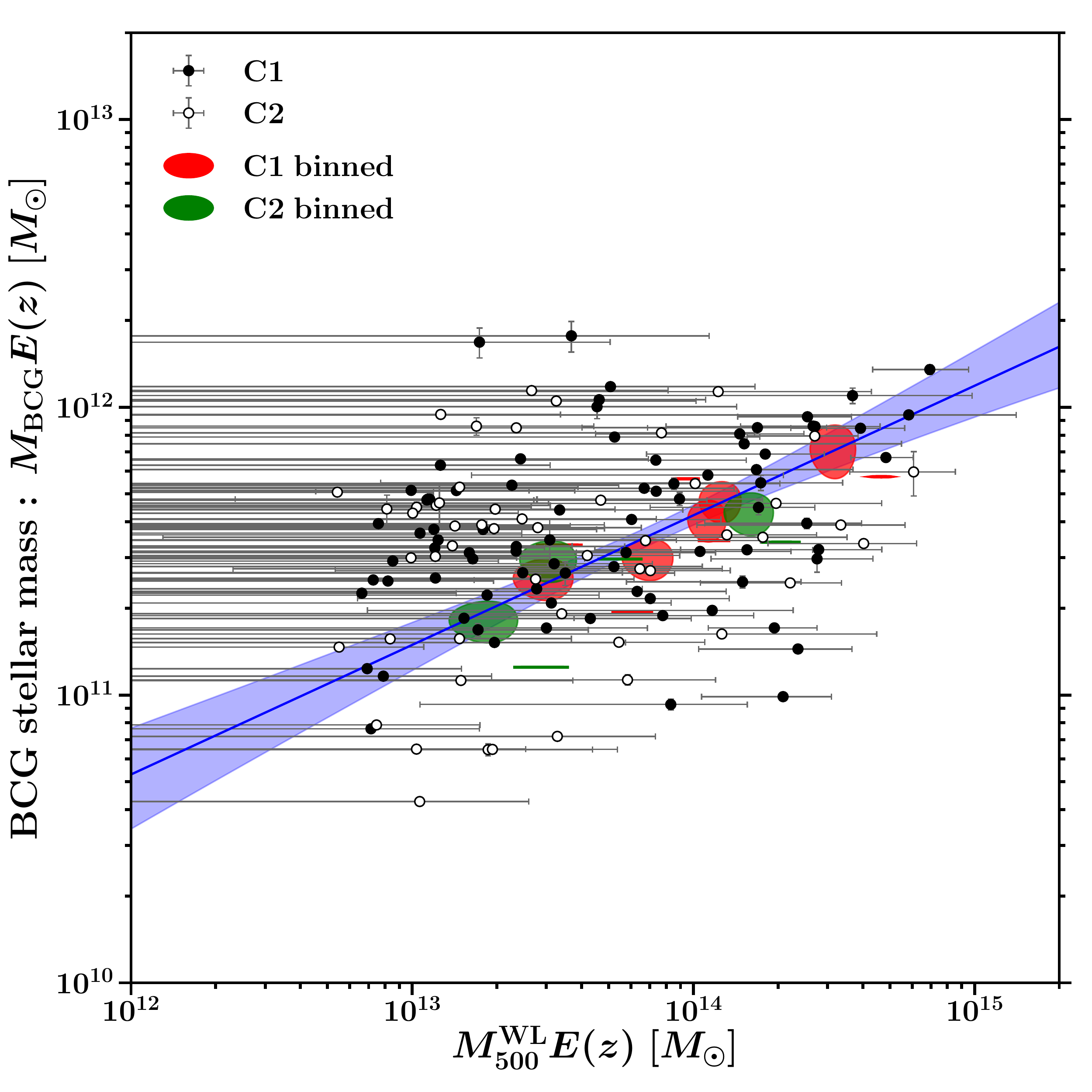}
    \caption{The BCG mass and weak-lensing mass relation. The solid blue line and region denote the estimate and $1\sigma$ uncertainty for the scaling-relation baseline, respectively. 
    The black and white circles and the red and green ellipses have the same meaning as Figure \ref{fig:mg-mwl}.
    Since the measurement errors of the BCG mass are much smaller than those of the WL masses, the shapes of the stacked quantities weighted with $\bm{\Sigma}_{\rm err}$ become lines along the $x$ axis. }
    \label{fig:mbcg-mwl}
\end{figure}

The estimated normalizations and slopes for the $L_X-M$, $M_g-M$, $M_*-M$, and $M_{\rm BCG}-M$ relations are shown in Table \ref{tab:chabrier_4d}. The posterior distribution is shown in Appendix \ref{sec:app_post} (Figure \ref{fig:Prob_slope}).
The slope of the X-ray luminosity, $1.38_{-0.18}^{+0.27}$, is $2.1\sigma$ higher than the prediction of the self-similar model ($\beta=1$).
Figures \ref{fig:mg-mwl}, \ref{fig:ms-mwl}, and \ref{fig:mbcg-mwl} show the resulting scaling relations of the gas, stellar, and BCG masses with the weak-lensing masses \citep[$M_{500}^{\rm WL}$;][]{2020ApJ...890..148U}, respectively. The scaling relations shown in the figures are described by  $\bm{Y}_Z=\bm{\alpha}-(\bm{\beta}/\beta_{\rm WL})\alpha_{\rm WL}+(\bm{\beta}/\beta_{\rm WL}) X_Z$.
For comparison, we also plot the direct observables and the stacked observables sorted by the X-ray luminosity. We did not fit using the stacked observables.
We stack 18 clusters in each subsample in ascending order of the C1/C2 X-ray luminosity, and the numbers of the remaining C1 and C2 clusters in the highest luminosity subsamples are 11 and 17, respectively. Since the stacked quantities are sorted by the X-ray luminosity, the subsample grouping is independent of any observables in the $x-$axis of Figures \ref{fig:mg-mwl}-\ref{fig:mbcg-mwl} or the $y-$axis of Figures \ref{fig:ms-mwl} and \ref{fig:mbcg-mwl}.
The stacked quantities are computed by
\begin{eqnarray}
      \langle \bm{v}\rangle =\Big(\sum_n (\bm{\Sigma}_{n})^{-1}\Big)^{-1}\sum_n (\bm{\Sigma}_{n})^{-1} \bm{v}_n
\end{eqnarray}
where $\bm{v}=\{x,\bm{y}\}$, $\bm{\Sigma}$ is the error matrix, $\bm{\Sigma}_{\rm err}$, or the composition matrix of $\bm{\Sigma}_{\rm err}+\bm{\Sigma}_{\rm int}$ for $\bm{v}$, and $n$ is the $n$-th cluster. 
The mean observables, weighted with the error matrix, show some scatter around the scaling-relation baselines, which exhibits intrinsic scatter. Such a feature is visible especially in the $M_{\rm BCG}-M$ relation with the largest intrinsic scatter.
We thus weight them with the composition matrix to compare with the baselines shown in blue in Figures \ref{fig:mg-mwl}-\ref{fig:mbcg-mwl}, and find that the stacked quantities are in good agreement with the baselines.
It also indicates the consistency of Bayesian inference among ${\bm \alpha}$, ${\bm \beta}$, and ${\bm \Sigma}_{\rm int}$ to explain the data.

We find that the slopes in the $M_g-M_{500}$ and $M_*-M_{500}$ relations are, $1.29_{-0.10}^{+0.16}$, and  $0.85_{-0.09}^{+0.12}$, steeper and shallower than the self-similar predictions ($\beta=1$), respectively. 
The significance levels of the deviations from unity are $\sim3\sigma$ and $1.5\sigma$, respectively. 
We find a shallower slope, $\beta_{\rm BCG}=0.49_{-0.10}^{+0.11}$, in the $M_{\rm BCG}-M_{500}$ relation, 
which indicates that the BCG stellar mass has only a weak dependence on the halo mass.

\subsection{Parent Population}

The resulting regression parameters for the parent population are
$\mu_{Z,0}=-1.02_{-0.20}^{+0.20}$, $\gamma_{\mu_Z}=3.53_{-0.62}^{+0.68}$,
$\sigma_{Z,0}=1.21_{-0.24}^{+0.26}$, and
$\gamma_{\sigma_Z}=-2.38_{-1.00}^{+0.82}$.
The mean mass and the standard deviation of the parent population increases and decreases with increasing redshift, respectively. 
Thus, the more massive clusters at higher redshifts are discovered by the XXL Survey,
as expected due to the X-ray dimming effect. The mass distributions at lower redshift is broader than those at higher redshift, indicating that it is easier to find clusters from a broad mass range at lower redshift.

\subsection{Baryon fractions}\label{subsec:fb}

\begin{figure} 
 \includegraphics[width=\hsize]{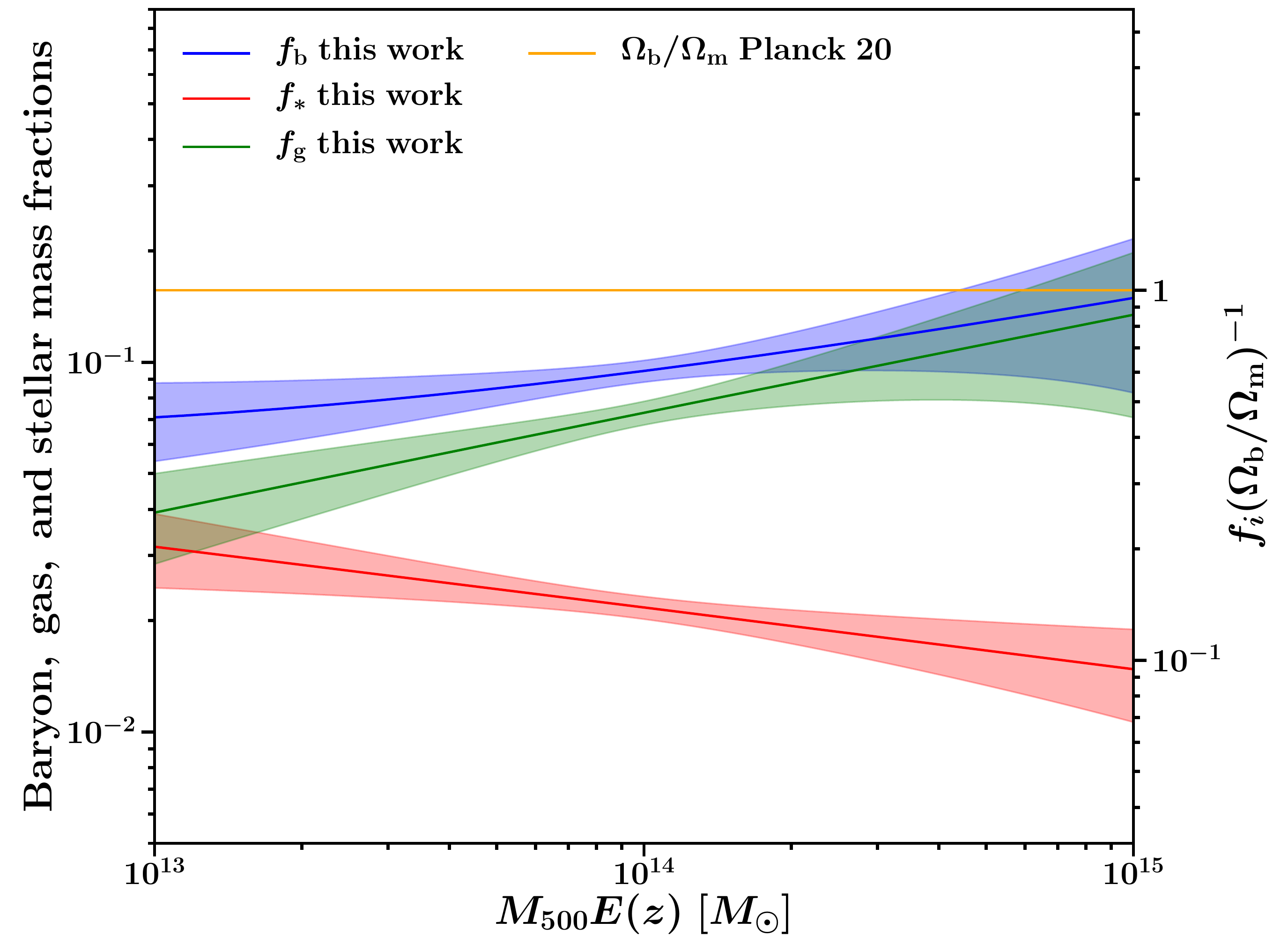}
    \caption{Baryon (blue), gas (green), and stellar (red) mass fractions as a function of the halo mass. The shade regions are the $1\sigma$ uncertainty. The orange horizontal line is $\Omega_{b}/\Omega_m$ \citep{2018arXiv180706209P}. 
    }
    \label{fig:fb}
\end{figure}

We convert the resulting scaling relations to the baryon, gas and stellar fractions as a function of the true halo mass ($M_{500}$) not the weak-lensing mass ($M_{500}^{\rm WL}$), $f_i=M_{i}(<r_{500})/M_{500}(<r_{500})$, where $i=\{b,g,*\}$ and $r_{500}$ is the overdensity radius of the true mass. Since the baselines, ${\bm Y}_Z(<r_{500}^{\rm WL})$, are computed by using $M_g(<r_{500}^{\rm WL})$ and $M_*(<r_{500}^{\rm WL})$ measured within the WL overdensity radii $r_{500}^{\rm WL}$, we convert to those measured within $r_{500}$. The details are described in Appendix \ref{app:aperture_cor}. 
The aperture correction depends on the baryonic mass density slope, the mass calibration, and the true mass (eq. \ref{eq:aperture_cor}).  
As for the stellar mass profile, we assumed that the stellar mass density profile follows the dark matter profile with the average concentration parameter $\langle c_{500}^{\rm WL}\rangle=2.8$ (Sec. \ref{subsec:Ms_est}). 
We assumed the King model of the electron number density follows $n_e\propto r^{-3\beta}$ with $\beta=2/3$ outside gas cores (Sec. \ref{subsec:Mg_est}). 
The stellar mass normalization with the aperture correction becomes $\sim 1.05$, $\sim 1.02$, and $\sim 0.99$ times that without the correction at $M_{500}E(z)\sim 10^{13} M_\odot$, $\sim 10^{14} M_\odot$, and $\sim 10^{15} M_\odot$, respectively. As for the gas mass, the aperture correction changes the normalization by $\sim 1.09$,  $\sim 1.04$, and $\sim 0.97$ times at $M_{500}E(z)\sim 10^{13} M_\odot$, $\sim 10^{14} M_\odot$, and $\sim 10^{15} M_\odot$, respectively.
Figure \ref{fig:fb} shows the resulting baryon, gas and stellar fractions.
Since the power-low mass calibration is validated in the true-mass range of $10^{13}M_\odot\simlt M_{500}E(z) \simlt 10^{15}M_\odot$, the lower and upper bounds of the $x$-axis in Figure \ref{fig:fb} are set to be $10^{13}M_\odot$ and $10^{15}\Msol$, respectively. 
It fully covers the true mass population at $0 \simlt z\simlt 1$.
We do not show the observables in the same figure because the quantity in the $x$-axis is not the weak-lensing mass but the true mass. 
Although the true masses can be statistically calculated by the mean relationship of the weak-lensing mass calibration (eq. \ref{eq:X_z}) and its intrinsic scatter, an actual weak-lensing mass bias or true mass of each cluster is unclear.

The uncertainties in Figure \ref{fig:fb} fully take into account the error covariance matrix of the linear regressions. The gas mass fraction, $f_g=M_g(<r_{500})/M_{500}$, increases as the halo mass increases, reaching 90 percent of $\Omega_b/\Omega_m=0.1564\pm0.0016$ \citep{2018arXiv180706209P} at $M_{500}E(z)\sim 10^{15}M_\odot$.
In contrast, the stellar mass fraction, $f_*=M_*(<r_{500})/M_{500}$, decreases as the halo mass increases. 
These treads are the same as the slope deviations from unity in the scaling relations (Sec. \ref{subsec:sc_result}).

The total baryon mass fraction, $f_b=f_g+f_*$, is $\sim 50$ percent of $\Omega_b/\Omega_m$ at $M_{500}E(z)\sim10^{13}M_\odot$, 
$\sim 60$ percent at $\sim10^{14}M_\odot$, and $\sim 100 \%$ at $\sim 10^{15}M_\odot$. The mass-dependent slope of $f_b$ on group scales of $\simlt 10^{14}M_\odot$ is less steep than that on massive clusters of $\simgt 5\times10^{14}M_\odot$.

When we use the Salpeter IMF, the baryon and stellar mass fractions at $\sim 10^{13}\Msol$ are $\sim 1.2$ and $\sim 1.5$ times higher than those derived by the Chabrier IMF, respectively. At $\sim10^{14}\Msol$, the baryon fraction increases only by $\sim1.1$ times. The overall trends do not significantly change by a choice of the IMF.

\subsection{BCG mass to total stellar mass ratio}

\begin{figure} 
\begin{center}
    \includegraphics[width=\hsize]{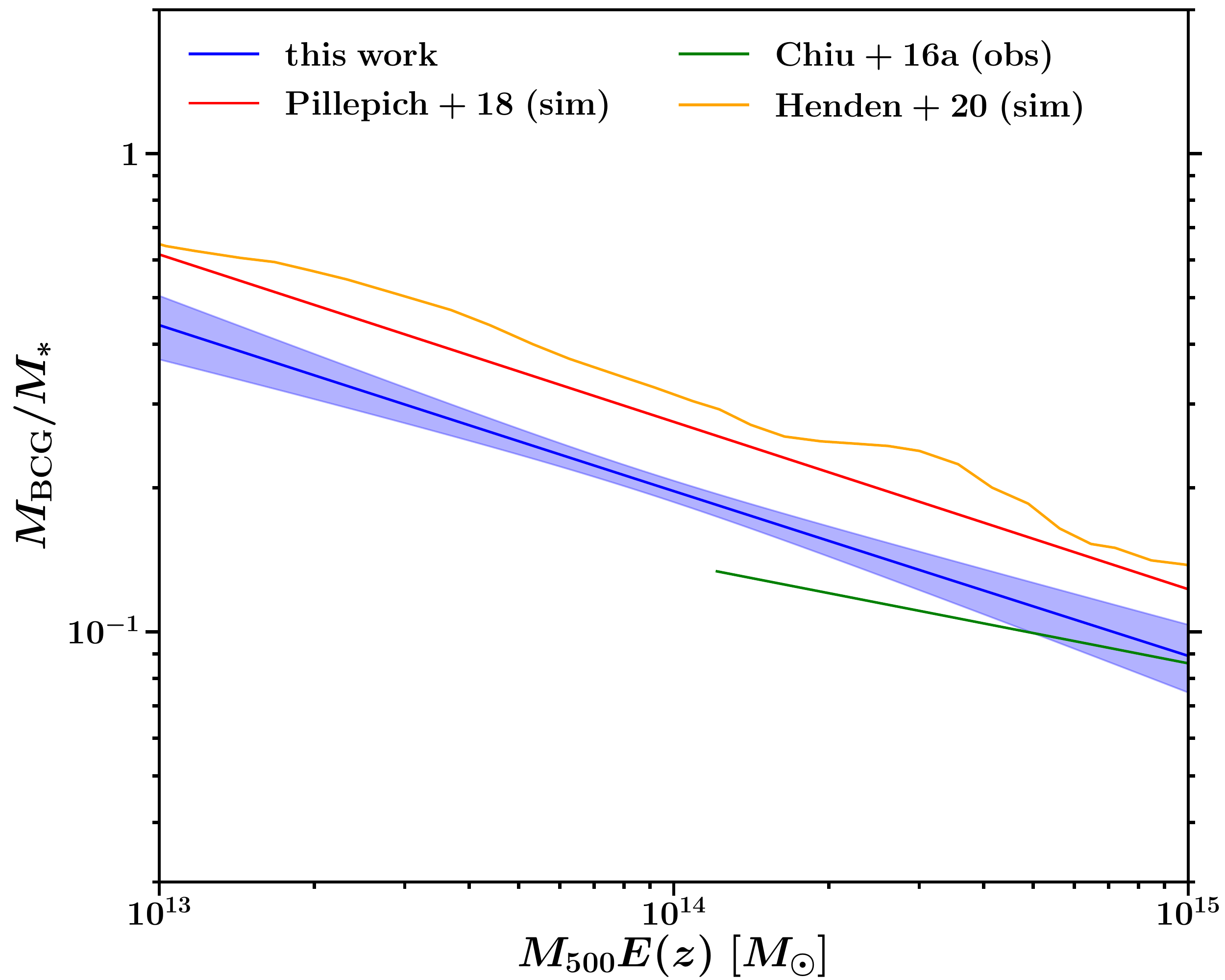}    
    \end{center}
    \caption{The BCG mass to total stellar mass ratio as a function of halo mass. 
The solid red and orange lines are the simulations of \citet{2018MNRAS.475..648P} and \citet{2020MNRAS.498.2114H}, respectively. We recompute $M_{\rm BCG}/M_{*}$ from the  $M_*-M$ and $M_{\rm BCG}-M$ lines \citep{2016MNRAS.455..258C}, shown as the solid green line.}
    \label{fig:fbcg_ms}
 
\end{figure}

We compute the BCG stellar mass to total stellar mass ratio as a function of the true mass (Figure \ref{fig:fbcg_ms}) from the $M_{\rm BCG}-M$ and $M_*-M$ relations.
Since the BCG stellar mass measurement is independent of the weak-lensing overdensity radius, it is independent of the aperture correction.
Since the errors of the linear regressions are correlated with each other, the error covariance matrix is taken into account to compute the errors of the ratio. The fraction of the BCG in the total mass is at most $\sim10$ percent at $M_{500}E(z)\sim 10^{15}\Msol$ and $\sim 20$ percent at $\sim 10^{14}\Msol$. However, the fraction at $\sim 10^{13}\Msol$ accounts for $\sim 45$ percent. Therefore, the BCG is a more dominant component of the stellar mass components on a group scale.

\subsection{Intrinsic covariance of baryon contents}\label{subsec:Sigma_int}

Another important property of the multivariate scaling relations is the intrinsic covariance.
Table \ref{tab:Sigma_int_4d} describes the resulting intrinsic covariance (see also Table \ref{tab:chabrier_4d}). The diagonal element, the lower off-diagonal element, and the upper off-diagonal element are intrinsic scatter, a pair correlation coefficient, and an off-diagonal element of the intrinsic covariance at fixed cluster mass, respectively. The posterior distribution is shown in Appendix \ref{sec:app_post} (Figure \ref{fig:Prob_cov}).

The intrinsic scatter in gas mass is $\sigma_g=0.39_{-0.08}^{+0.08}$, corresponding to $\sim0.17$ dex. The intrinsic scatter in the stellar mass, $\sigma_*=0.52_{-0.06}^{+0.09}\sim0.23$ dex, is larger than $\sigma_g$. The intrinsic scatter of the BCG mass is $\sigma_{\rm BCG}=0.70_{-0.05}^{+0.06}\sim0.30$ dex.  The largest intrinsic scatter is in BCG stellar mass, followed in order by stellar mass and gas mass; $\sigma_g<\sigma_*<\sigma_{\rm BCG}$. The scatter trend is visually confirmed in Figures \ref{fig:mg-mwl}-\ref{fig:mbcg-mwl}. The error-covariance-weighted means for the subsamples binned by the X-ray luminosity show some scatter in the scaling-relations (Figures \ref{fig:mg-mwl}-\ref{fig:mbcg-mwl}). Comparisons of numerical simulations and other observations are discussed in Sec. \ref{sec:dis}.

We find strong intrinsic correlation coefficients between stellar mass and BCG mass and between X-ray luminosity and gas mass; $r_{*,\rm{BCG}}=0.70_{-0.05}^{+0.06}$ and $r_{l,g}>0.97$. Other intrinsic correlation coefficients agree with no correlation within the errors; 
$r_{*,g}=0.24_{-0.33}^{+0.31}$, and $r_{{\rm BCG},g}=0.16_{-0.22}^{+0.17}$.
We also explored the possibility that a spurious positive or negative correlation could be caused by a finite sampling size.
We assess an accidental probability that 136 random pairs give the observed intrinsic correlation coefficient, following \cite{Okabe10c}.
The accidental probability, $\mathcal{P}(r\ge |r_{i,j}|)$, is specifically defined as follows; the correlation coefficient of the two random variables in a sample of 136 drawings is higher than the absolute value of the intrinsic correlation coefficient.
It corresponds to a probability of the null hypothesis that the two variables do not correlate with each other.
The resulting maximum $p$-values are $\mathcal{P}(r\ge |r_{*,\rm{BCG}}|)\sim \mathcal{O}(10^{-13})$,
$\mathcal{P}(r\ge |r_{*,g}|)\sim 3\times10^{-1}$, $\mathcal{P}(r\ge |r_{{\rm BCG},g}|)\sim 4\times10^{-1}$, and $\mathcal{P}(r\ge |r_{l,g}|)\sim 0$. respectively. We therefore reject a possibility of the accidental correlation between $M_{\rm BCG}$ and $M_*$ and between $L_X$ and $M_g$.

We also study the intrinsic correlation coefficient between BCG stellar mass and satellite galaxy mass defined by $M_{\rm sat}=M_*-M_{\rm BCG}$. The intrinsic correlation coefficient, $r_{\rm sat,BCG}=0.51_{-0.16}^{+0.10}$, is significant. 
We find that the intrinsic correlation coefficient between $M_{\rm sat}$ and $M_g$ is consistent with no correlation ; $r_{{\rm sat},g}=0.43_{-0.59}^{+0.30}$.

\subsection{C1 and C2 subsamples}\label{subsec:C1C2}

\begin{figure} 
\begin{center}
    \includegraphics[width=\hsize]{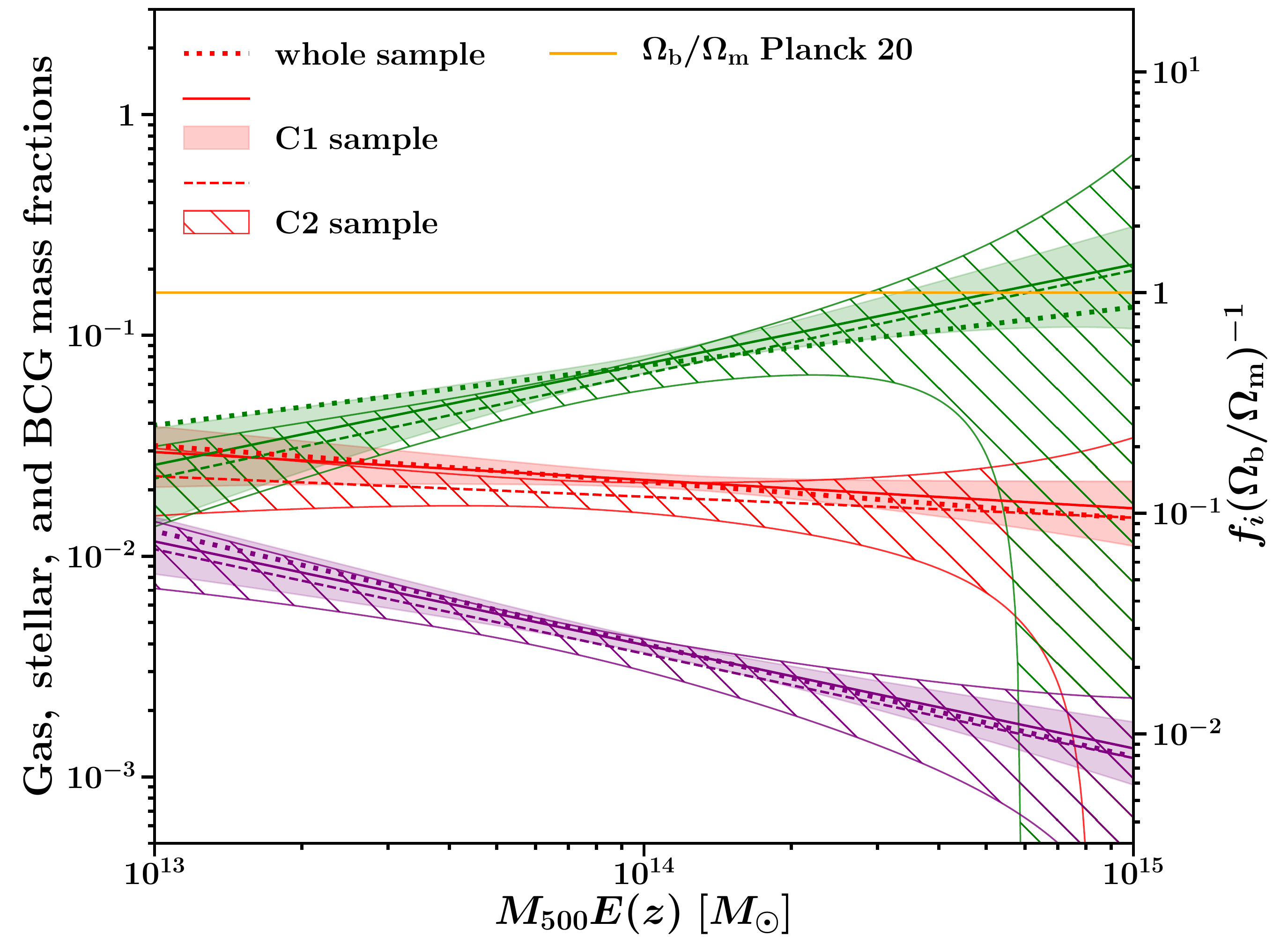}    
    \end{center}
    \caption{The gas (green), total stellar (red), and BCG (magenta) mass fractions as a function of halo mass. The solid, dashed, and dotted lines denote the Bayesian estimation for the C1 clusters, the C2 clusters, and the whole sample, respectively.
    The filled and hatched regions denote the $1\sigma$ uncertainties for the 83 C1 clusters and the 53 C2 clusters, respectively. 
    }
    \label{fig:fbcg_c1c2}
 
\end{figure}

Our sample comprises 83 C1 and 53 C2 clusters from the XXL DR2 sample \citep{2018A&A...620A...5A}. 
We here split the whole sample into the C1 and C2 subsamples. 
Since the mean luminosity for the C2 sample is lower than that for the C1 sample \citep{2018A&A...620A...5A}, we set the maximum threshold in $L_X$ as the highest X-ray luminosity among the C2 sample in the Bayesian analysis. 
Even when we remove the upper bound of the X-ray luminosity, the results do not significantly change.
The resulting regression parameters for the C1 and C2 subsamples are shown in top panel of Table \ref{tab:chabrier_sub}.
The baryon fractions for the whole sample agree with those for the C1 and C2 samples (Figure \ref{fig:fbcg_c1c2}), except for the low-mass end of the $M_g-M$ relation. The gas fractions between the whole sample and the C1 sample and between the whole sample and the C2 sample differ by $\sim1.1\sigma$ and $\sim1.5\sigma$ at $10^{13}M_\odot$, respectively. This small discrepancy is caused by the steeper C1 and C2 slopes of the $M_g-M$ relation. When we fix the unconstrained intrinsic scatter, $\sigma_g$, for the C1 and C2 samples with $\sigma_g=0.39$ obtained by the whole sample, we find that they agree within $1\sigma$. The determination of the slopes is associated with the intrinsic scatter.

The intrinsic covariances for the two subsamples are similar to those for the whole sample of clusters (top panel of Table \ref{tab:Sigma_int_sub}). In particular, we recover in both subsamples the order of the intrinsic scatters ($\sigma_g<\sigma_*<\sigma_{\rm BCG}$) and positive correlation coefficients $r_{*,\rm{BCG}}$ and $r_{l,g}$.
The intrinsic scatter of the total and BCG stellar components in the C2 sample is larger than that in the C1 sample. The differences of $\sigma_*$ and $\sigma_{\rm BCG}$ are at $3.3\sigma$ and $4.1\sigma$ levels, respectively.

\subsection{Subsample with central radio sources}\label{subsec:sub}

We split the sample into clusters with and without central radio sources associated with active galactic nucleus (AGN). 
 We search for central radio sources within $60$ arcsec from the BCGs using the Faint Images of the Radio Sky at Twenty-centimeters \citep[FIRST;][]{1997ApJ...475..479W} and TIFR GMRT Sky Survey \citep[TGSS;][]{2017A&A...598A..78I} surveys. We find central radio sources in 34 clusters. The fraction of central radio sources is 0.25 and almost constant over the redshift.  
 Average fractions to include radio sources within $60$ arcsec from random positions and random galaxies of which $z$-band magnitudes are brighter than 20 ABmag are only 0.04 and 0.09, respectively. 
  It is difficult to identify whether radio sources are associated with cluster members or not because of their extended distribution and a lack of their redshifts. A visual inspection of radio sources and optical distribution suggests that a contamination of radio overlapped at different redshifts is small. Even when we exclude two clusters whose central radio sources have a possibility to be overlapped with point sources at $z\sim1-1.5$, we find consistent results.

\begin{figure} 
\begin{center}
    \includegraphics[width=\hsize]{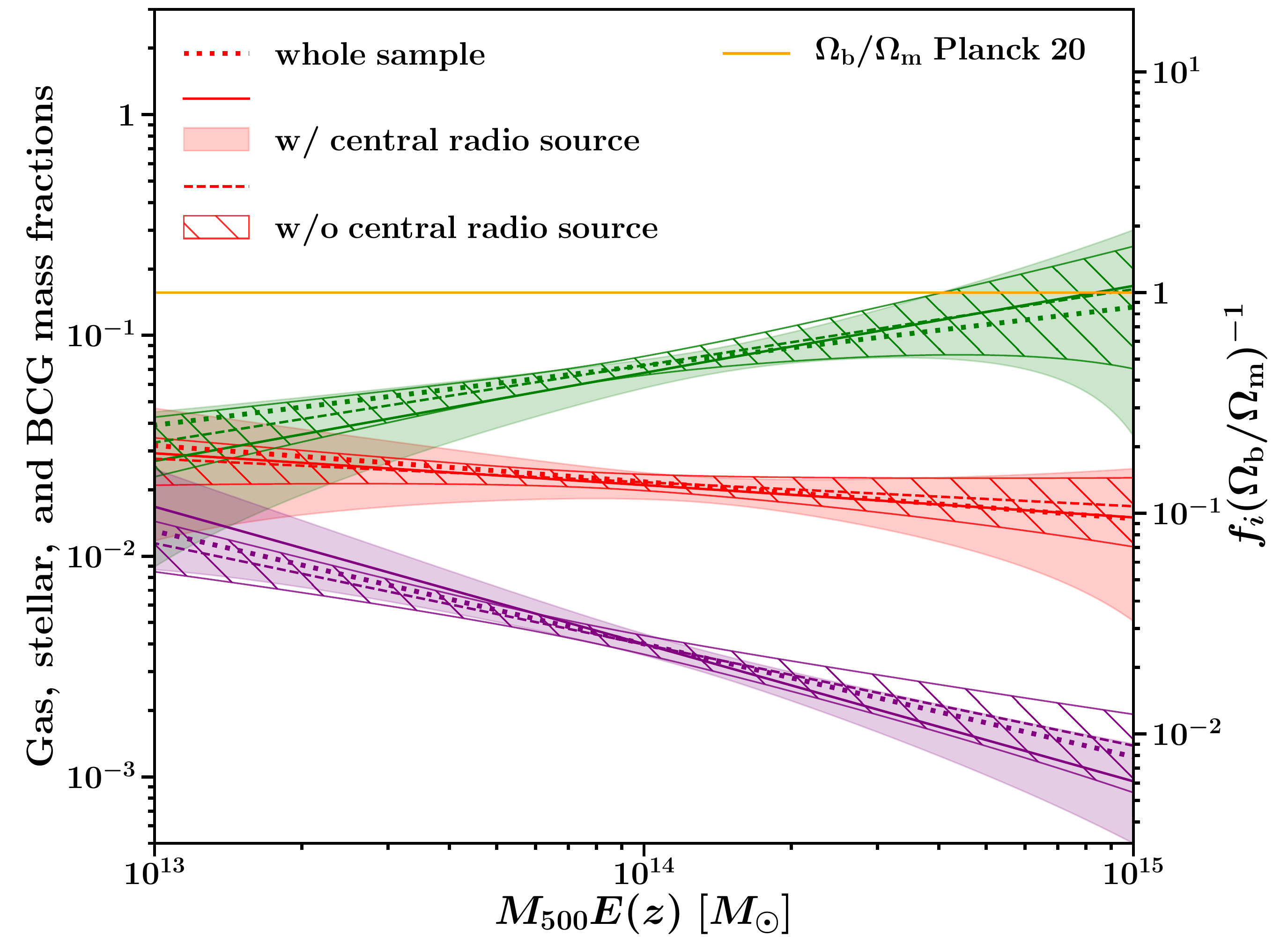}    
    \end{center}
    \caption{The gas (green), total stellar (red), and BCG (magenta) mass fractions as a function of halo mass. The solid, dashed, and dotted lines denote the Bayesian estimation for the 34 radio-AGN clusters, the 102 non-radio-AGN clusters, and the whole sample, respectively.
    The filled and hatched regions denote the $1\sigma$ uncertainties for the 34 radio-AGN clusters and the 102 non-radio-AGN clusters, respectively. 
    }
    \label{fig:fbcg_agn}
 
\end{figure}

We repeat the Bayesian analysis for the two subsamples with and without central AGNs. 
We refer to the former and latter samples as radio-AGN (R) clusters and non-radio-AGN (NR) clusters, respectively. 
The resulting regression parameters for the 34 radio-AGN clusters are similar to those of the 102 non-radio-AGN clusters (bottom panel Table \ref{tab:chabrier_sub}).
Since the errors are large, it is difficult to discriminate between the two subsamples, as shown in Figure \ref{fig:fbcg_agn}.
The intrinsic covariance between the baryon components for the two subsamples are similar to those for the whole sample of clusters (bottom panel of Table \ref{tab:Sigma_int_sub}).

\begin{table}[ht]
    \caption{ The top and bottom panels show the best-fit scaling relations for the 83 C1 clusters and the 53 C2 clusters and the 102 non-radio-AGN (NR) clusters and the 34 radio-AGN (R) clusters, respectively. The normalization, $\alpha$, and the slope, $\beta$, are defined by the linear regressions (eqs. \ref{eq:X_z} and \ref{eq:Y_Z}). $^\dagger$ : the results using a trivariate Gaussian prior as the WL mass calibration, as described in Sec. \ref{subsec:wl_calib}. NR and R expresses "non-radio-AGN" and "radio-AGN", respectively. The errors denote the $1\sigma$ uncertainty.}
    \begin{center}
    \scalebox{1.2}[1.2]{
    \begin{tabular}{c|rr}   
               & \multicolumn{1}{c}{$\alpha$}                  &  \multicolumn{1}{c}{$\beta$}          
        \\ \hline\hline
        {\footnotesize C1} : $M_{500}^{\rm WL}E(z)$             &   $-0.11_{-0.02}^{+0.02}$$^\dagger$   &  $1.09_{-0.02}^{+0.02}$$^\dagger$ \\
        $L_{X}E(z)^{-1}$             &  $0.30_{-0.18}^{+0.18}$   & $1.73_{-0.47}^{+0.34}$ \\ 
        $M_{*}E(z)$         & $0.78_{-0.09}^{+0.11}$&$0.89_{-0.13}^{+0.14}$\\
        $M_{\rm BCG}E(z)$   &$-0.93_{-0.08}^{+0.09}$& $0.53_{-0.13}^{+0.13}$  \\
        $M_{g}E(z)$     & $1.97_{-0.10}^{+0.10}$ &$1.48_{-0.25}^{+0.20}$\\ 
        \hline
        {\footnotesize C2} : $M_{500}^{\rm WL}E(z)$             &   $-0.11_{-0.02}^{+0.02}$$^\dagger$   &  $1.08_{-0.02}^{+0.02}$$^\dagger$ \\
        $L_{X}E(z)^{-1}$             &  $0.11_{-0.23}^{+0.28}$  &  $1.65_{-0.23}^{+0.34}$\\ 
        $M_{*}E(z)$         &  $0.60_{-0.15}^{+0.17}$& $0.92_{-0.14}^{+0.19}$\\
        $M_{\rm BCG}E(z)$   & $-1.02_{-0.16}^{+0.17}$&  $0.53_{-0.15}^{+0.18}$ \\
        $M_{g}E(z)$     &  $1.87_{-0.14}^{+0.16}$ &$1.50_{-0.15}^{+0.23}$\\ 
        \hline
        \hline
        {\footnotesize NR} : $M_{500}^{\rm WL}E(z)$             &   $-0.11_{-0.02}^{+0.02}$$^\dagger$   &  $1.09_{-0.02}^{+0.02}$$^\dagger$ \\
        $L_{X}E(z)^{-1}$             & $0.32_{-0.16}^{+0.18}$   & $1.48_{-0.23}^{+0.32}$\\ 
        $M_{*}E(z)$         &$0.75_{-0.09}^{+0.10}$&$0.91_{-0.10}^{+0.13}$\\
        $M_{\rm BCG}E(z)$   &$-0.92_{-0.10}^{+0.10}$& $0.54_{-0.12}^{+0.14}$   \\
        $M_{g}E(z)$     & $1.96_{-0.10}^{+0.11}$  &$1.37_{-0.14}^{+0.20}$\\ 
        \hline
        {\footnotesize R} : $M_{500}^{\rm WL}E(z)$             &   $-0.11_{-0.02}^{+0.02}$$^\dagger$   &  $1.09_{-0.02}^{+0.02}$$^\dagger$ \\
        $L_{X}E(z)^{-1}$             & $0.03_{-0.31}^{+0.26}$& $1.71_{-0.39}^{+0.43}$ \\ 
        $M_{*}E(z)$         &$0.72_{-0.15}^{+0.14}$& $0.87_{-0.20}^{+0.25}$\\
        $M_{\rm BCG}E(z)$   &$-0.92_{-0.12}^{+0.11}$& $0.38_{-0.17}^{+0.19}$\\
        $M_{g}E(z)$     &  $1.88_{-0.17}^{+0.14}$ &$1.42_{-0.21}^{+0.26}$ \\ 
    \end{tabular}
    }
    \end{center}
\label{tab:chabrier_sub}
\end{table}

\begin{table*}[ht]
    \caption{The top and bottom panels show the intrinsic covariance for the 83 C1 clusters and the 53 C2 clusters and the 102 non-radio-AGN (NR) clusters and the 34 radio-AGN (R) clusters, respectively. Each column is the same as in Table \ref{tab:Sigma_int_4d}.}
    \begin{center}
    \scalebox{1.2}[1.2]{
    \begin{tabular}{c|ccccc}
        &  $L_{X}E(z)^{-1}$ & $M_*E(z)$ & $M_{\rm BCG}E(z)$ & $M_gE(z)$ & \\ \hline
    \hline
    C1 : $L_{X}E(z)^{-1}$  & $0.38_{-0.17}^{+0.33}$ & $-0.02_{-0.03}^{+0.08}$ & $0.01_{-0.04}^{+0.06}$ & $0.06_{-0.04}^{+0.18}$  \\
        $M_*E(z)$     & $-0.15_{-0.39}^{+0.47}$ & $0.36_{-0.06}^{+0.09}$ & $0.06_{-0.03}^{+0.04}$ & $-0.01_{-0.03}^{+0.07}$ \\
        $M_{\rm BCG}E(z)$   & $0.04_{-0.29}^{+0.22}$ & $0.31_{-0.17}^{+0.14}$ & $0.54_{-0.05}^{+0.05}$ & $-0.01_{-0.03}^{+0.04}$ \\
        $M_gE(z)$  &   $>0.82$ &$-0.12_{-0.47}^{+0.61}$  & $-0.13_{-0.34}^{+0.30}$ & $<0.38$  \\
        \hline
     C2 :  $L_{X}E(z)^{-1}$  &  $0.25_{-0.08}^{+0.13}$ & $-0.05_{-0.06}^{+0.06}$ & $-0.04_{-0.08}^{+0.08}$ & $0.03_{-0.02}^{+0.03}$  \\
        $M_*E(z)$     &  $-0.31_{-0.33}^{+0.40}$&  $0.66_{-0.09}^{+0.08}$ & $0.44_{-0.11}^{+0.11}$ & $-0.01_{-0.05}^{+0.06}$ \\
        $M_{\rm BCG}E(z)$   & $-0.22_{-0.34}^{+0.38}$ &  $0.78_{-0.09}^{+0.06}$ &  $0.87_{-0.08}^{+0.07}$& $-0.01_{-0.07}^{+0.07}$ \\
        $M_gE(z)$  &    $0.71_{-0.43}^{+0.17}$ & $-0.06_{-0.43}^{+0.50}$  &  $-0.06_{-0.44}^{+0.48}$ &  $<0.21$\\
        \hline\hline
        NR : $L_{X}E(z)^{-1}$  & $0.70_{-0.27}^{+0.17}$ & $0.02_{-0.09}^{+0.15}$ & $0.07_{-0.10}^{+0.12}$ & $0.25_{-0.16}^{+0.17}$\\
        $M_*E(z)$     & $0.07_{-0.35}^{+0.32}$  &$0.49_{-0.07}^{+0.10}$& $0.25_{-0.07}^{+0.08}$ & $0.02_{-0.06}^{+0.11}$ \\
        $M_{\rm BCG}E(z)$   & $0.16_{-0.22}^{+0.17}$ & $0.68_{-0.13}^{+0.07}$ & $0.74_{-0.06}^{+0.07}$ & $0.03_{-0.06}^{+0.08}$ \\
        $M_gE(z)$  &   $>0.95$ & $0.11_{-0.44}^{+0.40}$ &  $0.12_{-0.26}^{+0.20}$ & $0.38_{-0.17}^{+0.11}$  \\
        \hline
      R :   $L_{X}E(z)^{-1}$  & $0.31_{-0.12}^{+0.21}$ & $-0.01_{-0.06}^{+0.07}$ & $-0.02_{-0.06}^{+0.07}$ & $0.04_{-0.02}^{+0.05}$  \\
        $M_*E(z)$     & $-0.04_{-0.43}^{+0.37}$ & $0.50_{-0.09}^{+0.11}$  & $0.13_{-0.06}^{+0.08}$ & $-0.00_{-0.04}^{+0.05}$ \\
        $M_{\rm BCG}E(z)$   & $-0.13_{-0.37}^{+0.37}$ &$0.53_{-0.20}^{+0.13}$& $0.53_{-0.07}^{+0.09}$ & $-0.03_{-0.03}^{+0.04}$ \\
        $M_gE(z)$  &   $0.83_{-0.33}^{+0.11}$  &  $-0.02_{-0.43}^{+0.49}$ & $-0.27_{-0.34}^{+0.42}$ &$0.17_{-0.03}^{+0.09}$\\
    \end{tabular}
    }
    \end{center}
\label{tab:Sigma_int_sub}
\end{table*}

\subsection{Redshift evolution} \label{subsec:zev}

We next investigate the redshift evolution of the baryon budget. We here define the observables independent of redshifts, as follows, 
\begin{eqnarray}
      x&=&\ln \frac{M_{500}^{\rm WL} }{10^{14}M_\odot} \label{eq:observable_xy} \\
      \bm{y}&=&\Bigr\{\ln \frac{L_X}{10^{43}\,{\rm erg s^{-1}}}, \ln \frac{M_*}{10^{12}M_\odot},\ln \frac{M_{\rm BCG} }{10^{12}M_\odot}, \ln \frac{M_g }{10^{12}M_\odot}\Bigl\}. \nonumber 
\end{eqnarray}
We assume the following redshift dependence of the scaling relations,
\begin{eqnarray}
    {\bm Y}_Z&=&{\bm \alpha} + {\bm \beta} Z +{\bm \gamma}\ln\left(\frac{E(z)}{E(0.3)}\right), \label{eq:Yz_Z_Ez} 
\end{eqnarray}
with $Z=\log (M_{500}/10^{14}\Msol)$. We repeat the Bayesian analysis for the multivariate scaling relations.  We assume that the redshift dependence of the $L_X-M$ relation follows a self-similar solution with $\gamma_l=2$ in eq. \ref{eq:Yz_Z_Ez} to infer the redshift-dependent parent population $p(Z,z)$. Table \ref{tab:camira_z} summarizes the 
resulting regression parameters.  
The resulting normalization and mass-dependent slopes are in good agreement with those for the the self-similar redshift evolution (Table \ref{tab:chabrier_4d}).
We find no redshift evolution in the $M_*-M$, $M_{\rm BCG}-M$, and $M_g-M$ relations which agree well with the self-similar redshift evolution $\gamma=0$.

\begin{table}[b]
    \caption{Regression parameters of the scaling relations with redshift evolution. The normalization, $\alpha$, the slope, $\beta$, and the redshift evolution, $\gamma$, as a function of the true mass are defined by the linear regression (eq. \ref{eq:Yz_Z_Ez}). The square bracket denotes the fixed value. $^\dagger$ : the results using a trivariate Gaussian prior as the WL mass calibration, as described in Sec. \ref{subsec:wl_calib}. The errors denote the $1\sigma$ uncertainty. 
    }
   \begin{tabular}{c|ccc}   
                & $\alpha$                  &  $\beta$                  &  $\gamma$ \\ \hline
        $M_{500}^{\rm WL}$   &  $-0.11_{-0.02}^{+0.02}$$^\dagger$ & $1.08_{-0.02}^{+0.02}$$^\dagger$ & [0] \\
        $L_X$   & $0.33_{-0.13}^{+0.14}$ & $1.28_{-0.17}^{+0.22}$ & [2] \\
        $M_{*}$   & $0.75_{-0.13}^{+0.13}$ & $0.80_{-0.11}^{+0.11}$ &$-0.25_{-0.49}^{+0.57}$\\
        $M_{\rm BCG}$  & $-1.03_{-0.15}^{+0.15}$ & $0.41_{-0.10}^{+0.11}$ & $-0.06_{-0.54}^{+0.55}$\\
        $M_{g}$       & $1.92_{-0.10}^{+0.10}$ & $1.23_{-0.10}^{+0.13}$ &  $0.39_{-0.32}^{+0.33}$\\
    \end{tabular}
\label{tab:camira_z}
\end{table}

\section{Discussion}\label{sec:dis}

\subsection{Systematics}\label{subsec:dis_b}

We recall the method of the multivariate-scaling-relations analysis of the baryon components. We set the vectors of baryons in ${\bm y}$, weak-lensing mass in $x$ (Sec. \ref{subsubsec:multi_scaling}) and the error covariance matrix (Sec. \ref{subsubsec:errormatrix}).
The weak-lensing mass and true mass are statistically related through a power-law relation with intrinsic scatter based on a prior motivated by numerical simulations (Sec. \ref{subsec:wl_calib}). Our Bayesian method (Sec. \ref{subsec:linear_sigma}) simultaneously computes 
the linear regression parameters (${\bm \alpha}$ and ${\bm \beta}$), the intrinsic covariance ($\bm{\Sigma}_{\rm int}$), and the parent population of the true mass ($p(Z,z)$). In the regression analysis, it is vitally important to control
regression dilution effect and selection effect (see details in Appendix \ref{app:a}). The two effects are simultaneously calibrated by the estimated parameters of the assumed parent population (Sec. \ref{subsec:linear_sigma} and Appendix \ref{app:a}). The shape of the parent population depends on the weak-lensing mass distribution as well as 
the X-ray luminosity which is used as an approximated tracer of our cluster finder.
This subsection discusses possible sources of systematics in the Bayesian regression analysis.

\subsubsection{Performance of Bayesian analysis}\label{subsubsub:dis_perfomance}

We assess the reliability of our Bayesian analysis using mock simulations computed with the error matrix similar to the observational one (see details in Appendix \ref{app:a}). We define a multiplicative error and an additive error in the relation $\bm{\theta}_{\rm output}=m \times \bm{\theta}_{\rm input}+c$, where $\bm{\theta}_{\rm input}$ and $\bm{\theta}_{\rm output}$ are the input and the output parameters. The resulting multiplicative and additive errors in the simulation of $p(Z,z)$ are $\langle m\rangle=0.989\pm0.018$ and $\langle c\rangle=-0.004\pm0.011$ averaged over all the parameters, indicating that our code recovers well the input parameters. The uncertainties for the estimated parameters of the multivariate scaling relations are larger than the accuracy of the recovery of the input parameters. In the case of the large measurement errors of the weak-lensing masses, it is important to carefully and fully consider the error correlations. If larger errors of two observables are not correlated, it is difficult to constrain the intrinsic covariance (Figure \ref{fig:hibrecs2} in Appendix \ref{app:a}).
The stacked observables sorted by X-ray luminosities and inverse-weighted with $\bm{\Sigma}_{\rm err}+\bm{\Sigma}_{\rm int}$ (Figure \ref{fig:mg-mwl}-\ref{fig:mbcg-mwl}; Sec. \ref{subsec:sc_result}) are in good agreement with the baselines described by $\bm{\alpha}$ and $\bm{\beta}$, which ensures the consistency between the independent regression parameters to represent the data. 

\subsubsection{Parent population}\label{subsubsec:Pz}

We employed the Gaussian distribution for the parent population, $p(Z,z)$, for the logarithm of the true mass. This functional formulae differs from the XXL X-ray luminosity function \citep{2016A&A...592A...2P,2018A&A...620A...5A,2018A&A...614A..72V}. We infer the shape parameters ($\mu_Z(z)$ and $\sigma_Z(z)$) of $p(Z,z)$ by the hierarchical Bayesian modelling, which
keeps a flexibility to approximately describe an unknown parent population or a halo mass function of \citet{Tinker08}. It effectively corrects both the regression dilution effect and the selection bias (see Appendix \ref{app:a} and \ref{app:b}).
Furthermore, our results are not affected by cluster mass function and the tension in $\sigma_8$ measurements between {\it Planck} early universe experiment and nearby universe observations \citep{Pratt19}.
We emphasize that the purpose to introduce $p(Z,z)$ is not to constrain cosmological parameters or to accurately determine the mass function but to correct the above two effects in the analysis of the multivariate scaling relations (Appendix \ref{app:a} and \ref{app:b}).

The XXL selection function behind the XXL cluster catalog uses the actual surface brightness profile, namely, core radius and total count-rate, to avoid contamination by X-ray point sources \citep{2016A&A...592A...2P,2018A&A...620A...5A,2018A&A...614A..72V}.
The total X-ray luminosity is computed by integrating the X-ray surface brightness distribution.
We can easily infer $\sigma_Z(z)$ and $\mu_Z(z)$ through the $L_X-M$ relation, which is sufficient to constrain the regression parameters of the baryon contents for the current sample, as seen in Figures \ref{fig:mg-mwl}-\ref{fig:mbcg-mwl} and Appendix \ref{app:a}. 
To accurately measure the $p(Z,z)$ distribution, we could use multiple Gaussian distributions with different weights. However, when we used the double Gaussian distribution, we were not able to constrain the parameters of the second Gaussian distribution (Sec. \ref{subsec:linear_sigma}). Therefore, the single Gaussian is sufficient to describe the multivariate scaling relations for the current sample. With a larger sample of clusters and/or small measurement errors of the weak-lensing masses, we would require a more sophisticated model such as the cluster mass function combined with the XXL selection function.

\begin{figure}
    \includegraphics[width=\hsize]{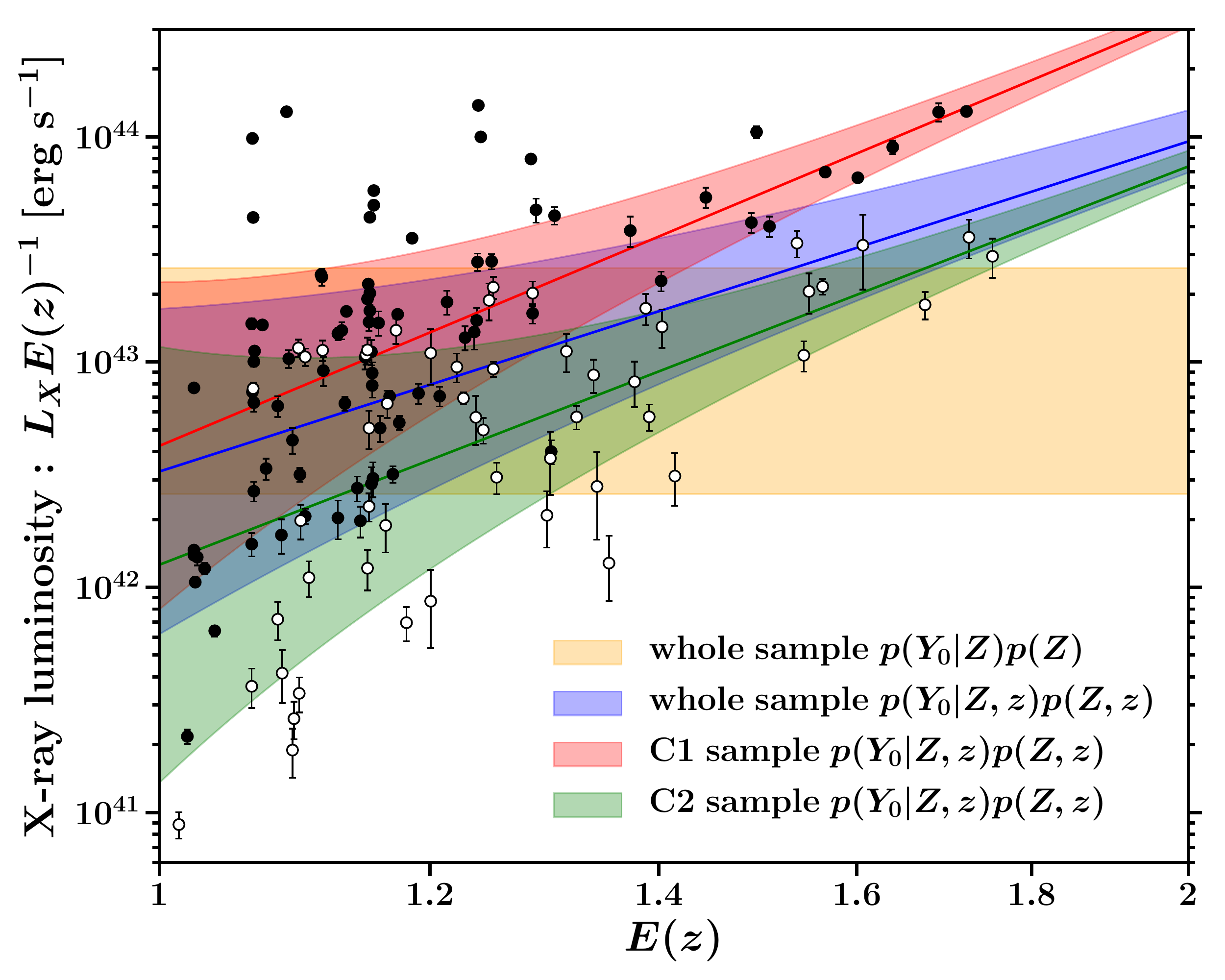}
    \caption{The X-ray luminosity versus $E(z)$. Black and white circles are the C1 and C2 clusters, respectively. The color regions denote the standard deviation ranges of the X-ray luminosity population expected by the resulting parent population of true mass $p(Z)$. The solid lines are the mean of the X-ray luminosity population. Orange color is the redshift-independent case of $p(Y_0|Z,z)p(Z)$. Blue, red, and green colors are the whole, C1, and C2 samples with the redshift-dependent case of $p(Y_0|Z,z)p(Z,z)$.
    }
    \label{fig:Pz}
\end{figure}

Since the X-ray selection is affected by cosmological dimming and the selected cluster masses depend on the redshift, we introduced the redshift-dependent mean $\mu_Z(z)$ and standard deviation $\sigma_Z(z)$ of $p(Z,z)$ as hyper-parameters. 
Figure \ref{fig:Pz} shows the resulting X-ray luminosity population as a function of $E(z)$ which is computed by $p(Y_0|Z,z)p(Z,z)$. 
The resulting models of the X-ray luminosity population estimated by the true mass distributions of $p(Z,z)$ agree with the data distribution.
The resulting models indicate that the XXL cluster catalog covers a wide range of X-ray luminosities at low redshifts and comprises only the most X-ray luminous clusters at high redshifts. The X-ray luminosity population for the C1 clusters is shifted to a higher value compared to that for the C2 clusters. The whole X-ray luminosity population is distributed around the intermediate position between the two C1 and C2 X-ray population.

As an alternative modelling, we here assume a redshift-independent Gaussian distribution $p(Z)={\mathcal N}(\mu_Z,\sigma_Z)$ in the Bayesian analysis, where the mean ($\mu_Z$) and standard deviation ($\sigma_Z$) are free parameters independent of the redshift. We refer it to as $\rm no\mathchar`-z$.
We compare the models using Akaike's information criterion (AIC) and Bayesian information criterion (BIC). The AIC and BIC are defined by ${\rm AIC}=2N_{\rm para}-2\ln {p}_{\rm max}$ and ${\rm BIC}=N_{\rm para}\ln{N_{\rm data}}-2\ln p_{\rm max}$, respectively. Here $N_{\rm para}$ is the number of parameters, $N_{\rm data}$ is the number of data points, and $p_{\rm max}$ is the maximum value of the posterior probability (eq. \ref{eq:likelihood}). The first terms in both the AIC and BIC describe a penalty of over-fitting by increasing the number of parameters in the model. The AIC, derived by relative entropy, measures relative loss among given different models. A low AIC value means that a model is considered to be closer to the truth. The BIC is derived by the framework of Bayesian theory to maximize the posterior probability of a model given the data. In other words, a model with the lowest BIC is preferred to be the truth. The AIC and BIC are based on different motivations and thus they provide complementary information. When we compare with two models, the model with the lower value is preferred. The difference between the models are significant according to both the AIC and BIC,
 $\Delta {\rm AIC}={\rm AIC}_{\rm no\mathchar`-z}-{\rm AIC}=+43$ and $\Delta {\rm BIC}={\rm BIC}_{\rm no\mathchar`-z}-{\rm BIC}=+35$. The orange region in Figure \ref{fig:Pz} does not match with the data distribution. Therefore, the redshift-dependent parent population is preferable.

We next assume a second-order redshift dependence for $\mu_Z(z)=\mu_{Z,0}+\gamma_{\mu_Z}\ln E(z)+\gamma_{\mu_Z,2}(\ln E(z))^2$ and $\ln \sigma_Z(z)=\ln \sigma_{Z,0}+\gamma_{\sigma_Z}\ln E(z)+\gamma_{\sigma_Z,2}(\ln E(z))^2$. 
The result does not significantly change and, thus, the resulting AIC and BIC become larger (worse) than those of our main result due to the penalty from the increased number of parameters; $\Delta {\rm AIC}=+3$ and $\Delta {\rm BIC}=+13$.
The first-order dependence is sufficient to describe the data.

The code could in principle estimate the $\mu_Z(z)$ and $\sigma_Z(z)$ parameters without a correction for the Malmquist bias.
We perform the Bayesian analysis using a subset of $\bm{y}=\{y_*,y_{\rm BCG},y_g\}$ in order to understand the impact of the tracer of cluster finders in the multivariate scaling relation analysis.
The resulting $(\sigma_Z(z))^2$ becomes higher by $\sim7$ \%, and consequently the mass-dependent slopes, ${\bm \beta}$, become
shallower by $\sim5$ \% in $\beta_*$, $\sim7$ \% in $\beta_{\rm BCG}$, and $\sim5$ \% in $\beta_g$. This change is caused by the relationship between the variance in the parent population and the slope (eq. \ref{eq:app_b3}), which is described in Appendix \ref{app:b}. Changes in the normalization is less than $1$ \%. 
Although the overall results do not change, 
the simultaneous treatment of the X-ray luminosity approximately related to the cluster finding can more properly estimate the parent population in the computation of the multivariate scaling relations.

\subsubsection{Systematics by weak-lensing mass calibration}\label{subsubsec:dis_WLcalib}

In the limit of low S/N weak-lensing signals, the errors of the weak-lensing masses are mainly caused by the number of background galaxies, rather than intrinsic halo properties of the halo non-sphericity, subhalos and its surrounding large-scale structure.
Indeed, \citet{2020ApJ...890..148U} have shown that the measurement errors using synthetic weak-lensing data of the analytic NFW model are comparable to those using cosmological simulations. 
We independently introduced the bias and scatter in weak-lensing mass measurement based on dark-matter-only simulations \citep{2020ApJ...890..148U}. 
Higher-mass halos tend to have less spherical structure 
because they are the most recent forming systems and thus 
growth of halos may not have yet erased the information about initial condition and formation process.
\cite[e.g.][]{2002ApJ...574..538J,2006MNRAS.367.1781A}. 
The abundance of subhalos in high-mass halos is larger than in low-mass halos.
Therefore, the weak-lensing mass calibration inherent in the intrinsic halo properties depends on cluster masses \citep{2020ApJ...890..148U}. 

Since it is difficult to calibrate weak-lensing masses for individual clusters
in response to their own properties, we employed the statistical approach of the weak-lensing mass calibration (eq. \ref{eq:X_z}).
The overdensity radii are accordingly changed from the weak-lensing overdensity radii by the weak-lensing mass calibration. 
We adapted the weak-lensing overdensity radius as the measurement radius for the gas mass and the stellar mass because the individual true masses are not clarified.
Since the weak-lensing overdensity radii are only $-6\pm4$ percent smaller than those of the true mass, it is negligible compared to the statistical errors.

In Sec \ref{subsec:fb}, we estimated how much the gas mass and the stellar mass are changed by the aperture correction which is caused by the difference in the measurement radii of $r_{500}^{\rm WL}$ and $r_{500}$. Since the mass bias becomes larger with decreasing the true mass (Fig. \ref{fig:masscalib}), the aperture correction makes the stellar mass and the gas mass $5\%$ and $9\%$ higher at $M_{500}E(z)\sim10^{13}M_\odot$, respectively. In contrast, the aperture correction at the high mass end of $\sim10^{15}$ is less than a few percent.

When we fix the estimated 
values of $\alpha_{\rm WL},\beta_{\rm WL},\sigma_{\rm WL}$ of the weak-lensing mass calibration ignoring their uncertainties, the measurement errors of the gas mass scaling relation at $M_{500}E(z)\sim10^{15}\Msol$ and the intrinsic correlation coefficient are reduced by $10-30$ percent, while the other errors are not significantly changed. When we remove the prior of $\sigma_{\rm WL}$, we obtain $\sigma_{\rm WL}=0.14_{-0.06}^{+0.09}$ which is consistent with the mass calibration determined by the 639 simulated clusters.

We also investigate the multivariate scaling relation with the mass calibration with the $\tanh$ function (Sec. \ref{subsec:wl_calib}). In that case, we replace $M_{500}^{\rm WL}$ by the calibrated mass, $M_{500}^{\rm calib}$, in the $X$ quantity and treat $X=Z$ with a fixed $\sigma_X=\sigma_{\rm WL}$. 
The resulting regression parameters are consistent with those of our main result (Table \ref{tab:chabrier_4d}) within $1\sigma$ errors. When we scale the $y_*$ and $y_{g}$ observables by $\delta y\simeq \delta x$ according to the mass calibration, the result does not change significantly.

We fix the intrinsic correlation coefficients of $r_{{\rm WL},i}=0$ in the analysis.
When we treat them as free parameters, the fitting results are almost the same. Therefore, an over-fitting by increasing the number of free parameters occurs.
A penalty of the over-fitting increases the information criterion, especially $\Delta {\rm BIC}=+17$. We thus prevent over-fitting by fixing the correlations in the analysis.

\subsubsection{Systematics by error correlation}

Since the member galaxies are sparsely distributed, we adopted a single error correlation coefficient $r_{\rm{WL},*}^{\rm err}$ computed over the whole sample. Intrinsic covariance might be affected by this treatment. We therefore assess how much the intrinsic covariance is changed by our choice of $r_{\rm{WL},*}^{\rm err}$. We first pick a uniform random number from a range of [$r_{\rm{WL},*}^{\rm err}-0.1$,$r_{\rm{WL},*}^{\rm err}+0.1$] for each cluster and we find that the resulting intrinsic covariance is consistent with our reference results (Table \ref{tab:Sigma_int_4d}). Therefore, our treatment does not significantly impact the results.  Next, we use $r_{\rm{WL},*}^{\rm err}=-0.5,0,0.5$ which are lower than 0.873 (Sec. \ref{subsubsec:errormatrix}), and we find that all the results of $r_{*,g}$ become negative or no correlation in contrast to our positive result (Table \ref{tab:Sigma_int_4d}), 
and the uncertainties for $r_{{\rm BCG},*}$ and $r_{*,g}$ become larger by 2.2 and 1.6 times, respectively. The other parameters are not significantly affected by the assumption.
The change of $r_{*,g}$ is caused by $r_{*,g}^{\rm err}=r_{\rm{WL},*}^{\rm err}r_{{\rm WL},g}^{\rm err}$. Therefore, an improper treatment can give rise to spurious anti-correlation between gas mass and stellar mass.

\subsubsection{Blue galaxies and the intracluster light} \label{subsubsec:blueICL}

We counted the total stellar masses of the red galaxies selected by the color-magnitude planes using the XXL centers and redshifts.
In general, cluster members are composed of red and blue galaxies which are distributed in the inner and outer regions, respectively \citep[e.g.][]{1993ApJ...407..489W,2004MNRAS.351..125D,2018PASJ...70S..24N}. Red galaxies would be the dominant component of the cold baryon within $r_{500}$ which is roughly about half of virial radius. As for blue galaxies distributed at outer radii, there is the possibility of an over-subtraction of background component.
We estimate how much stellar mass is changed when including blue galaxies. We first select galaxies from 
the MIZUKI photometric redshift catalog \citep{2020arXiv200301511N,HSC2ndDR,2018PASJ...70S...9T,MIZUKI} with  criterion of $|z_{\rm ph}-z_{c}|<0.05(1+z_{c})$ and $M_*>10^{10}\Msol$, where $z_{\rm ph}$ is a photometric redshift. 
We pick-up galaxies which are not identified in the red galaxy catalog but in the photometric catalog and refer them to as blue galaxies. When we include blue galaxies, the total stellar masses for individual clusters are changed by $15\pm17$ percent. We repeat the Bayesian analysis for the red and blue galaxies and obtain
$\alpha_*=0.70_{-0.08}^{+0.09}$, $\beta_*=0.81_{-0.10}^{+0.13}$, and $\sigma_*=0.49_{-0.06}^{+0.10}$. 
The baseline for the red and blue galaxies agrees with that for the red galaxies within the uncertainties (Table \ref{tab:chabrier_4d}). The intrinsic correlation coefficients ( $r_{*,{\rm BCG}}=0.57_{-0.10}^{+0.08}$ and $r_{*,g}=0.29_{-0.26}^{+0.34}$) are not significantly changed, either.

The tidal stripping of stars from interacting galaxies and the merger of small galaxies with central brightest cluster galaxies make a diffuse intracluster light (ICL). In particular, extended low-surface brightness envelope forms around the central galaxies. However, it is very difficult to observationally detect such a weak excess of the ICL component from the image background because of over-subtraction. The HSC-SSP data is currently not adequate for the study of the ICL.
Since we adopted the \texttt{cmodel} magnitude, we did not include the ICL component. 
\citet{2018MNRAS.475.3348H} have studied how much the \texttt{cmodel} photometry underestimates stellar masses for massive galaxies at $z<0.5$. They evaluated a difference between stellar masses estimated by the \texttt{cmodel} magnitude and surface mass density profiles out to 100 kpc without imaging stacking. The stellar mass within $100$ kpc corresponds to the total stellar mass because $100$ kpc aperture covers $5-10$ times of effective radii.
\citet{2018MNRAS.475.3348H} found that a median $M_*$ with the \texttt{cmodel} magnitudes underestimate the stellar masses only for massive galaxies ($M_*>10^{11.6}\Msol\sim 4\times10^{11}\Msol$) by $\sim 0.1-0.15$ dex.
Based on their results, we expect that the total stellar+ICL masses within $100$ kpc aperture around the massive galaxies like BCGs would be $\sim 1.3-1.4$ times higher than our \texttt{cmodel} estimates.

\subsection{Baryon budget}\label{subsec:dis_b2}

This subsection is focused on the discussion of our measurements of baryon budgets of the clusters.

\subsubsection{Baryon fractions}\label{subsubsec:dis_fb}

We found that the gas and stellar mass fractions increase and decrease with increasing halo mass (Figure \ref{fig:fb} and Sections \ref{subsec:sc_result} and \ref{subsec:fb}), respectively. This trend can be explained by a halo mass dependence of the star formation efficiency. 
The star formation efficiency in low-mass clusters and groups is expected to be higher than that in high-mass clusters.
In addition, tidal interactions among galaxies and  the removal of the gas reservoir of galaxies by ram-pressure are  more inefficient in low-mass clusters than in high-mass ones.  Therefore, a larger fraction of the gas in low-mass clusters is consumed to form stars through cooling, while galaxy formation tends to be inhibited in high-mass clusters.

AGN feedback is also important to determine the baryon budget, because it
heats the surrounding gas and suppresses star formation and more or less modifies radial distribution of gas, especially in low-mass clusters.
Some AGNs especially in centeral galaxies are energetic enough to expel the gas material of stars out from the relatively shallow potential well of the low-mass clusters.
The expelled gas in low-mass clusters is difficult to be re-accreted.
Since the star formation activity does not  change the total baryon fraction because of the mass conservation \citep[e.g.][]{Kravtsov05}, the total baryon fraction without AGN feedback is expected to be constant against the halo mass. However, the gas redistribution by AGN feedback could change halo mass dependence on the baryon and gas mass fractions. Therefore, the degree of balance between star formation and all the AGN activities throughout the entire cluster history controls the mass dependence of the baryon contents. 
The result that the total baryon fraction reaches to the cosmic mean baryon fraction $\Omega_b/\Omega_m$ at high-mass halos of $10^{15}\Msol$ indicates that the high-mass halos are close to a closed-box in which the total baryon is confined. On the other hand, since $f_b\sim0.5(\Omega_b/\Omega_m)$ at low-mass clusters of $\sim10^{13}\Msol$, the low-mass halos are likely to be an open-box in which baryons are not conserved. This is likely caused by AGN feedback.

We investigated the baryon fractions for the clusters which currently host radio AGN activity and those for the other clusters (Figure \ref{fig:fbcg_agn} and Sec. \ref{subsec:sub}). We do not find a significant difference of the baryon fraction in response to the current AGN activity. 
In general, AGN activity is a transient phenomenon, whereas gas ejection from the potential well depends on the total integrated non-gravitational energy. 
It implies that the cumulative quantities such as the gas and stellar masses are insensitive to the current AGN activity. However, a larger sample is essential to further constrain the parameters.

We also found good agreement of the baryon fractions between the C1 and C2 clusters (Figure \ref{fig:fbcg_c1c2}), though the likelihood function of the XXL selection for the C1 class is different from that of the C2 class. This is promising for XXL X-ray cluster counts analyses of cosmological parameters.

\subsubsection{Intrinsic covariance in baryon content}\label{subsubsec:dis_Sigmaint}

Clusters move around the baselines in the scaling relations due to mass accretion, mergers, cooling, and AGN feedback.
Since the baryonic evolution is an order of sound-crossing time, their positions in the scaling-relations instantly change. Their statistical properties are observed as intrinsic covariance. 
If all the baryons were confined within the halo (closed-box), the intrinsic correlation coefficient between the gas mass and the stellar mass is expected to be negative because of $\delta f_b= 0 = \delta f_g + \delta f_*$. 
As we mentioned above, the anti-correlation appears only for the case of the improper treatment of the error covariance matrix. We found no evidence that the intrinsic correlation coefficient between gas mass and stellar mass in the whole sample is correlated or anti-correlated. It is generally very difficult to accept the null hypothesis that the true correlation is zero under a finite uncertainty. 
With the $N$ sample, the constraint has to satisfy with $|r_{*,g}|\simlt 0.168(N/136)^{-1/2}$ so that the $p$-value of the null hypothesis can be higher than 5 percent. The required uncertainty is about half of the current constraint.  
However, the margin of the error includes the null correlation.
Our result does not contradict with the open-box scenario suggested by the total baryon fraction (Sec \ref{subsubsec:dis_fb}).

The $M_{\rm BCG}- M_{500}$ relation showed the largest intrinsic scatter and a weak-mass dependence, which implies a presence of another factor besides the halo mass in the BCG mass growth. 
We found significant correlations between $M_*$ and $M_{\rm BCG}$ and between $M_{\rm sat}$ and $M_{\rm BCG}$ (Sec \ref{subsec:Sigma_int}), 
qualitatively suggesting that the BCGs co-evolve with the satellite galaxies.

Intrinsic correlation between soft-band $L_X$ and $M_g$ is close to unity which is naturally explained by the X-ray emissivity.

The C2 clusters have larger intrinsic scatter of total and BCG stellar masses than the C1 clusters. The discrepancies for the former and latter cases are $3.3\sigma$ and $4.1\sigma$, respectively.
Since the C2 class has lower X-ray luminosity (lower masses), the stellar properties in low X-ray clusters are likely to be more diverse from cluster-to-cluster.

\subsection{Comparison of numerical simulations}

Recent cosmological hydrodynamic simulations
\citep[e.g.][]{2011MNRAS.413..691Y,2011MNRAS.412.1965M,2013MNRAS.431.1487P,2014MNRAS.440.2290M,2014MNRAS.441.1270L,2015MNRAS.452.1982W,2016MNRAS.457.4063S,2017MNRAS.465.2936M,2017MNRAS.471.1088B,2018MNRAS.478.2618F,2020MNRAS.498.2114H,2020MNRAS.493.1361F} studied stellar mass and gas distributions in clusters and/or groups. The simulations include the effect of cooling, AGN feedback, star formation, and SN feedback and compare them with the results of non-radiative simulations. 
Since gas distributions are radially modified by AGN feedback, the scaling relations depends on 
overdensity radius \citep[e.g.][]{2011MNRAS.413..691Y,2018MNRAS.478.2618F}. Thus, when we compare numerical simulation with observations, it is important to choose the same overdensity as observations (i.e. $\Delta=500$). 
Results of numerical simulations depend on the different AGN models  \citep[e.g.][]{2011MNRAS.412.1965M,2014MNRAS.441.1270L,2016MNRAS.459.2973S}. 

\subsubsection{Scaling relations and mass-dependent slopes}\label{subsubsec:slope_sim}

\begin{figure*} 
   \includegraphics[width=0.33\hsize]{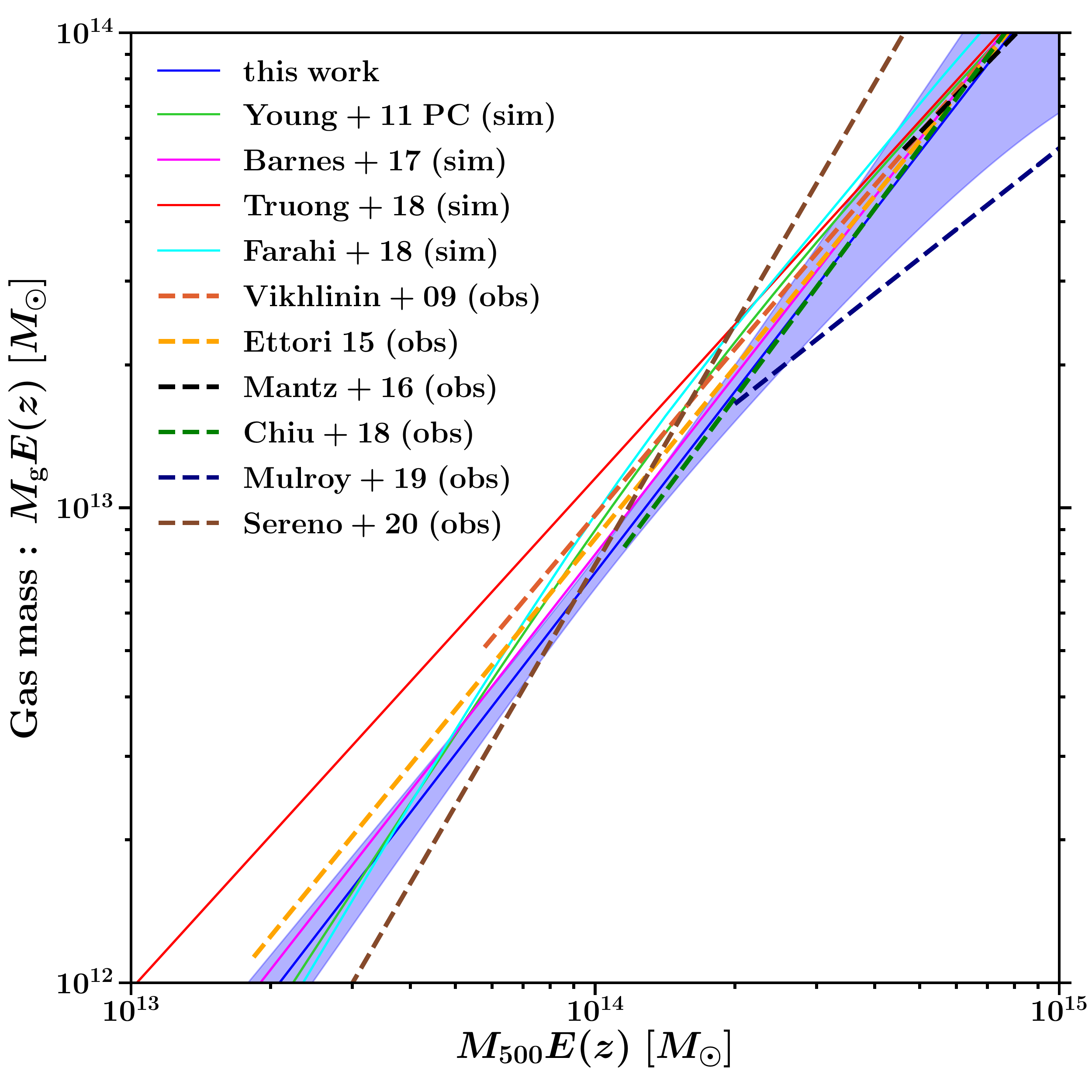} 
     \includegraphics[width=0.33\hsize]{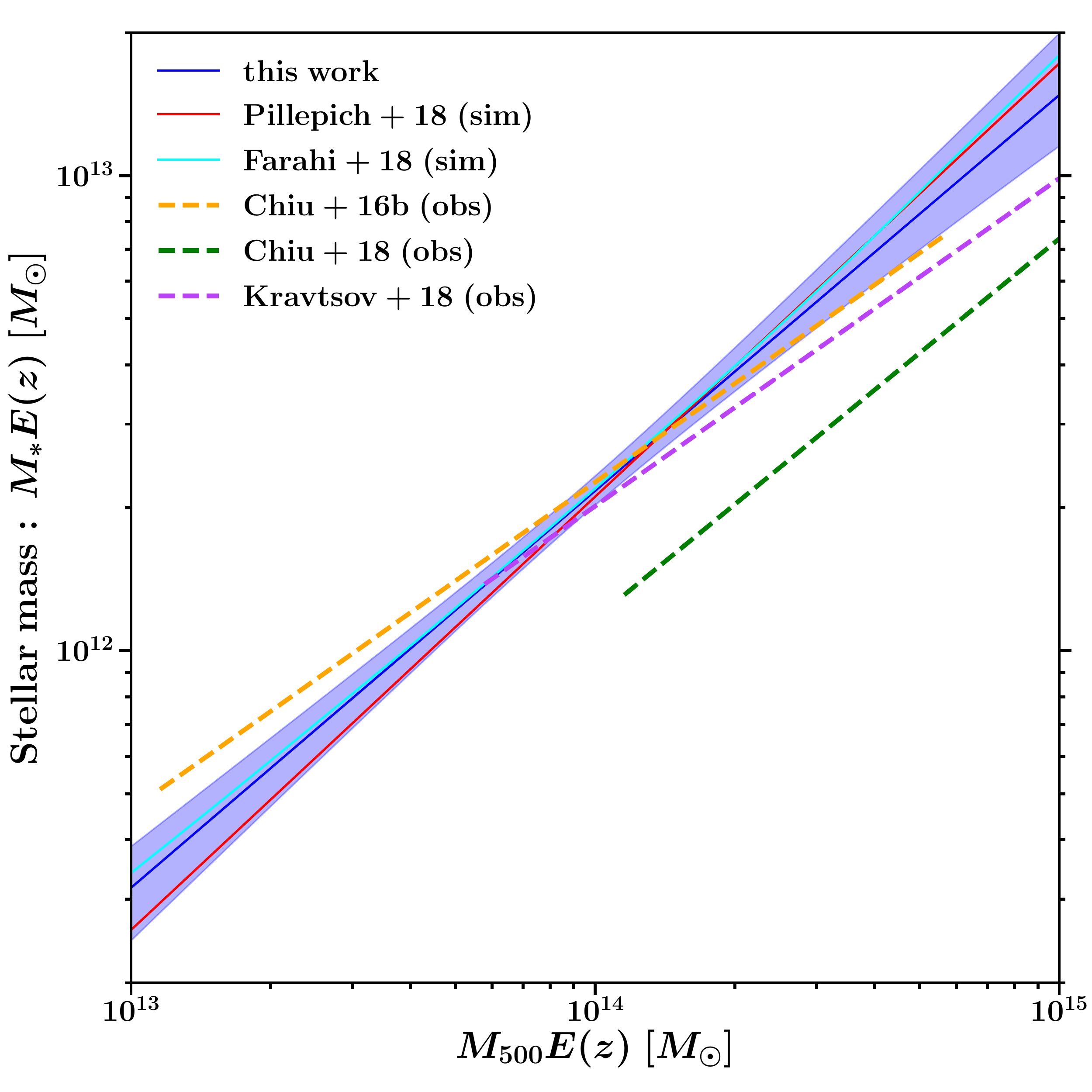} 
       \includegraphics[width=0.33\hsize]{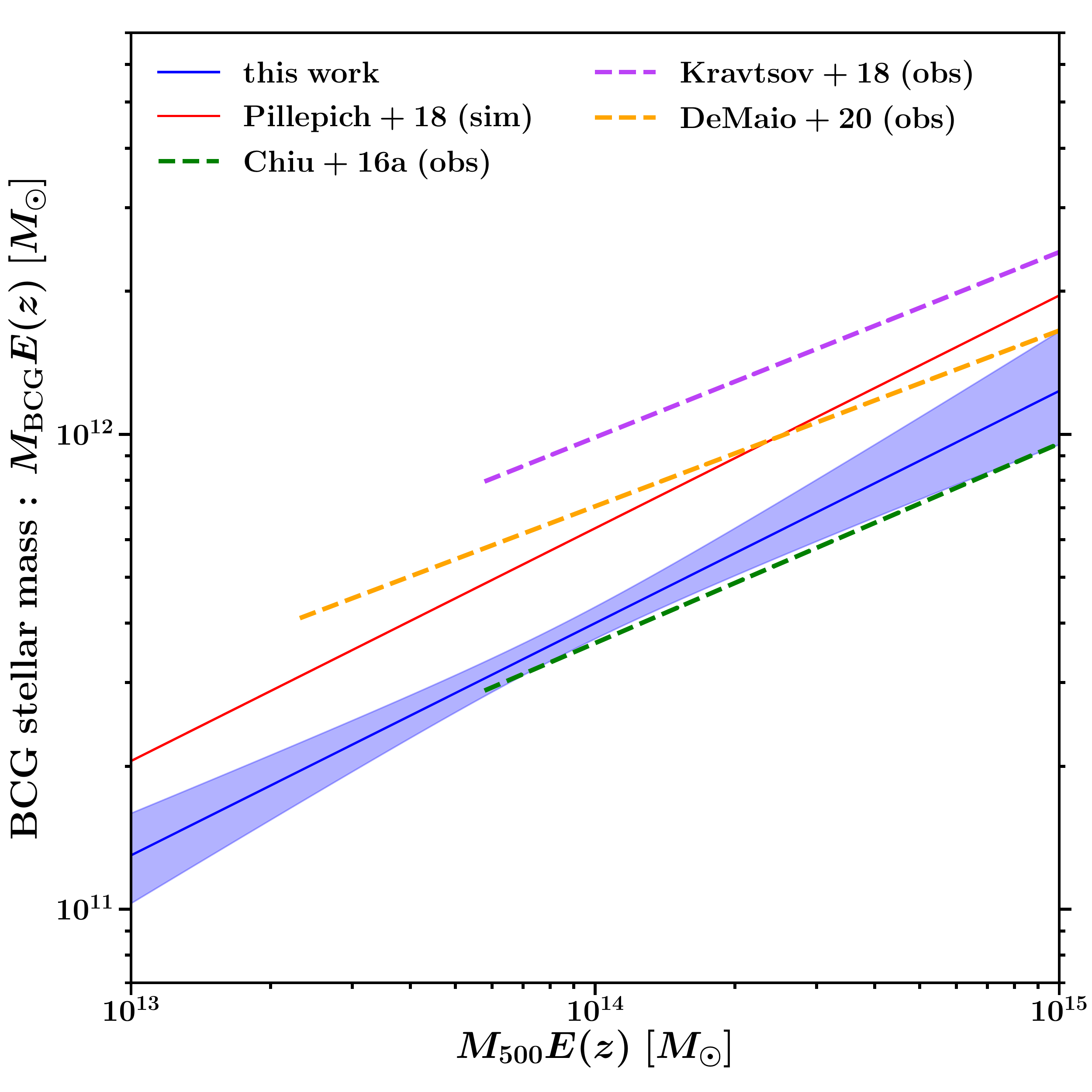} 
    \caption{Comparison of numerical simulations and observations in literature. From left to right, the gas, stellar, and BCG mass scaling relations. The solid blue line and regions denote the estimate and $1\sigma$ uncertainty, respectively. 
    Our results and numerical simulations denote the true masses in the all panels, while the masses for the previous observations except for \citet{2020MNRAS.492.4528S} are not (mass proxies, lensing masses or hydrostatic masses).
    In the left panel, the solid green, magenta, red and light-blue lines are simulations  \citep{2011MNRAS.413..691Y,2017MNRAS.465..213B,2018MNRAS.474.4089T,2018A&A...620A...8F}, and the dashed orange, yellow, black, green, dark blue, and brown lines are previous observations \citep{Vikhlinin09a,2015MNRAS.446.2629E,2016MNRAS.463.3582M,2018MNRAS.478.3072C,2019MNRAS.484...60M,2020MNRAS.492.4528S}, respectively. The line lengths of the observations denote their mass ranges. In the middle panel, the solid red and light-blue lines and the dashed yellow, green, magenta lines are simulations \citep{2018MNRAS.475..648P,2018A&A...620A...8F} and previous observations \citep{2016MNRAS.458..379C,2018MNRAS.478.3072C,2018AstL...44....8K}, respectively. In the right panel, the solid red line and the dashed green, magenta, and yellow lines are a simulation \citep{2018MNRAS.475..648P} and observations \citep{2016MNRAS.455..258C,2018AstL...44....8K,2020MNRAS.491.3751D}, respectively. }
    \label{fig:compare_scaling}
\end{figure*}

We compare our results with some of the simulations in Figures \ref{fig:compare_scaling}.
Simulations results are rescaled to $z=0.3$, close to the median 
redshift of the XXL clusters, assuming self similar evolution.
We find that our normalization and slope in $M_g-M$ and $M_*-M$ relations broadly agree with 
those of numerical simulations over two orders of magnitude in mass.

\begin{figure}
    \includegraphics[width=\hsize]{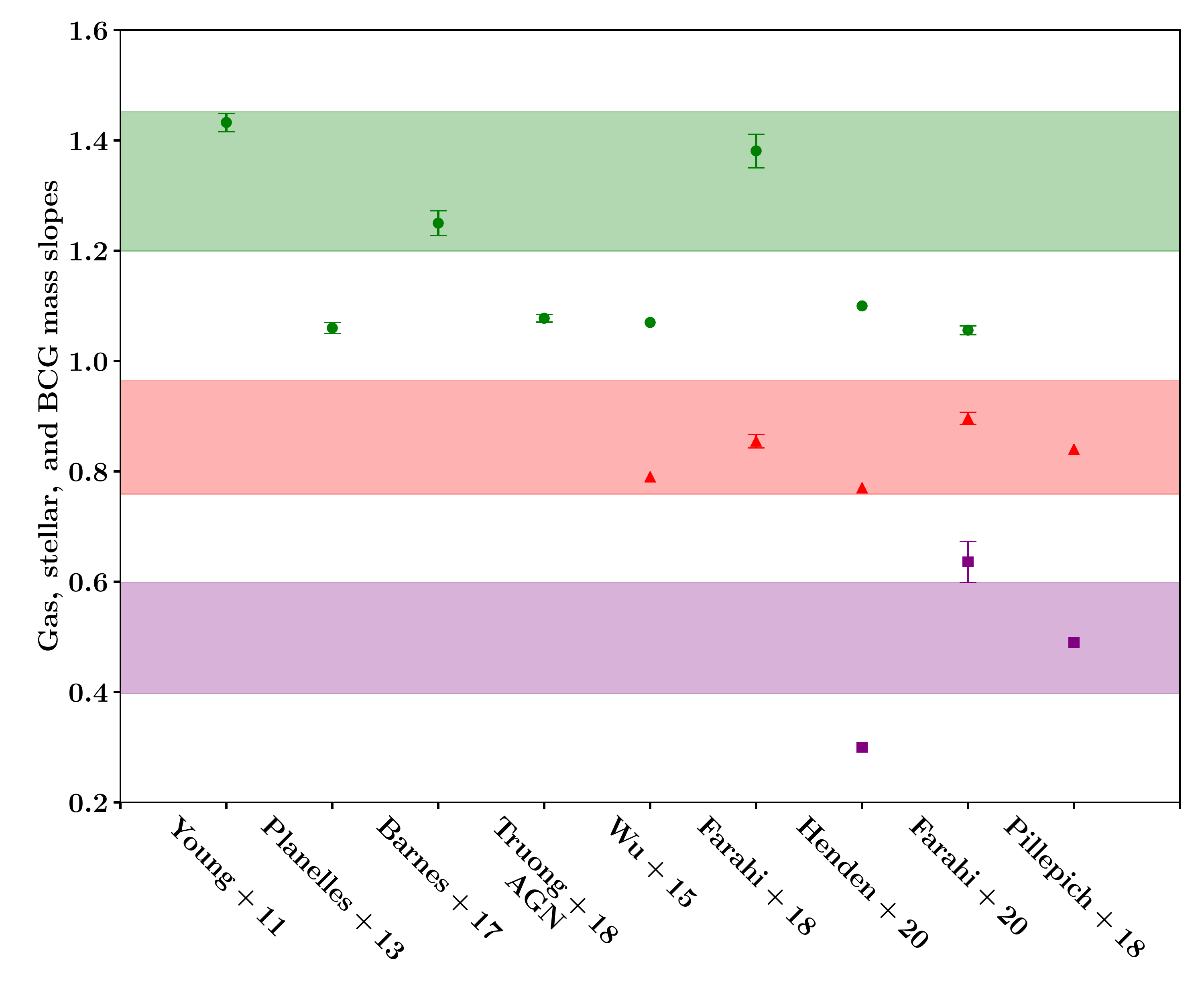}
    \caption{Comparison of mass-dependent slopes $\bm{\beta}$ with those of numerical simulations \citep{2011MNRAS.413..691Y,2013MNRAS.431.1487P,2017MNRAS.465..213B,2018MNRAS.474.4089T,2015MNRAS.452.1982W,2018MNRAS.478.2618F,2020MNRAS.498.2114H,2020MNRAS.493.1361F,2018MNRAS.475..648P}. Green, red, and magenta colour regions denote the $1\sigma$ uncertainties of gas, stellar, and BCG mass slopes, respectively. }
    \label{fig:beta}
\end{figure}

We compare the slopes in the scaling relations 
at $\Delta=500$ (Figure \ref{fig:beta}). The slopes of some numerical simulations depend on the halo mass. For instance, \citet{2020MNRAS.493.1361F} showed that the slopes of the gas mass and stellar mass scaling relations at massive clusters ($M_{500}\sim10^{15}M_\odot$) are close to unity and becomes steeper and shallower with decreasing mass, respectively. We therefore estimate the average value and scatter with a weight of the resulting parent population $p(Z,z)$ to fairly compare with their values in our mass range. 
We use the results being as close as possible to our median redshift and consider the redshift dependence of $p(Z,z)$.
The gas mass slopes of numerical simulations 
\citep{2011MNRAS.413..691Y,2013MNRAS.431.1487P,2017MNRAS.471.1088B,2015MNRAS.452.1982W,2018MNRAS.474.4089T,2018MNRAS.478.2618F,2020MNRAS.498.2114H,2020MNRAS.493.1361F} are higher than predicted by the self-similar model ($\beta=1$). 
Some simulations are slightly steeper than the self-similar expectation ($1<\beta_g<1.1$) while others have a clear higher slope ($\beta_g\simgt 1.2$).
The simulation results are not converged. Our results  agree with the former case \citep[e.g.][]{2011MNRAS.413..691Y,2017MNRAS.465..213B,2018MNRAS.478.2618F}. 
The slopes of the stellar to total mass relation
\citep{2015MNRAS.452.1982W,2018MNRAS.478.2618F,2020MNRAS.498.2114H,2020MNRAS.493.1361F,2018MNRAS.475..648P} are less than unity and agree with our results.  The steep gas slope and the shallow stellar slope are consistent with the physical interpretation that the star formation efficiency is higher in low-mass systems than in high-mass ones (Sec \ref{subsubsec:dis_fb}).

The $M_{\rm BCG}-M$ relation was studied in several numerical simulations \citep[e.g.][]{2014MNRAS.441.1270L,2015MNRAS.451.2703C,2018MNRAS.475..648P,2020MNRAS.493.1361F,2020MNRAS.498.2114H}. 
The evolution of BCGs depends on both star formation efficiency and AGN feedback. In addition, since the BCG is located near cluster center, galaxy-galaxy mergers are an important process for its fast growth. 
Using the UniverseMachine simulation \citep{2019MNRAS.488.3143B}, \citet{2020MNRAS.493..337B} showed the shallow slope and large intrinsic scatter in the $M_{\rm BCG}-M_{200}$ relation. It is explained by
that $M_{\rm BCG}$ is a function of not only halo mass but also halo formation time.

We compare the BCG-total mass slope with numerical simulations at $\Delta=500$. The BCG mass slopes \citep{2020MNRAS.498.2114H,2020MNRAS.493.1361F,2018MNRAS.475..648P} are much shallower than those of the $M_*-M$ and $M_g-M$ relations. We use stellar mass within three dimensional aperture of 30kpc \citep{2018MNRAS.475..648P} for a comparison. Although the BCG slopes show a diversity, they are similar to ours (Figure \ref{fig:beta}). 
We compare with \citet{2018MNRAS.475..648P} because the other two simulations do not show the normalization.
They showed that the normalization of the BCG mass depends on the three-dimensional aperture size.
The normalization with 100 kpc radius is about twice larger than that with 30 kpc. This is caused by the ICL component at the BCG outskirts. They also mentioned that the ICL stellar mass outside $30$ kpc accounts for $\sim40$ percent of the total stellar mass of central galaxies and their surrounding ICL at $M_{200}\sim10^{13}\Msol$ and $\sim80$ percent at $M_{200}\sim10^{15}\Msol$.
We here compare with the normalization measured with 30kpc radius which is the minimum radius discussed in \citet{2018MNRAS.475..648P} and covers the measurement regions of the \texttt{cmodel} magnitude.
The normalization of the BCG mass to total stellar mass ratio of the numerical simulations (Figure \ref{fig:fbcg_ms}; see also Figure \ref{fig:compare_scaling} ) is constantly offset from our baseline by $\sim1.4$ times. We found that our BCG stellar mass estimates agree with those  estimated by the CFHT photometry \citep{2016MNRAS.462.4141L}, as shown in Sec. \ref{subsec:mbcg_obs}. 
When we multiply the BCG mass by 1.3 because of the underestimation of the \texttt{cmodel} magnitude \citep[][and Sec. \ref{subsubsec:blueICL}]{2018MNRAS.475.3348H}, the discrepancy is improved. However, if we accordingly change the aperture size to 100 kpc,  a factor 2-3 discrepancy between the observations and the simulation still remains. We leave for future work to understand the normalization offset.

\subsubsection{Intrinsic covariance in scaling relations}

Intrinsic covariance is one of benchmarks to understand cluster evolution. 
\citet{2018MNRAS.478.2618F} computed how slope, normalization and intrinsic scatter 
change by a halo mass, using both BAHAMAS \citep{2017MNRAS.465.2936M} and MACSIS \citep{2017MNRAS.465..213B} simulations. We weight their intrinsic scatters by $p(Z,z)$
to derive representative values for our sample and obtain $\langle \sigma_g\rangle=0.20\pm0.02$ and $\langle \sigma_*\rangle=0.24\pm0.02$ at $z=0.5$, where the errors are
the $1\sigma$ range over our mass range. \citet{2020MNRAS.493.1361F} obtained the intrinsic scatter of $\sigma_g=0.065\pm0.003$, $\sigma_*=0.098\pm0.004$, and $\sigma_{\rm BCG}=0.333\pm0.015$ using IllustrisTNG simulations \citep{2018MNRAS.475..648P}. 
There is  a discrepancy  between different numerical simulations.
Although their scatter is somewhat lower than our results, the ascending order of intrinsic scatter of each baryon component is the same as our results: $\sigma_g<\sigma_*<\sigma_{\rm BCG}$.

Some numerical simulations \citep[e.g.][]{2015MNRAS.452.1982W,2018MNRAS.478.2618F,2020MNRAS.493.1361F} showed that intrinsic correlation coefficient $r_{*,g}$ at a fixed total mass is negative.
\citet{2015MNRAS.452.1982W} showed a strong negative rank correlation $r_{*,g}=-0.69$ between the deviations of the gas and stellar mass fractions from their baselines, although their definition is different from ours. The negative correlation appears in a wide overdensity range $\Delta=2500-10$. 
They proposed a closed-box scenario where the intrinsic correlation coefficient $r_{*,g}$ is anti-correlated.
\citet{2018MNRAS.478.2618F} also found that intrinsic 
correlation coefficient changes with the halo mass $M_{500}$.
The intrinsic correlation at $z=0.5$ is nearly zero at $M_{500}\sim10^{13}\Msol$, and it is negetive at $\simgt10^{14}\Msol$.
They proposed that non-correlation at $\sim10^{13}\Msol$ is caused by an open baryon box scenario in which the total baryon in low-mass clusters is not conserved by AGN feedback and proposed the closed-box scenario for the negative correlation for high-mass clusters. We recompute $r_{*,g}$ at $M_{500}$ from Figure 5 in \citet{2018MNRAS.478.2618F} with a weight of $p(Z,z)$ and obtain $\langle r_{*,g}\rangle=-0.21\pm0.05$, where the second quantity is the $1\sigma$ range.  \citet{2020MNRAS.493.1361F} also found a similar result $-0.255\pm0.074$ measured at $M_{200}$.  
The probability of accidental correlations from 136 random pairs to realize the simulated result is $\mathcal{O}(10^{-4})$. Although we do not find such a negative correlation, a difference between their and our results is only $1.4\sigma$ with our uncertainty.

\citet{2020MNRAS.495..686A} found a positive intrinsic correlation between central galaxy and total stellar masses at $M_{200}$. The intrinsic correlation coefficient weighted with $p(Z,z)$, $\langle r_{*,\rm{BCG}}\rangle=0.44\pm0.07$, $0.44\pm0.04$, $0.64\pm0.01$ and $0.40\pm0.05$, varies according to the simulation schemes.
\citet{2020MNRAS.493.1361F} found $r_{*,\rm{BCG}}=0.273\pm0.052$ at $M_{200}$. Although the overdensity definitions are different, numerical simulations and our observation suggest that the mass growths of the total stellar mass and the BCG mass are correlated.

\subsubsection{Redshift evolution}

\citet{2020MNRAS.498.2114H} investigated a redshift evolution in scaling relations. They found that the gas mass and the stellar mass at fixed halo mass increases and decreases with increasing redshift as $M_g\propto (1+z)^{0.41\pm0.14}$ and $M_*\propto (1+z)^{-0.51\pm0.08}$, respectively, and the BCG  mass weakly depends on the redshift $M_{\rm BCG}\propto (1+z)^{-0.15\pm0.10}$. They concluded that the gas redshift evolution is attributed to the effectiveness of gas expulsion by AGN feedback with decreasing redshift. \citet{2017MNRAS.466.4442L} also found that the gas mass evolves with redshift as $M_g\propto E(z)^{0.576\pm 0.066}$.
On the other hand, \citet{2013MNRAS.431.1487P} showed that redshift evolution for $f_g$ and $f_*$ are negligible. 
\citet{2018MNRAS.474.4089T} investigated a redshift evolution in X-ray scaling relations and they did not find a significant redshift evolution in the $M-M_g$ relation. Redshift evolution differs by different numerical simulations. 
Since our measurement errors of $\gamma_g$ and $\gamma_*$ are large, we cannot discriminate differences between numerical simulations.

\begin{figure*} 
\begin{center}  
      \includegraphics[width=0.45\hsize]{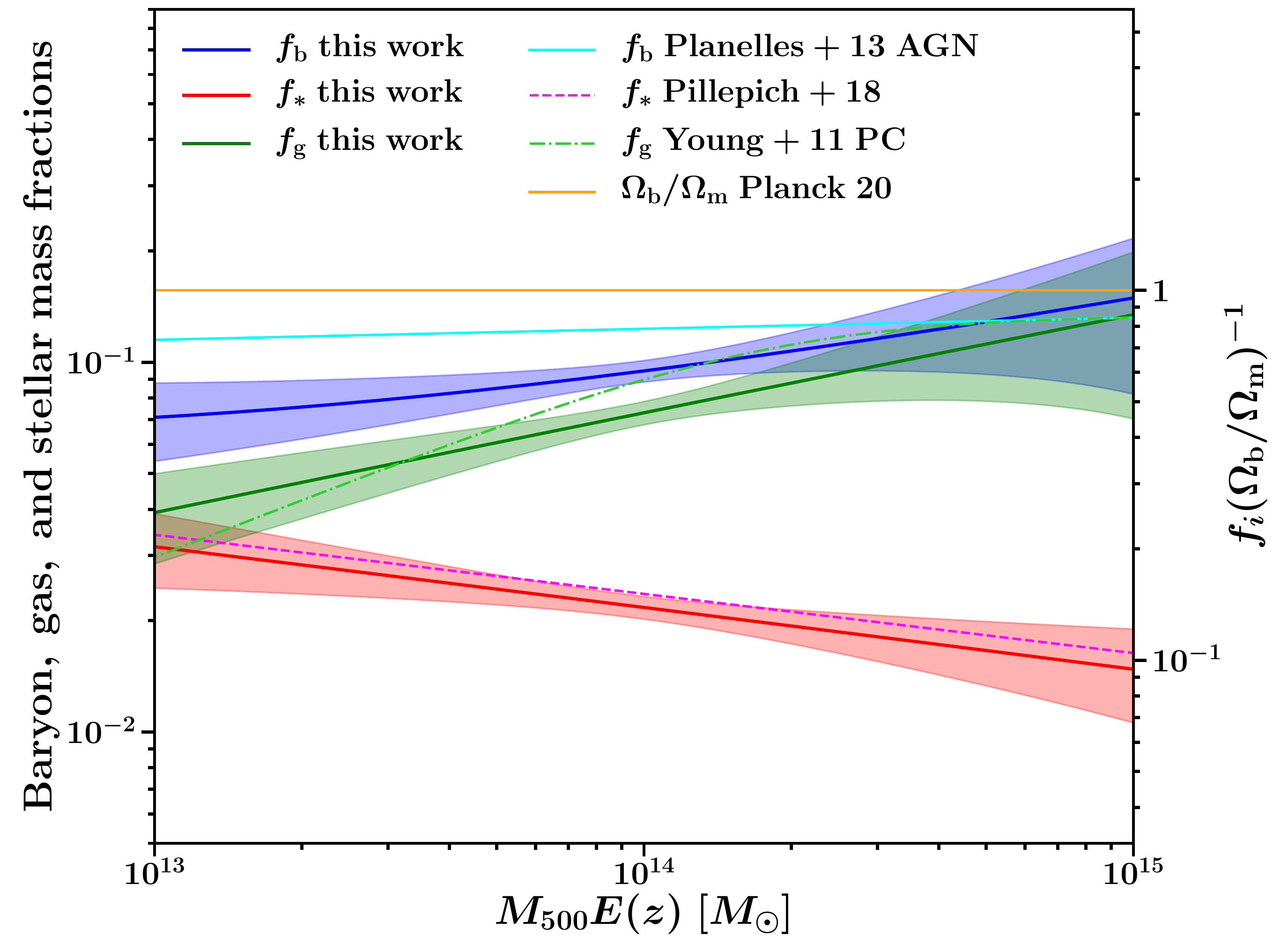} \includegraphics[width=0.45\hsize]{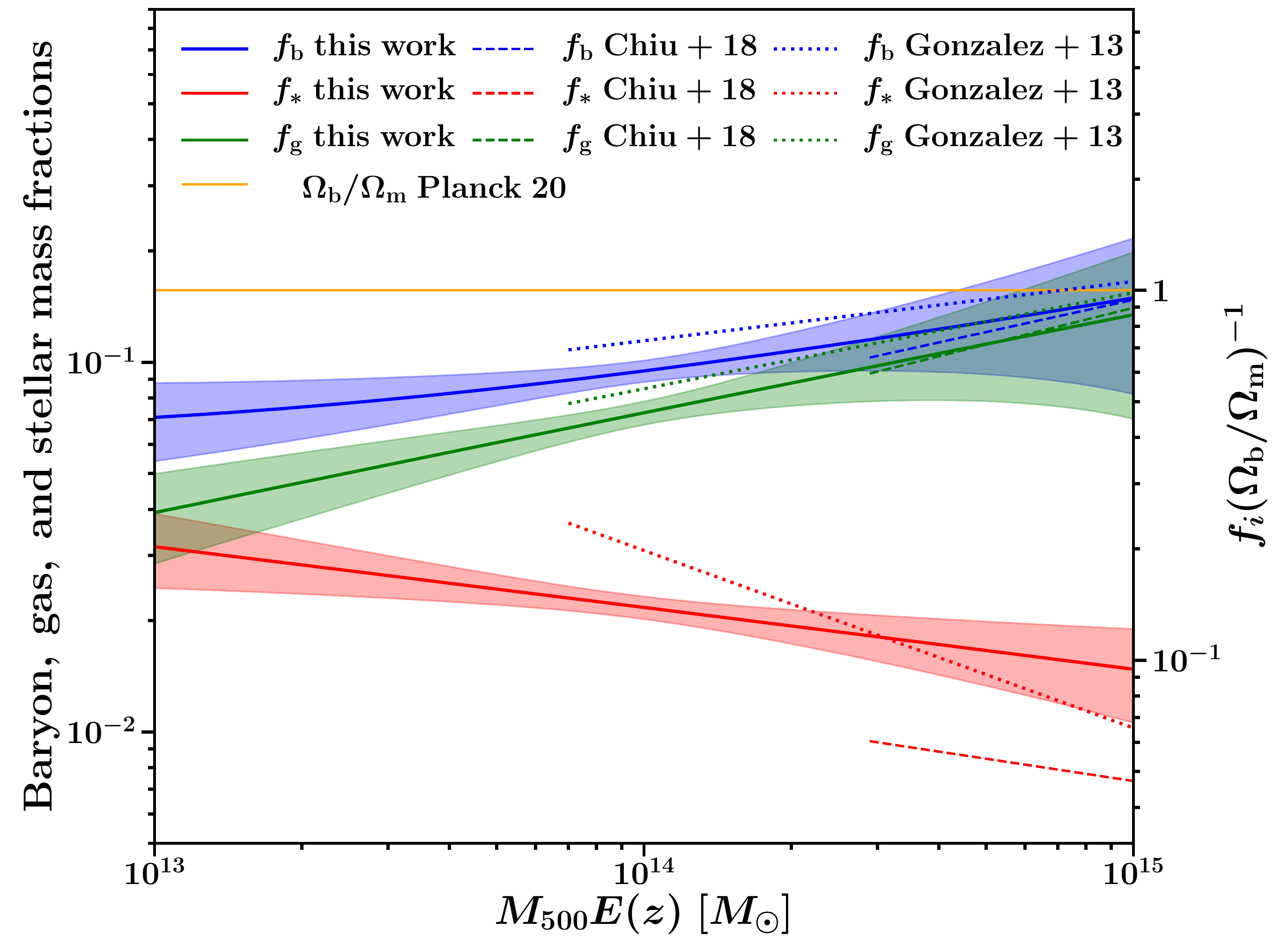}
\end{center}
    \caption{Comparisons of numerical simulations (left) and previously observational studies (right). The solid blue, red and green lines are $f_b$, $f_*$ and $f_g$ of this work, respectively. The orange horizontal line is $\Omega_{b}/\Omega_m$ \citep{2018arXiv180706209P}. In the left panel, the solid light-blue, dashed magenta and long-dashed light-green lines are $f_b$ \citep{2013MNRAS.431.1487P}, $f_*$ \citep{2018MNRAS.475..648P} and $f_g$ \citep{2011MNRAS.413..691Y}, respectively. In the right panel, the dashed blue, red, and green lines are $f_b$, $f_*$ and $f_g$ for the SPT clusters \citep{2018MNRAS.478.3072C}. The dotted blue, red, and green lines are $f_b$, $f_*$ and $f_g$ for the Abell clusters \citep{2013ApJ...778...14G}.  } 
    \label{fig:comparison}
  
\end{figure*}

\subsection{Comparison of observations}

Gas and stellar masses in clusters were measured by various previous papers and projects 
\citep[e.g.][]{Lin03,Lin04,Okabe10c,2012ApJ...745L...3L,2013ApJ...778...14G,2013A&A...555A..66L,2016A&A...592A..12E,2016ApJ...816...98Z,2016MNRAS.458..379C,2017ApJ...851..139L,2018AstL...44....8K,2018MNRAS.478.3072C,2019MNRAS.484...60M,2019NatCo..10.2504F,2020MNRAS.492.4528S}. Since each study adopted a different approach (Table \ref{tab:compare}), it is important to discuss differences in cluster sample, cluster mass measurement, and fitting procedure. 

First, nowadays selection effects are more and more important. A cluster catalog can be constructed from
optical \citep[e.g.][]{Rykoff14,Oguri14b,Rozo16,Oguri18,2019MNRAS.485..498M}, X-ray
\citep[e.g.][]{2004A&A...425..367B,Piffaretti11,2018A&A...620A...5A}, thermal SZ
effect \citep[e.g.][]{PlanckSZ14,2021ApJS..253....3H,2021arXiv210205033B} or weak-lensing
observations \citep[e.g.][]{Miyazaki07,Miyazaki18,2021PASJ..tmp...56O}. 
The Malmquist bias should be properly treated in fitting. Redshift ranges 
also vary with surveys.
After all considerations, further differences could depend on intrinsic selection effects inherent in cluster astrophysics and/or different observational techniques.

Second, cluster mass measurements are one of important sources of
systematic errors. To date, hydrostatic equilibrium mass, weak-lensing mass, or mass derived through scaling relations with mass proxies are used in the literature. A deviation from hydrostatic equilibrium, a lensing mass bias or intrinsic scatter in scaling relations \citep[e.g.][]{Pratt19} should be considered in fitting methods \citep[e.g.][]{Sereno16}.

 Third, when the gas mass and stellar mass are measured within the same overdensity radii of the mass measurement, an error correlation between baryonic observables and mass should be considered in fitting \citep[e.g.][]{Okabe10c}. 

\begin{table*}[ht]
    \caption{Summary of baryon fraction studies in literature and this study. $^a$:cluster finders. 
    X, SZ, opt, and mix represent X-ray, SZ, optical, and mixed (X/SZ/opt) clusters, respectively. Parentheses are samples defined by the XXL Survey, the ROSAT All-Sky Survey(RASS), XMM–Newton–Blanco Cosmology Survey (BCS), South Pole Telescope(SPT),  or random sample (R), respectively. 
    $^b$:sample size.
    $^c$:cluster mass measurement. WL, HE, and proxy represent weak-lensing mass, hydrostatic equilibrium mass, and mass estimated by scaling relations, respectively. 
    $^d$:typical mass in unit of $M_\odot$. The mass ranges are shown in Figures \ref{fig:compare_scaling} and \ref{fig:comparison}. 
    $^e$ correction of mass bias in linear regression. $^f$: scaling relation components (gas mass, total stellar mass, and BCG stellar mass). Some papers include other observables in multivariate scaling relations, which is denoted by $^\dagger$. $^g$: consideration of selection effect in linear regression. $^h$:consideration of the parent population. $G$, K, and mass represent Gaussian distribution, multiple Gaussian distributions using the Bayesian regression code of \citet{2007ApJ...665.1489K} and mass function, respectively. $^i$: aperture-induced error correlation between observable and measured mass. $^\ddagger$ denotes a different approach.  $^j$:intrinsic covariance matrix for baryoninc components, $\bm{\Sigma}_{\rm int}$. full, diag, and partial denotes the full matrix, the diagonal terms, and the matrix  including fixed elements. }
    \vspace{1em}\hspace{-2em}\begin{tabular}{c|cccccccccc}
         & \shortstack{Clusters$^a$\\~} & \shortstack{\footnotesize{Size}$^b$\\~}& \shortstack{\footnotesize{Mass}$^c$\\~} & \shortstack{\footnotesize{Typical} \footnotesize{mass}$^d$\\ $[M_\odot]$} & \shortstack{Mass\\bias$^e$} & \shortstack{Contents$^f$\\~} &  \hspace{-1em}\shortstack{\footnotesize{Selection} \\ \footnotesize{effect}$^g$} \hspace{-1em} & \shortstack{$p(Z)$$^h$} & \shortstack{\footnotesize{Aperture}\\ \footnotesize{$\bm{\Sigma}_{\rm err}$}$^i$} & \shortstack{$\bm{\Sigma}_{\rm int}$$^j$\\~}\\\hline 
         \bf{this work} & $\rm \bf{X(XXL)}$ & $\footnotesize{\bf{136}}$ & \bf{WL} & \footnotesize{\bm{$\sim 10^{14}$}}  &  \bf{yes} &\hspace{-1em} \footnotesize{\bm{$M_{g}/M_*/M_{\rm BCG}$}} \hspace{-1em}&\bf{yes} & $\bm{G}$ & \bf{yes} & \bf{full} \\          
         \cite{2020MNRAS.492.4528S} & $\rm X(XXL)$ & $\footnotesize{118}$ & WL & \footnotesize{$\sim 10^{14} $} & yes & \footnotesize{$M_g$} $^\dagger$ & yes & $G$ & - & full \\
          Eckert et al. (2016) & $\rm X(XXL)$ & $\footnotesize{95}$ & proxy & \footnotesize{$\sim10^{14}$} & no & \footnotesize{$M_g$} & yes & K & no & diag \\
         \cite{2016MNRAS.463.3582M} & $\rm X(RASS)$ & $\footnotesize{27}$ & WL & \footnotesize{$\sim8\times 10^{14}\,$} & yes & \footnotesize{$M_g$} $^\dagger$ & yes & mass & no$^\ddagger$ & partial \\
         \cite{2019MNRAS.484...60M} & $\rm X(RASS)$ & $\footnotesize{41}$ & WL & \footnotesize{$\sim 6\times 10^{14}$} & yes & \footnotesize{$M_g$} $^\dagger$ & yes & mass & yes & full\\
         \cite{2019NatCo..10.2504F} & $\rm X(RASS)$ & $\footnotesize{41}$ & WL & \footnotesize{$\sim 6\times 10^{14}$} & yes & \footnotesize{$M_g$} $^\dagger$ & yes & mass & yes & full\\
         \cite{Vikhlinin09a}  & $\rm X(RASS)$ & $\footnotesize{85}$ & proxy & \footnotesize{$\sim3\times 10^{14}$} & no & \footnotesize{$M_g$} & yes & mass & no & diag  \\
         \cite{2016MNRAS.455..258C} & SZ(SPT)& $\footnotesize{54}$ & proxy & \footnotesize{$\sim 6\times 10^{14}$} & no & \footnotesize{$M_g/M_*/M_{\rm BCG}$} & no & no & no & diag\\
         \cite{2018MNRAS.478.3072C} & SZ(SPT) & $\footnotesize{91}$ & proxy & \footnotesize{$\sim 5\times 10^{14}$} & no & \footnotesize{$M_g/M_*$} & yes & mass & no & diag \\
         \cite{2016MNRAS.458..379C} & $\rm X(BCS)$& $\footnotesize{46}$ & proxy & \footnotesize{$\sim 8\times 10^{13}$} & no & \footnotesize{$M_*$} & yes & mass & no & diag \\
         \cite{2013ApJ...778...14G} & opt(R) & $\footnotesize{15}$ & proxy & \footnotesize{$\sim 2\times 10^{14}$} & no & \footnotesize{$M_g/M_*$} & no & no & no & diag\\
         \cite{2015MNRAS.446.2629E}  & $\rm X$(R) & $\footnotesize{59}$ & HE & \footnotesize{$\sim3\times 10^{14}$} & no & \footnotesize{$M_g$} & no & no & yes & diag  \\
         \cite{2018AstL...44....8K} & opt(R) & $\footnotesize{9}$ & proxy & \footnotesize{$\sim5\times 10^{14}$} & no & \footnotesize{$M_*/M_{\rm BCG}$} & no & no & no & diag \\
         \cite{2016MNRAS.462.4141L} & $\rm X(XXL)$ & $\footnotesize{85}$ & proxy & \footnotesize{$\sim 2\times 10^{14}$} & no & \footnotesize{$M_{\rm BCG}$} & no & no & no & no \\
         \cite{2020MNRAS.491.3751D} & mix(R) & $\footnotesize{23}$ & proxy/WL & \footnotesize{$\sim 10^{14}$} & no & \footnotesize{$M_{\rm BCG}$} & no & no & no & no \\
    \end{tabular}
\label{tab:compare}
\end{table*}

\subsubsection{Gas and stellar mass fractions}  \label{subsubsec:obs}

The left and middle panels of Figure \ref{fig:compare_scaling} compare our results with the $M_g-M$ and $M_*-M$ relations from literature. Most of the previous papers analyzed several tens of clusters. Table \ref{tab:compare} summarizes mass measurements and fitting method. In Figure \ref{fig:compare_scaling}, we use the Chabrier IMF for a comparison of stellar masses.  The $x$- range of each line explicitly describes the mass range of each sample ($M_{500}\simgt 10^{14}\Msol$). 
We multiply the best-fit lines of the literature by $E(z)$ when the literature uses $M_g$ and $M_{500}$ instead of $M_gE(z)$ and $M_{500}E(z)$.
Approaches of the previous papers 
can differ from ours.
Nevertheless, the scaling relations broadly agree with our results. 
We stress the uniqueness of this study: the large sample of the 136 clusters with the nearly two orders of magnitude in mass including low-mass clusters of $\mathcal{O}(10^{13}\Msol)$ and our Bayesian analysis method fully considering the error covariance matrix, the selection effect, and the weak-lensing mass calibration.

We compare our $M_g-M$ relation with the previous XXL papers \citep{ 2016A&A...592A..12E,2020MNRAS.492.4528S}. \citet{2016A&A...592A..12E} studied the gas mass fraction for the 100 XXL brightest cluster sample \citep{2016A&A...592A...2P}. They used a temperature as mass proxy calibrated with a mass and temperature scaling relation based on a CFHTLenS  \citep{2012MNRAS.427..146H,2013MNRAS.433.2545E,2016A&A...592A...4L}. We have updated weak-lensing masses to the HSC-SSP shape catalog \citep{HSCWL1styr,2020ApJ...890..148U}. Since overdensity radii are also changed, we accordingly re-estimate gas mass and properly propagate the errors of weak-lensing mass in the scaling relation analysis. We also correct for the weak-lensing mass calibration \citep{2020ApJ...890..148U}. In the end, our main result of the $M_g-M$ relation is about $30$ percent higher than that of \cite{2016A&A...592A..12E} at $M_{500}\sim10^{14}\Msol$. This point was already discussed in \citet{2020MNRAS.492.4528S}. 
\citet{2020MNRAS.492.4528S} carried out the Bayesian analysis for the 118 XXL subset clusters using the HSC-SSP weak-lensing mass \citep{2020ApJ...890..148U} and the gas mass measured within different radius. Their measurement radius is computed by an iterative procedure using the
surface brightness profile and the $f_g-M_{500}$ relation from \citet{2016A&A...592A..12E}. Their gas mass slope, $1.55\pm0.30$, is consistent with ours within errors, although their line in Figure \ref{fig:compare_scaling} seems to be steeper. 
More conservatively, we repeat the regression analysis for the 118 clusters used in \citet{2020MNRAS.492.4528S} and find $\alpha_g=1.96_{-0.08}^{+0.07}$ and $\beta=1.14_{-0.10}^{+0.12}$. The difference in the slope is $1.4\sigma$, where $\sigma$ is the error from \citet{2020MNRAS.492.4528S}. The normalization of \citet{2020MNRAS.492.4528S} is $\sim 4_{-4}^{+1}$ times higher than ours at $M_{500}E(z)=10^{15}M_\odot$ and $\sim 0.4_{-0.1}^{+0.6}$ times of ours at $10^{13}M_\odot$. The gas mass fraction of \citet{2020MNRAS.492.4528S} becomes higher than the cosmic baryon fraction $\Omega_b/\Omega_m$ at $M_{500}\simgt 4\times10^{14}M_\odot$, while our result is lower at $M_{500}\simlt 10^{15}M_\odot$. The two measurements are marginally consistent within the large error of \citet{2020MNRAS.492.4528S}.

We compare with gas mass scaling relations from Weighing the Giants \citep[WtG;][]{2016MNRAS.463.3582M} and Local Cluster Substructure Survey \citep[LoCuSS;][]{2019MNRAS.484...60M}. 
The two surveys select their cluster samples from ROSAT All Sky Survey catalogues \citep[RASS;]{1998MNRAS.301..881E,2000MNRAS.318..333E,2004A&A...425..367B} and comprise some of the most massive clusters ($M_{500}\sim 4\times10^{14}M_\odot$). The redshift ranges for WtG and LoCuSS are redshifts $0<z<0.5$ and $0.15<z<0.3$, respectively.
They carried out analyses of multivariate scaling relations based on individual weak-lensing masses, taking into account both selection effect and cluster mass function. We cannot find an explicit description about 
an aperture-induced error correlation between two baryoninc observables (Sec \ref{subsubsec:errormatrix}; e.g. $r_{*,g}^{\rm err}$). 

For WtG, the result of \citet{2016MNRAS.463.3582M} is similar to ours. However, their slope using cluster data 
is $\beta_g=1.00\pm0.01$. Their measurement error seems to be extremely small. Their slope is $\sim 3\sigma$ lower than ours where $\sigma$ is our measurement error.

For LoCuSS, \citet{2019MNRAS.484...60M} gives shallower slope $\beta_g=0.7\pm0.1$, which is a $\sim 6\sigma$ difference from ours. When they applied the Bayesian code of \citet{2007ApJ...665.1489K} to single $M_g-M$ scaling relation, the slope $\beta_g=0.99^{+0.13}_{-0.14}$ becomes consistent with the slope of WtG. Therefore, the shallow gas mass slope of their main result might be related to their fitting code. When we adopt the result of \citet{2007ApJ...665.1489K} for \citet{2019MNRAS.484...60M}, the gas mass slopes for massive clusters are close to unity. Numerical simulations \citep[][see also Sec \ref{subsubsec:slope_sim}]{2018MNRAS.478.2618F} pointed out that the gas mass slope for very massive clusters $\sim 10^{15}M_\odot$ is close to unity and becomes steeper with decreasing mass. 
The difference of slopes between ours and WtG/LoCuSS can be ascribed to
the difference of halo mass. Since the two papers \citep{2016MNRAS.463.3582M,2019MNRAS.484...60M} did not include stellar masses in their analyses, we cannot compare with them.

We next compare to the stellar mass fractions from literature. 
The XMM–Newton–Blanco Cosmology Survey \citep[XMM-BCS;][]{2016MNRAS.458..379C} estimated the $M_*-M_{500}$ relation for a sample of 46 clusters based on masses estimated through a scaling relation and found a shallow slope,  $M_*\propto M_{500}^{0.69\pm0.15} (1+z)^{-0.04\pm0.47}$, with negligible redshift evolution. 
\citet{2017ApJ...851..139L} studied the $M_*-M_{200}$ relation for optically selected clusters \citep[CAMIRA;][]{Oguri18} using HSC-SSP data and found that the relation for the optical clusters are similar to those of X-ray selected clusters \citep{2012ApJ...745L...3L}.
\citet{2013ApJ...778...14G} studied gas mass and stellar mass fraction of optical Abell clusters \citep{1989ApJS...70....1A} in the local Universe, with a hydrostatic-mass-based scaling relation.  
Their photometry includes ICL, but their stellar mass fraction is about half of ours. 
\citet{2018MNRAS.478.3072C} studied baryon budget in 91 SZ clusters ($0.2<z<1.25$) selected by the South Pole telescope (SPT) using  $M_{500}$ masses obtained through a scaling-relation and found $M_*\propto M_{500}^{0.80\pm0.12}(1+z)^{0.05\pm0.27}$ (see also Figure \ref{fig:compare_scaling}). 
Their stellar mass slope agrees with ours, though the normalization of their stellar mass fraction is about $70$ percent lower than ours.
\citet{2019ApJ...878...72D} studied stellar mass fraction for massive ($M_{500}\simgt 2\times10^{14} M_\odot$) and high-redshift ($z\simgt 0.93$) clusters selected by an infrared survey, the Massive and Distant Clusters of the WISE Survey (MaDCoWS) and the SPT Survey. Their stellar mass fraction is consistent with optically-selected clusters \citep{2013ApJ...778...14G}, and thus lower than ours. They also found that the difference of averaged stellar mass fraction between infrared and SZ selected sample is not significant.

The right panel of Figure \ref{fig:comparison} shows the gas and stellar mass fractions of the SZ \citep{2018MNRAS.478.3072C} and optically selected clusters  \citep{2013ApJ...778...14G} for a comparison of different cluster finders (Table \ref{tab:compare}). Their mass ranges are higher than $\sim 10^{14}\Msol$. 
The gas mass fraction in the SZ and optically selected clusters are similar to ours.
We rescale the stellar mass fraction of \citet{2013ApJ...778...14G} 
which used the Salpeter IMF by the Chabrier IMF. Although they include the component of the ICL, the stellar mass fraction for the optical clusters is lower than those in the XXL and SPT clusters.
The gas mass fractions are similar to each other irrespective of the cluster finding methods,
 while the stellar mass fractions are slightly different.

\subsubsection{$M_{\rm BCG}-M$ relation}\label{subsec:mbcg_obs}

Literature results\citep{2016MNRAS.455..258C,2016MNRAS.462.4141L,2018AstL...44....8K,2020MNRAS.491.3751D} show that the mass-dependence slope in the $M_{\rm BCG}-M_{500}$ relation is $\beta_*\sim 0.4-0.6$. Analyses can differ by methodology: mass measurements, photometric measurements, inclusion or exclusion of the ICL. Nevertheless, the previous studies are comparable to our result within a factor 2-3
(Figure \ref{fig:compare_scaling}).  Other papers \citep{2012MNRAS.427..550L,2016MNRAS.460.2862B,2016ApJ...816...98Z,2017ApJ...851..139L,2019A&A...631A.175E} using $M_{200}$ also reported such a shallow slope.

\cite{2016MNRAS.455..258C} measured the $M_*-M$ relation without the ICL component. Both the slope, $0.42\pm0.07$, and the normalization are similar to ours. 

\cite{2020MNRAS.491.3751D} measured the BCG stellar masses including the ICL component for 42 galaxy groups and clusters at $z=0.05-1.75$, using {\it Hubble Space Telescope} (HST) data. The slope, $0.37\pm0.05$, is similar to ours, while the normalziation is $\sim 1.2-1.8$ times higher than ours at $10^{15}M_\odot$ and $2\times 10^{13}M_\odot$. They also found that the stellar envelope masses at $10\,{\rm kpc}< r < 100$ kpc are $\sim50-80\%$ of the stellar masses within $r<100$ kpc at $M_{500}\sim2\times10^{13}-10^{15}M_\odot$. The normalization offset would be partially due to the ICL component.

\cite{2018AstL...44....8K} studied the $M_{\rm BCG}-M$ relation for nine nearby clusters at $z\simlt0.1$, using the SDSS DR8. They measured light profiles using raw images in order to account for the ICL component at the BCG outskirts and covert the stellar mass using the mass and light relation. The best-fit baseline for the nine clusters is shown by magenta dashed line in the right panel of Figure \ref{fig:compare_scaling}. Their slope, $0.39\pm0.17$, is similar to ours, while their normalization is about twice higher than ours. We compute the BCG masses for their clusters using the same data and find the weighted geometric mean ratio, $\langle M_{\rm BCG}^{\rm our}/M_{\rm BCG}^{K}\rangle =0.42\pm0.01$. When we use the SDSS DR16, $\langle M_{\rm BCG}^{\rm our}/M_{\rm BCG}^{K}\rangle =0.53\pm0.02$. The discrepancy between the normalization can be explained by a difference of treatment of the ICL component. \citet{2018AstL...44....8K} have also pointed out that their $r$-band luminosities are twice or more higher than those estimates by the \texttt{cmodel} magnitude, which is consistent with our comparison.

\citet{2016MNRAS.462.4141L} investigated the $M_{\rm BCG}-M$ relation for the XXL 100 brightest cluster sample \citep{2016A&A...592A...2P} using the scaling-relation-based masses and found a steep slope $\beta_{\rm BCG}=1/(0.84\pm0.09)=1.19_{-0.12}^{+0.14}$. It disagrees with our result and the aforementioned papers. We found that our BCG identifications are the same as them in the common 35 clusters and the average relationship of the BCG masses with the Salpeter IMF is $M_{\rm BCG}^{\rm Lavoie}=(1.09\pm0.07)M_{\rm BCG}$.
Therefore, this discrepancy is likely caused by the different fitting methods. 
Since they used the BCG mass as the $x$ quantity and ignored their measurement error at each fitting run, 
their underestimation of the slope ($1/\beta_{\rm BCG}=0.84$) in their fitting method could be caused by the regression dilution effect (Appendix \ref{app:a} and \ref{app:b}). We also mention that the $\chi^2$ minimization using the likelihood function without the determinant term is not adequate for scaling relation analysis because the determinant (eq. \ref{eq:likelihood}) includes the parameters of the slopes and intrinsic scatter and one cannot ignore it \citep[see details;][]{Okabe10c}.  We carried out their $\chi^2$ minimization method that they randomly pick up values from the normal distribution with a mean of the observables and a standard deviation of the measurement errors, and then confirmed to recover their steep slope.
When we properly treat the measurement errors and the regression dilution effect in our code, we find $\beta_{\rm BCG}=0.72_{-0.17}^{+0.24}$ for their sample. The value is still stepper but does not conflict with our result.

\citet{2018A&A...620A..13R} studied the luminosity function of 142 XXL selected clusters and found by the $\chi^2$ minimization that the median BCG magnitude is brighter with both redshift and richness. The median BCG luminosity is proportional to $(1+z)^{1.12\pm0.28}$ and $\lambda^{0.24\pm0.08}$, recomputed from Table 4 of \citet{2018A&A...620A..13R},  where $\lambda$ is the cluster richness. 
Although it is difficult to fairly compare with our results because the observables are not the same, the shallow slope of mass dependence is similar to our results and a difference of redshift evolution is only $\sim2\sigma$.

\subsubsection{Intrinsic covariance}

\citet{2019NatCo..10.2504F} found an anti-correlation between intrinsic scatter of the gas mass and the K-band luminosity of galaxies from 41 LoCuSS clusters of average mass  $M_{500}\sim4\times10^{14}M_\odot$.  The pairwise correlation coefficient is $-0.56_{-0.28}^{+0.36}$. We did not find significant negative correlation coefficient at $M_{500}\simgt 10^{13}\Msol$. 
We compute the probability of accidental anti-correlation for their 41 clusters and find $\mathcal{P}(r\ge |r_{*,g}|)<0.2$, which is not significant. 
We compute a probability to accidentally realize their correlation coefficient using our 136 clusters, and obtain $<2\times10^{-2}$. However, we do
 not find such a negative correlation. Since we cannot rule out a possibility that the intrinsic correlation 
 depends on halo mass, we need a larger sample of clusters for a more precise measurement. 

To our knowledge, the positive correlation coefficients between the BCG and total stellar mass and between the stellar mass of BCG and satellite galaxies (Sec. \ref{subsec:Sigma_int}) have not been reported in previous observational studies.

\section{Summary}\label{sec:sum}

We carried out Bayesian analysis for the multivariate scaling relations of the baryonic components of the 136 XXL groups and clusters over a wide range of nearly two orders of magnitude in mass at $0\simlt z\simlt 1$.We combined the HSC-SSP weak-lensing mass measurements \citep{2020ApJ...890..148U}, the XXL X-ray gas measurements, and the HSC-SSP and SDSS photometry. Bayesian regression simultaneously and consistently takes into account weak-lensing mass calibration, selection effect, error covariance matrix, and intrinsic covariance. The analysis constrains well the slopes, normalizations, and intrinsic covariance among the baryonic component masses. 
Our method models the parent population together with the scaling relations.
Thanks to this modeling, we can correct for selection effect and regression dilution effect. The method does not deal with cosmological inference and it is not affected by uncertainties in the estimation of, e.g. $\sigma_8$.

The slope of the gas mass and cluster mass scaling relation is $1.29_{-0.10}^{+0.16}$ steeper than predicted by the self-similar model ($\beta=1$), while the slope of the stellar mass is $0.85_{-0.09}^{+0.12}$ shallower (Table \ref{tab:chabrier_4d}, Figures \ref{fig:fb} and \ref{fig:beta}, and Sec. \ref{subsec:sc_result}).  As shown in  Figure \ref{fig:fb} and Sec. \ref{subsec:fb}, the gas mass fraction increases from $f_g(\Omega_b/\Omega_m)^{-1}\sim0.3$ at $M_{500}E(z)\sim10^{13}\Msol$ to $\sim0.9$ at $\sim10^{15}\Msol$. The stellar mass fraction decreases from $f_*(\Omega_b/\Omega_m)^{-1}\sim0.2$ at $M_{500}E(z)\sim10^{13}\Msol$ to $\sim0.1$ at $\sim10^{15}\Msol$. Accordingly, the total baryon fraction increases $f_b(\Omega_b/\Omega_m)^{-1}\sim0.5$ at $M_{500}E(z)\sim10^{13}\Msol$ to $\sim0.6$ at $\sim10^{14}\Msol$ and $\sim1.0$ at $\sim10^{15}\Msol$. 
The low baryon fraction implies that clusters in our mass range are likely to be open-box systems. 
The baryon, gas mass, and stellar mass fractions as a function of $M_{500}$ agree with previous numerical simulations and some previous observations (Figure \ref{fig:comparison} and Sec \ref{subsubsec:slope_sim} and \ref{subsubsec:obs}). 
Our analysis can differ form previous works for the treatment of the aperture radius and the mass calibration.
The slope of the BCG stellar mass is $0.49_{-0.10}^{+0.11}$ shallower than the other two slopes, indicating a weak-mass dependence (Table \ref{tab:chabrier_4d} and Sec. \ref{subsec:sc_result}). We do not find a significant evidence of redshift revolution in the scaling relations (Table \ref{tab:camira_z} and Sec \ref{subsec:zev}) because of their large errors.

The intrinsic scatter is ranked as $\sigma_g<\sigma_*<\sigma_{\rm BCG}$, as numerical simulations \citep{2018MNRAS.475..648P,2018MNRAS.478.2618F}.
We found a positive intrinsic correlation coefficient between stellar mass and BCG stellar mass (Table \ref{tab:Sigma_int_4d} and Sec \ref{subsec:Sigma_int}), which is statistically significant, in agreement with numerical simulations \citep{2020MNRAS.495..686A,2020MNRAS.493.1361F}. 
The intrinsic correlation between gas and stellar mass shows no positive nor negative correlation (Table \ref{tab:Sigma_int_4d} and Sec \ref{subsec:Sigma_int}), but the statistical significance is marginal.

We do not find a significant difference between the clusters with and without central radio sources (Table \ref{tab:chabrier_sub}, Figure \ref{fig:fbcg_agn}, and Sec \ref{subsec:C1C2}). It implies that the cumulative quantity such as the gas and stellar masses are insensitive to the current radio AGN activity. The intrinsic scatter of the total and BCG stellar masses in the C2 clusters with lower mass is larger than that of the C1 clusters with higher mass  (Table \ref{tab:Sigma_int_sub}).

This paper comprises the largest sample of X-ray selected clusters over a wide range of nearly two orders of magnitude in mass. Studies of the mass-dependence of the intrinsic covariance and the local slope are out of scope and they could be at reach of the final sample of the XXL clusters with the XXL selection function depending on the cosmological parameters. Studies on subsamples divided by mergers \citep[e.g.][]{2019PASJ...71...79O,2021MNRAS.501.1701O} will be also done. Analyses of optically-selected \citep[e.g.][]{Rykoff14,Oguri14b,Rozo16,Oguri18,2019MNRAS.485..498M}, SZ \citep[e.g.][]{PlanckSZ14,2021ApJS..253....3H,2021arXiv210205033B} and shear-selected \citep[e.g.][]{Miyazaki18,2021PASJ..tmp...56O} clusters with more than 100 sample sizes are essential to understand the selection effect of cluster finders on baryonic physics.


\section*{Acknowledgments}
We gratefully thank the anonymous referee for careful reading and helpful comments.
The Hyper Suprime-Cam (HSC) collaboration includes the astronomical communities of Japan and Taiwan, and Princeton University.  The HSC instrumentation and software were developed by the National Astronomical Observatory of Japan (NAOJ), the Kavli Institute for the Physics and Mathematics of the Universe (Kavli IPMU), the University of Tokyo, the High Energy Accelerator Research Organization (KEK), the Academia Sinica Institute for Astronomy and Astrophysics in Taiwan (ASIAA), and Princeton University.  Funding was contributed by the FIRST program from the Japanese Cabinet Office, the Ministry of Education, Culture, Sports, Science and Technology (MEXT), the Japan Society for the Promotion of Science (JSPS), Japan Science and Technology Agency  (JST), the Toray Science  Foundation, NAOJ, Kavli IPMU, KEK, ASIAA, and Princeton University.

This paper makes use of software developed for the Large Synoptic Survey Telescope. We thank the LSST Project for making their code available as free software at  http://dm.lsst.org

This paper is based [in part] on data collected at the Subaru Telescope and retrieved from the HSC data archive system, which is operated by Subaru Telescope and Astronomy Data Center (ADC) at NAOJ. Data analysis was in part carried out with the cooperation of Center for Computational Astrophysics (CfCA), NAOJ. We are honored and grateful for the opportunity of observing the Universe from Maunakea, which has the cultural, historical and natural
significance in Hawaii.

The Pan-STARRS1 Surveys (PS1) and the PS1 public science archive have been made possible through contributions by the Institute for Astronomy, the University of Hawaii, the Pan-STARRS Project Office, the Max Planck Society and its participating institutes, the Max Planck Institute for Astronomy, Heidelberg, and the Max Planck Institute for Extraterrestrial Physics, Garching, The Johns Hopkins University, Durham University, the University of Edinburgh, the Queen's University Belfast, the Harvard-Smithsonian Center for Astrophysics, the Las Cumbres Observatory Global Telescope Network Incorporated, the National Central University of Taiwan, the Space Telescope Science Institute, the National Aeronautics and Space Administration under grant No. NNX08AR22G issued through the Planetary Science Division of the NASA Science Mission Directorate, the National Science Foundation grant No. AST-1238877, the University of Maryland, Eotvos Lorand University (ELTE), the Los Alamos National Laboratory, and the Gordon and Betty Moore Foundation.

XXL is an international project based around an XMM Very Large Programme surveying two 25 deg$^2$ extragalactic fields at a depth of $\sim6 \times 10^{-15}\, {\rm erg\,cm^{-2}\,s^{-1}}$ in the [0.5-2] keV band for point-like sources. The XXL website is http://irfu.cea.fr/xxl. Multi-band information and spectroscopic follow-up of the X-ray sources are obtained through a number of survey programmes, summarised at http://xxlmultiwave.pbworks.com/.

This work was supported by Core Research for Energetic Universe in Hiroshima University (the MEXT
 program for promoting the enhancement of research universities, Japan).
MS acknowledges financial contribution from contract ASI-INAF n.2017-14-H.0 and INAF `Call per interventi aggiuntivi a sostegno della ricerca di main stream di INAF'.
K.U. acknowledges support from the Ministry of Science and Technology
of Taiwan (grants MOST 106-2628-M-001-003-MY3 and MOST
109-2112-M-001-018-MY3) and from the Academia Sinica Investigator
Award (grant AS-IA-107-M01).
SE acknowledges financial contribution from the contracts ASI-INAF Athena 2019-27-HH.0,
``Attivit\`a di Studio per la comunit\`a scientifica di Astrofisica delle Alte Energie e Fisica Astroparticellare''
(Accordo Attuativo ASI-INAF n. 2017-14-H.0), INAF mainstream project 1.05.01.86.10, and
from the European Union’s Horizon 2020 Programme under the AHEAD2020 project (grant agreement n. 871158).


\bibliographystyle{apj}
\bibliography{my,baryon,hsc_tech,hscrefs,xxlpapers}

\newpage
\newpage 

\appendix
\renewcommand{\thesection}{\Alph{section}}
\section{Linear regression for multivariate scaling relations}\label{app:a}

Let us consider an observed dataset $\{x_n,{\bm y}_n\}_{n=1}^{N}$, composed of $D+1$ variables for $N$ sampled clusters, where $\bm y = \{y_0,y_1,\ldots,y_{D}\}$. 
The true variables of $\{X_n,{\bm Y}_n\}_{n=1}^{N} \, (\bm Y = \{Y_0,Y_1,\ldots,Y_{D}\}$) for the $n$-th cluster sample are related to the observables by $p(x_n,\bm{y}_n|X,\bm{Y}_n)=\mathcal{N}(\{x_n,\bm{y}_n\}\big| \{X_n, \bm{Y}_n\}, \bm{\Sigma}_{{\rm err},n})$ with Gaussian error covariance matrix, $\bm{\Sigma}_{\rm err}$. In cluster scaling relation studies, the $x$ value can be a logarithm on the weak-lensing, hydrostatic mass or mass estimates from a scaling relation.
They are affected by scatter and/or bias from the true mass because of some observational systematics \citep[e.g.][]{Becker11,Zhang10,2012MNRAS.425.2287H,Okabe16b,Pratt19,2020ApJ...890..148U}. 
Following \citet{Sereno16},  we introduce the latent variable $Z$ as the true mass \citep[see also][]{2019MNRAS.484...60M,2019NatCo..10.2504F}.  The linear regression equation between $\{X,{\bm Y}\}$ and $Z$ are expressed by $X_Z=\alpha_X +\beta_X Z$ and $\bm{Y}_Z=\bm{\alpha}+\bm{\beta}Z$, respectively. Given the intrinsic covariance matrix \citep[$\bm{\Sigma}_{\rm int}$,  e.g.][]{Okabe10c}, 
the conditional probability, $p(X_n,\bm{Y}_n|Z_n,\bm \theta)$, is a multivariate normal density distribution of $p(X_n,\bm{Y}_n|Z_n,\bm \theta)=\mathcal{N}(\{X_n, \bm{Y}_n\} \big| \{\alpha_X +\beta_X Z_n, \bm{\alpha}+\bm{\beta}Z_n\},\bm{\Sigma}_{\rm int})$, where $\bm{\theta}$ denotes the parameters of the distribution. The element of the intrinsic covariance matrix is specified by
\begin{eqnarray}
      \bm{\Sigma}_{\rm int}=\left(
\begin{array}{ccc}
\sigma_X^2 & r_{X Y_i}\sigma_X\sigma_{Y_i} & r_{X Y_j}\sigma_{X}\sigma_{Y_j}\\
r_{X Y_i}\sigma_X\sigma_{Y_i} & \sigma_{Y_i}^2 &  r_{Y_i Y_j}\sigma_{Y_i}\sigma_{Y_j} \\
 r_{X Y_j}\sigma_{X}\sigma_{Y_j} & r_{Y_i Y_j}\sigma_{Y_i}\sigma_{Y_j} &  \sigma_{Y_j}^2 \\
\end{array}
\right) \label{eq:Sigma_int}
\end{eqnarray}
where $\sigma_X$ and $\sigma_{Y_i}$ are the intrinsic scatter and $r_{XY_i}$ and $r_{Y_iY_j}$ are the intrinsic correlation coefficient between $X$ and $Y_i$ and between $Y_i$ and $Y_j$, respectively.  All the elements in the intrinsic covariance should satisfy the condition of covariance correlation matrix: all the eigenvalues are positive. 
Since the intrinsic scatter is a positive quantity, we use logarithmic quantities, $\ln \sigma_X$ and $\ln \sigma_{Y_i}$, in actual computations to avoid boundary artifacts at zero.

Linear regression of multi-wavelength datasets from a survey must take into account for two systematic effects.
First, the slopes can be underestimated by the measurement errors of the $x$ value, so-called, regression dilution effect \citep{1996ApJ...470..706A,2007ApJ...665.1489K,Sereno16}. 
Second, selection effects, e.g. Malmquist bias, affects sample selected above an observational threshold \citep[e.g.][]{Mantz10,Sereno16,2016MNRAS.463.3582M,2019MNRAS.484...60M,2019MNRAS.485.4863M}. We follow the mathematical formulation of \cite{Sereno16} to overcome the above two problems. 
We introduce the parent population, $p(Z|\bm{\theta})$, assuming the Gaussian distribution, $\mathcal{N}(\mu_Z, \sigma_Z)$, where $\mu_Z$ and $\sigma_Z$ are hyper-parameters. In a generalized case, the parent population can be the summation of multiple Gaussian distributions: $p(Z|\btheta)=\sum_i \pi_i \mathcal{N}(\mu_{Z,i},\sigma_{Z,i})$, where $\pi_i$ is the normalization satisfying with $\sum_i \pi_i=1$.
The standard deviation, $\sigma_Z$, effectively corrects for the regression dilution effect \citep{1996ApJ...470..706A,2007ApJ...665.1489K,Sereno16}. 
The total parameters are $\bm{\theta}=\bm{\alpha},{\bm \beta},\bm{\Sigma}_{\rm int},\mu_Z,\sigma_Z$. 
The selection bias is modelled by truncating the probability distribution with the threshold of $y_{{\rm th},0,n}$ on a tracer $y_0$ for cluster finders, where the subscript $n$ denotes the $n$-th cluster. The Bayesian chain rule gives the likelihood function of the $D$-dimensional scaling relations, as follows, 

\begin{table*}[ht]
\begin{equation}
  p(x,\bm y|\bm \theta)=\frac{\displaystyle \prod_n^{N} \int_{-\infty}^{\infty} \!\!\! dX_n \int_{-\infty}^{\infty} \!\!\!d^{D}{\bm Y}_n\int_{-\infty}^{\infty}\!\!\!  dZ_n\ p(x_n,\bm{y}_n|X_n,{\bm Y}_n)p(X_n,{\bm Y}_n|Z_n,\bm \theta)p(Z_n|\bm{\theta})}{\displaystyle \prod_n^{N}  \int_{y_{\mathrm{th},0,n}}^{\infty}\!\!\! dy_{0,n}\!\!\int_{-\infty}^{\infty}\!\!\! dx_n\!\!  \int_{-\infty}^{\infty}\!\!\! d^{D-1}\bm y_n\!\! \int_{-\infty}^{\infty} \!\!\! dX_n\!\! \int_{-\infty}^{\infty} \!\!\! d^{D}{\bm Y}_n 
    \!\! \int_{-\infty}^{\infty} \!\!\! dZ_n\ p(x_n,{\bm y}_n|X_n, {\bm Y}_n)p(X_n,{\bm Y}_n|Z_n,\bm \theta)p(Z_n|\bm{\theta})} \label{eq:likelihood}
\end{equation}
\end{table*}
\FloatBarrier

where $n$ denotes the $n$-th cluster, $d^{D-1}\bm{y}_n=dy_{n,1}dy_{n,2}\cdots dy_{n,D}$, and $d^{D}\bm{Y}_n=dY_{n,0}dY_{n,1}\cdots dY_{n,D}$. 
Given Bayes' theorem, a conditional probability given the observables is expressed by $p(\bm{\theta}|x,\bm{y})\propto p(x,\bm{y}|\bm{\theta})p(\bm{\theta})$ where $p(\bm{\theta})$ is the prior distribution of the parameters. The method takes into account the intrinsic covariance between cluster properties ($\bm{\Sigma}_{\rm int})$, selection effect ($\sigma_Z$, $\mu_Z$, $y_{\rm th,0}$),  weak-lensing mass calibration ($\alpha_X,\beta_X,\sigma_X$), and observational error covariance matrix ($\bm{\Sigma}_{\rm err})$, for the purpose of the analysis of the multivariate scaling relations. 

We adopt the Markov Chain Monte Carlo (MCMC) method and a biweight estimate of the posterior distributions.  We use a flat prior $[-10^4,10^4]$ on $\bm{\theta}=\bm{\alpha}$, each parameter of the intrinsic covariance of $\bm{\Sigma}_{\rm int}$ (eq. \ref{eq:Sigma_int}), and $\mu_Z$. We employ a Student's $t_1$ distribution with one degree of freedom on $\bm{\beta}$ so that slope angles become uniformly distributed. A non-informative prior distribution on the variance of the parent distribution, $\sigma_Z^2$, follows a scaled inverse $\chi^2$ distribution as a conjugate prior satisfying that posterior distributions have the same probability distribution family as the prior distribution.

When we consider redshift evolution in linear regression and $p(Z)$, the linear regression forms $\alpha_X+\beta_XZ+\gamma_X\ln E(z)$ and $\bm{\alpha}+\bm{\beta}Z+\bm{\gamma}\ln E(z)$ and the parameters of $p(Z,z)$ are described by $\mu_Z(z)=\mu_{Z,0}+\gamma_{\mu_Z}\ln E(z)$ and $\ln \sigma_Z(z)=\ln \sigma_{Z,0}+\gamma_{\sigma_Z}\ln E(z)$. Since we assume no error in cluster redshifts, the regression dilution effect disappears. The prior of  redshift-dependent slopes, $\gamma$, is a Student's $t_1$ distribution with one degree of freedom.

There is also a practical problem of numerical errors in computing the inverse of the composition matrix of the intrinsic covariance and the measurement errors. As aforementioned, all the parameters in the intrinsic covariance and the composition matrix should satisfy the condition of the covariance correlation matrix. We compute the eigenvalues of the two matrices by the parameters randomly and independently proposed in each step. If there is at least one negative eigenvalue, we re-propose new parameters. Otherwise, the inverse matrix calculation has large numerical errors especially in higher dimensions, giving poorly-constrained posterior distributions.

\begin{figure*}
\begin{center}  
       \includegraphics[width=0.45\hsize]{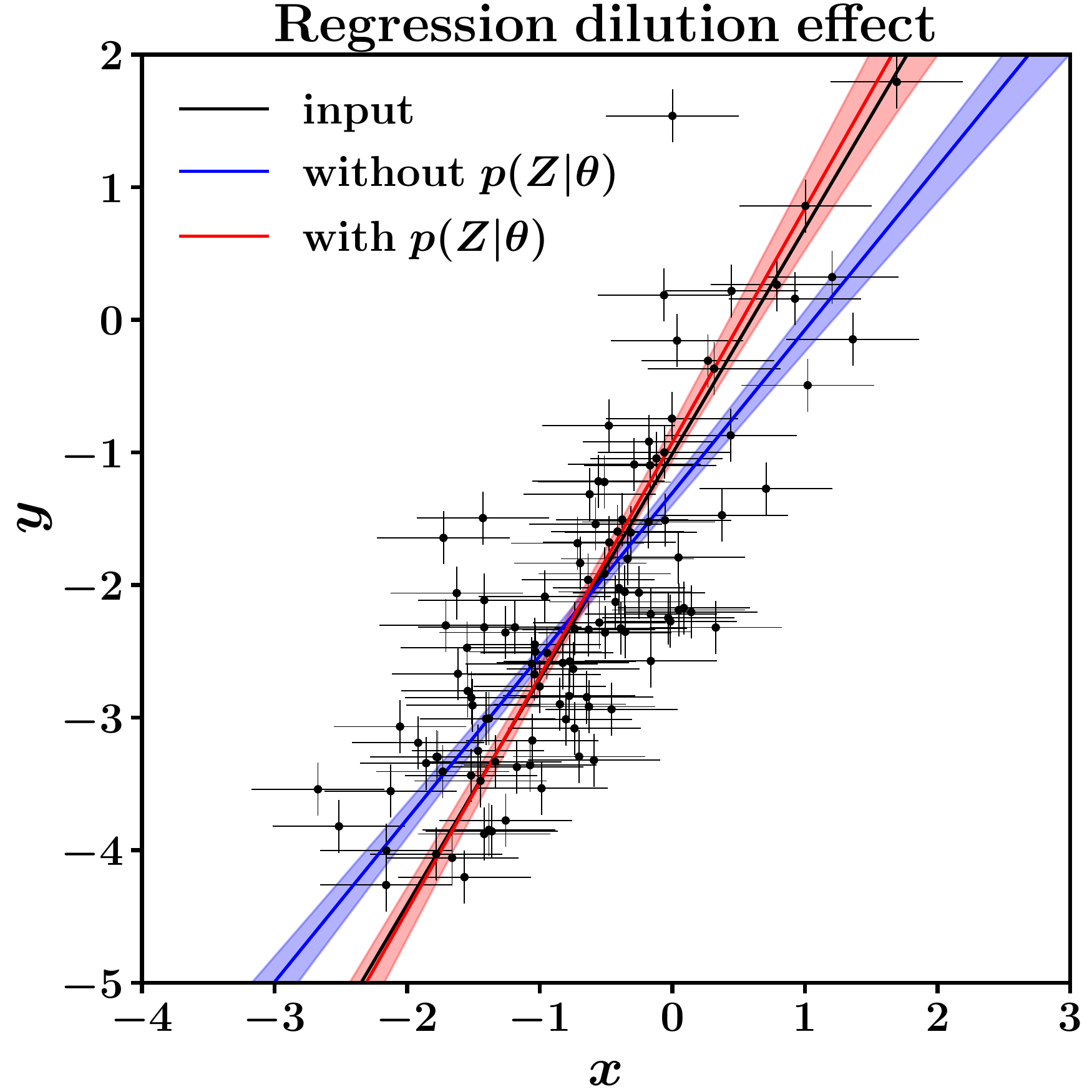} \includegraphics[width=0.45\hsize]{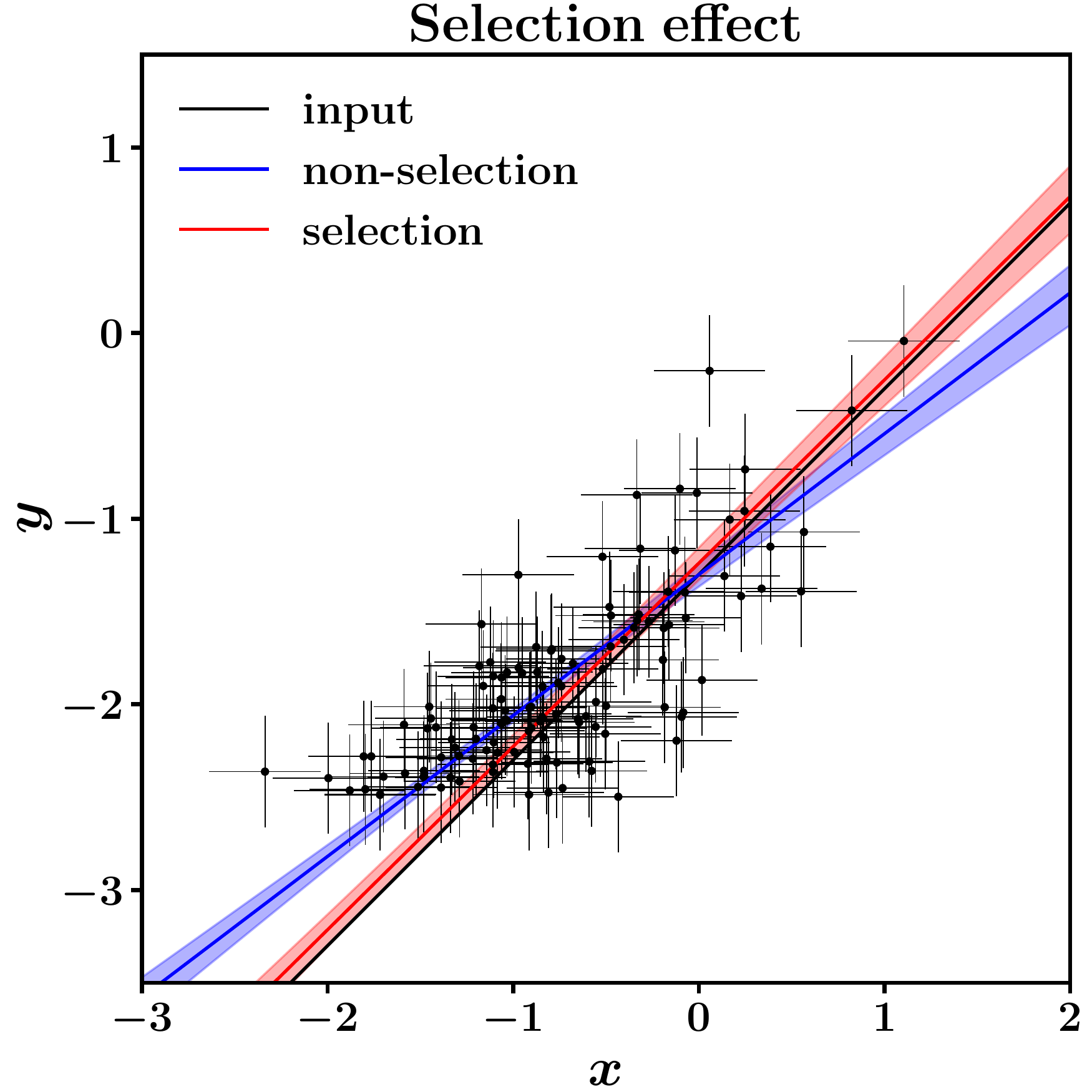}
\end{center}
    \caption{The regression dilution effect (left) and selection effect (right) of the mock simulations. The solid black lines denote the input scaling relations. The solid red lines and transparent regions are the estimate and $1\sigma$ uncertainty by our code. The solid blue lines and transparent regions are the estimate and $1\sigma$ uncertainty
    without a consideration of the regression dilution effect or the selection effect. When we improperly treat these effects, the slopes are underestimated. The data in the right panel is selected above the threshold of $y=-2.5$. 
    } 
    \label{fig:hibrecs}
\end{figure*}

Figure \ref{fig:hibrecs} demonstrates the regression dilution effect and the selection effect in 1D scaling relations from mock simulations including errors and intrinsic scatter.
The left panel uses data independent of any selection processes 
and the right panel uses a sample catalog selected from a mass function 
through a mass-observable scaling relation with the truncated threshold on the $y$ quantity.
The code recovers well the input parameters, while the computation without $p(Z|\bm{\theta})$ (regression dilution effect) or with $y_{{\rm th},0,n}\rightarrow -\infty$ (selection effect) underestimates the slopes.

\begin{figure} 
\begin{center}  
  \includegraphics[width=\hsize]{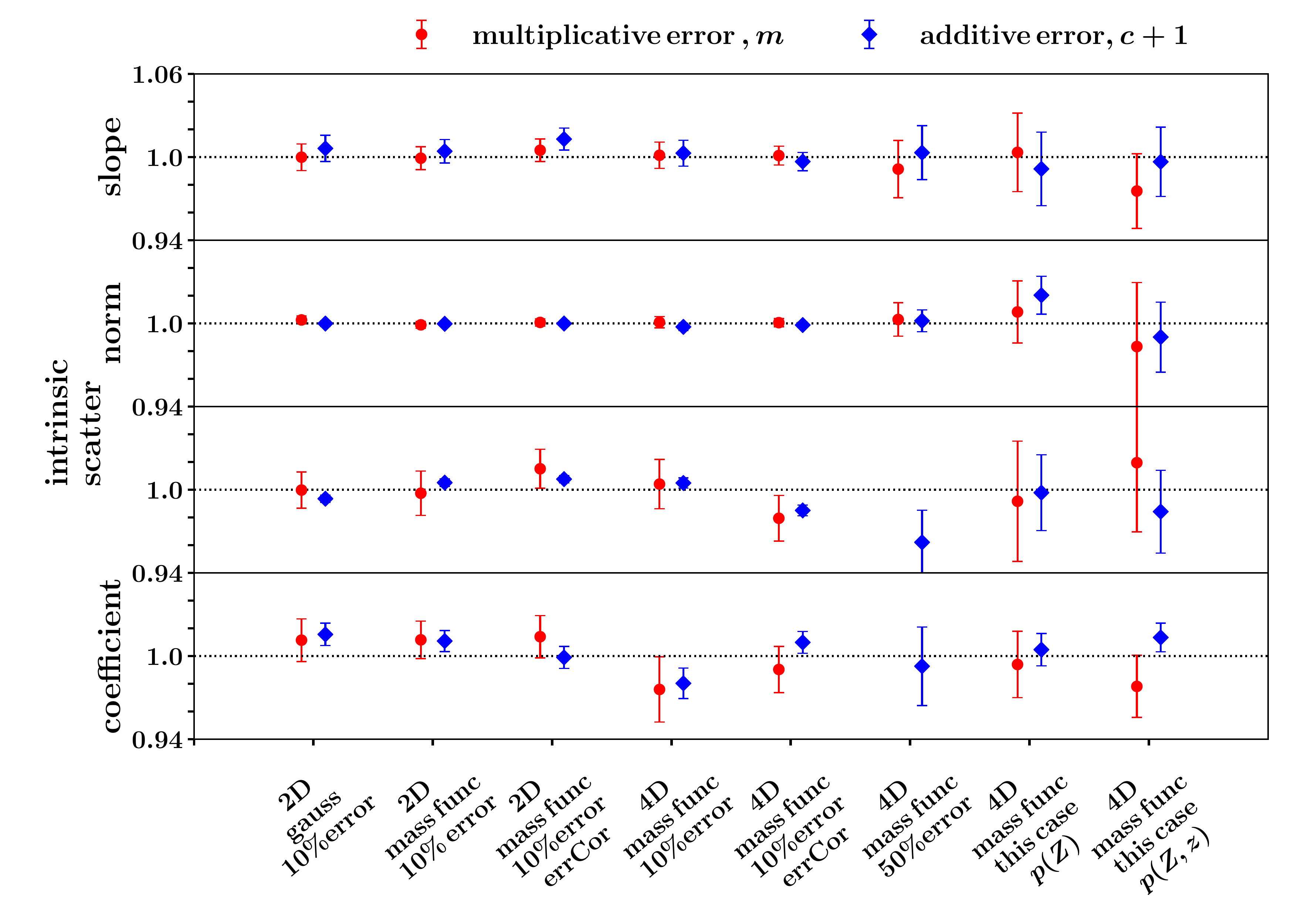}
\end{center}
    \caption{A performance of our Bayesian code of slope, normalization, intrinsic scatter and intrinsic correlation coefficient (from top to bottom). A multiplicative error ($m$; red circles) and an additive error ($c$; blue diamonds) are specified in an equation $\bm{\theta}_{\rm output} = m \times \bm{\theta}_{\rm input} + c$. Each error size corresponds to the error per a parameter.
    We consider 2D and 4D scaling relations with a combination of 10 percent and 50 percent measurement errors in $\{x,\bm{y}\}$ and two parent populations. The parent population of samples is assumed to be either a halo mass function \citep{Tinker08} at $z=0.3$ or Gaussian distribution. Each subset complies 500 simulations. One simulation is composed of $\sim 100$ clusters. The sample clusters are drawn through a tracer $y_0$. Input parameters of scaling relations are uniformly distributed in $\alpha=[0,1], \beta=[0.3,1.7], \sigma_{\rm int}=[0.05,0.4]$, and $r=[-0.8,0.8]$. 
    The code can constrain well slope and normalization, irrespective of a parent distribution. The intrinsic scatter and correlation coefficient can be constrained when the measurement errors are smaller than intrinsic scatter. In the opposite case, neither intrinsic scatter nor correlation coefficient is constrained with a large $1\sigma$ uncertainty.  When we employ the measurement error matrix similar to this study ($p(Z)$ and $p(Z,z)$), the input parameters are recovered.    
    } 
    \label{fig:hibrecs2}
\end{figure}

We further assess the performance of the newly developed code using eight mock simulations, each of which is composed of 500 realizations. 
The first six simulations use a $2D$ or $4D$ scaling relation with different setup parameters. Individual measurement errors in $\{x,{\bm y}\}$ are randomly assigned from $\sigma^{\rm err}=[0.05,0.2]$ or $\sigma^{\rm err}=[0.4,0.6]$, we refer them to as $10\%$ and $50\%$, respectively. 
We assumed two cases with the error correlation $r^{\rm err}=0$ or $r^{\rm err}=[-1,1]$. 
Each cluster is drawn through a tracer $y_0$ from a parent population of Gaussian distribution or a halo mass function of \citet{Tinker08} at $z=0.3$. 
Each parameter combination are referred in the $x$ label of Fig. \ref{fig:hibrecs2}.
We simulate $\sim100$ clusters per run, corresponding to the case of this study.  
In each simulation, the input parameters of the scaling relations are randomly drawn from a uniform distribution in $\alpha_{\bm Y}=[0,1], \beta_{\bm Y}=[0.3,1.7], \sigma_{\bm Y}=[0.05,0.4]$, and $r_{\bm Y}=[-0.8,0.8]$, respectively. The range of intrinsic scatter, $\sigma_{\bm Y}$, is lower than $50\%$ measurement error, and higher than or comparable to $10\%$ error.
We fix external mass calibration parameters ; $\alpha_X=0$, $\beta_X=1$, $\sigma_X=0.1$, and $r_{X,i}=0$. 

We quantify the performance of our code by 
$\bm{\theta}_{\rm output}=m \times \bm{\theta}_{\rm input}+ c$, where $\bm{\theta}_{\rm input}$ is the input parameters, $\bm{\theta}_{\rm output}$ is the output parameter, $m$ is a multiplicative error, and $c$ is an additive error. The results are shown in Figure \ref{fig:hibrecs2}. The code can constrain both slope and normalization, irrespective of the parent distribution. 
While intrinsic scatter and correlation coefficients are constrained well when the measurement errors are smaller than intrinsic scatter, they cannot be constrained when the measurement errors are larger than intrinsic scatter.

We choose the parameters for the last two simulations as similar to measurement errors of this study;  
$\sigma_x^{\rm err}=[0.75,1.25]$, $\sigma_{y_0}^{\rm err}=0.1$, $\sigma_{y_1}^{\rm err}=[0.2,0.7]$, $\sigma_{y_2}^{\rm err}=0.02$, and $\sigma_{y_3}^{\rm err}=[0.4,1]$. The error correlation coefficient is set to be $r_{x,y_1}^{\rm err}=0.85$, $r_{x,y_3}^{\rm err}=0.95$, and $r_{y_1,y_3}^{\rm err}=[0.64,0.97]$ and the others are fixed to zero. The input parameters of the scaling relations are randomly drawn from a uniform distribution in $\alpha=[0,1], \beta=[0.3,1.7], \sigma_{\rm int}=[0.4,0.8]$, and $r_{\rm int}=[-0.8,0.8]$, respectively. The intrinsic scatter is larger than that employed in the first six simulations. We fix $\alpha_X=\ln 0.89$, $\beta_X=1.09$, $\sigma_X=0.21$, and $r_{X,Y_i}=0$. As for the parent population, we adopt the two cases of no-redshift dependence $p(Z)$ and redshift dependence $p(Z,z)$. In the latter case, the random redshift is uniformly distributed in $z=[0.1,1]$ and the mass function is computed at the given redshift. Then, the sample is constructed with a selection cut similar to the observation (Figure \ref{fig:Pz}). Although the measurement errors for the $x$ quantity are large, the input parameters, including intrinsic scatter and correlation coefficients, are recovered because the error correlations are all taken into account.

\section{Regression dilution effect}\label{app:b}

We briefly explain the regression dilution effect \citep[e.g.][]{1996ApJ...470..706A}.
We first consider the two true variables of $\{X_n,Y_n\}$ for the $n$-th cluster of the sample size $N$, where we assume $\alpha_X=0$, $\beta_X=1$, and $\sigma_X=0$ and thus, $p(X)=p(Z)$.
We assume that the two variables follow a simple linear regression, specified by
\begin{eqnarray}
      Y_n=\alpha_Y + \beta_Y X_n + \varepsilon_n, \label{eq:app_b1}
\end{eqnarray}
where $\varepsilon_n$ is a random variable drawn from the intrinsic scatter $\mathcal{N}(0,\sigma_Y$) of the parameter $Y$.

The regression parameters, $\alpha_Y$ and $\beta_Y$, are described by the first- and second-order moments of eq. (\ref{eq:app_b1}), 
\begin{eqnarray}
      \alpha_Y&=&E(Y_n)-\beta_YE(X_n)=\mu_Y-\beta_Y \mu_X,\\
      \beta_Y&=&\frac{(1/N)\sum_n(X_n-\mu_X)(Y_n-\mu_Y)}{(1/N)\sum_n (X_n-\mu_X)^2}=\frac{\sigma_{XY}}{\sigma_X^2},
\end{eqnarray}
where $\mu_X=E(X_n)$ and $\mu_Y=E(Y_n)$ are the average quantities, $\sigma_X^2$ is a variance of the parent population $p(X)$, and $\sigma_{XY}$ is the covariance of the bivariate $X$ and $Y$ distribution.

We next consider a measurement error model on $X$,
\begin{eqnarray}
      Y_n&=&\alpha + \beta x_n+\varepsilon_n\\
      x_n&=&X_n+\eta_n
\end{eqnarray}
where $x_n$ is the observed quantity and $\eta_n$ is the measurement error drawn from $\mathcal{N}(0,\sigma_x^{\rm err}$), irrespective of $X_n$. The regression parameters become
\begin{eqnarray}
       \beta&=&\frac{(1/N)\sum_n(x_n-E(x_n))(Y_n-\mu_Y)}{(1/N)\sum_n (x_n-E(x_n))^2}, \nonumber \\
       &=& \frac{(1/N)\sum_n(X_n+\eta_n-\mu_X)(Y_n-\mu_Y)}{(1/N) \sum_n (X_n+\eta_n-\mu_X)^2 },\nonumber \\
       &=&\frac{(1/N)\sum_n (X_n-\mu_X)^2 \beta_Y}{(1/N)\sum_n (X_n-\mu_X)^2+(\sigma_{x}^{\rm err})^2}, \nonumber \\
       &=&\frac{\sigma_X^2 }{\sigma_X^2+(\sigma_x^{\rm err})^2}\beta_Y, \label{eq:app_b2}\\
       \alpha&=& \mu_Y-\frac{\sigma_X^2 }{\sigma_X^2+(\sigma_x^{\rm err})^2}\beta_Y\mu_X, \nonumber \\
       &=& \alpha_Y + \left(1-\frac{\sigma_X^2 }{\sigma_X^2+(\sigma_x^{\rm err})^2}\right) \beta_Y\mu_X
\end{eqnarray}
If the sample size is infinite ($\sigma_X\rightarrow \infty$), the regression parameters,  $\alpha$ and $\beta$, coincide with $\alpha_X$ and $\beta_X$, respectively. 
However, in the case of the finite sample size, the slope directly estimated by the observables is underestimated by $(1+(\sigma_x^{\rm err})^2/\sigma_X^2)^{-1}$. When we consider the measurement error and the variance of the parent population in the regression analysis, we correct the measured value, $\beta$, by $(1+(\sigma_x^{\rm err})^2/\sigma_X^2)$. The Bayesian forward modeling enables us to directly recover $\beta_Y$.

We next introduce the observed quantity $y_n$, as follows
\begin{eqnarray}
      y_n&=&\alpha'+\beta'x_n+\varepsilon_n \\
       x_n&=&X_n+\eta_n \\
      y_n&=&Y_n+\xi_n
\end{eqnarray}
where $\xi_n$ is the measurement error of $y_n$. The slope parameter becomes
\begin{eqnarray}
       \beta'&=&\frac{(1/N)\sum_n(x_n-E(x_n))(y_n-E(y_n))}{(1/N)\sum_n (x_n-E(x_n))^2}, \nonumber \\
       &=& \frac{(1/N)\sum_n(X_n+\eta_n-\mu_X)(Y_n+\xi_n-\mu_Y)}{(1/N) \sum_n (X_n+\eta_n-\mu_X)^2 },\nonumber \\
              &=& \frac{(1/N)\sum_n(X_n-\mu_X)(Y_n-\mu_Y)+(1/N)\sum_n \eta_n\xi_n}{(1/N) \sum_n (X_n-\mu_X)^2+(1/N)\sum_n \eta_n^2 },\nonumber \\
       &=&\frac{(1/N)\sum_n (X_n-\mu_X)^2 \beta_Y+\sigma_{xy}^{\rm err}}{(1/N)\sum_n (X_n-\mu_X)^2+(\sigma_{x}^{\rm err})^2}, \nonumber \\
       &=&\frac{\sigma_X^2 }{\sigma_X^2+(\sigma_x^{\rm err})^2}\beta_Y+\frac{\sigma_{xy}^{\rm err} }{\sigma_X^2+(\sigma_x^{\rm err})^2}, \label{eq:app_b3}
\end{eqnarray}
where $\sigma_{xy}^{\rm err}$ is the error correlation between the $x$ and $y$ quantities.  
If there is no error correlation ($\sigma_{xy}^{\rm err}=0$), eq. \ref{eq:app_b3} coincides with eq. \ref{eq:app_b2} and thus the regression dilution effect depends on the error of the $x$ quantity. If the errors are correlated, it is important to implement the error covariance matrix in the regression analysis (eq. \ref{eq:likelihood}).

As described above, $\sigma_x^{\rm err}$ in the finite sample gives rise to the regression dilution effect, while $\sigma_y^{\rm err}$ does not as long as it is not correlated with $\sigma_x^{\rm err}$. 
Therefore, it is vitally important to infer $\sigma_X^2$ and $\mu_X$ to accurately estimate the regression coefficient parameters. Here, we do not assume any specific distributions of the $X$ quantity, but require only $\sigma_X^2$ and $\mu_X$. In our case, they correspond to $\sigma_Z^2$ and $\mu_Z$ of the logarithm distribution of the true mass. 
Hence, we do not necessarily need a full description of $p(Z)$ but effective estimations of $\sigma_X$ and $\mu$ in the regression analysis.
In other words, the introduction of $\mu_Z$ and $\sigma_Z$ in our scaling relation analysis is neither for cosmological purposes nor for an accurate measurement of the mass function. 
This paper assumes the Gaussian distribution, $\mathcal{N}(\mu_Z,\sigma_Z)$ to correct for the regression dilution effect. The Gaussian distribution enables a fast computation. 
They are determined as the hyper-parameters to respond flexibly to an unknown population distribution (Appendix \ref{app:a}).

\section{Aperture correction} \label{app:aperture_cor}

In the above regression analysis, we use the ${\bm Y}$ quantities measured within the observed overdensity radii, $r^X$,  and obtain the scaling relation of ${\bm Y}_Z={\bm Y}_Z(<r^X)={\bm \alpha}+{\bm \beta}Z$. When we study the correlation between the true mass and the observables measured within the true mass overdensity, $r^Z$, we convert ${\bm Y}_Z(<r^X)$ into ${\bm Y}_Z(<r^Z)$. The original baseline, ${\bm Y}_Z(<r^X)$, can be expanded in series up to the first order of a radius deviation, $\delta r^Z$, as follows,
\begin{eqnarray}
    {\bm Y}_{Z}(<r^X)= {\bm Y}_Z(<r^Z)+\frac{\partial {\bm Y}_Z(<r^Z)}{\partial r^Z} \delta r^Z.
\end{eqnarray}
The new baseline is thus specified by 
\begin{eqnarray}
    {\bm Y}_{Z}(<r^Z)&=& {\bm Y}_Z(<r^X) - \frac{\partial {\bm Y}_Z(<r^Z)}{\partial r^Z} \delta r^Z \nonumber \\
    &=& {\bm \alpha}+\delta {\bm Y}_Z+{\bm \beta}Z, \label{eq:Yz_rz}
\end{eqnarray}
where $\delta {\bm Y}_Z$ is the aperture correction term in the scaling relation of interest.
We consider the logarithmic quantity, $Y_i\propto \ln {\mathcal O}_i$ and $Z\propto \ln M(<r^Z)$, where ${\mathcal O}_i$ and $M$ are the observable and the true mass, respectively. Assuming that the averaged density profile for the observable, $\mathcal{O}_i$, follows a power law distribution $\rho_i \propto r^{-p}$, we find $\delta Y_Z = -(3-p)(\delta r^Z/r^Z)=-(1-p/3) \delta M(<r^Z)/M(<r^Z)$. We recall the mass calibration relation between the observed mass and the true mass, $X=\alpha_X+\beta_X Z$, and then obtain $\delta M(<r^Z)/M(<r^Z)=e^{\alpha_X+(\beta_X-1)Z }-1$. In short, the aperture correction term becomes  
\begin{eqnarray}
      \delta {\bm Y}_Z= \left(1-\frac{{\bm p}}{3}\right)\left(1-\exp\left[\alpha_X+(\beta_X-1)Z\right]\right). \label{eq:aperture_cor}
\end{eqnarray}
The normalization, ${\bm \alpha}+\delta {\bm Y}_Z$, in the scaling relation (eq. \ref{eq:Yz_rz}) slightly changes as a function of the true mass.

\section{Posterior distribution}\label{sec:app_post}

\begin{figure*}
\begin{center}
    \includegraphics[width=\hsize]{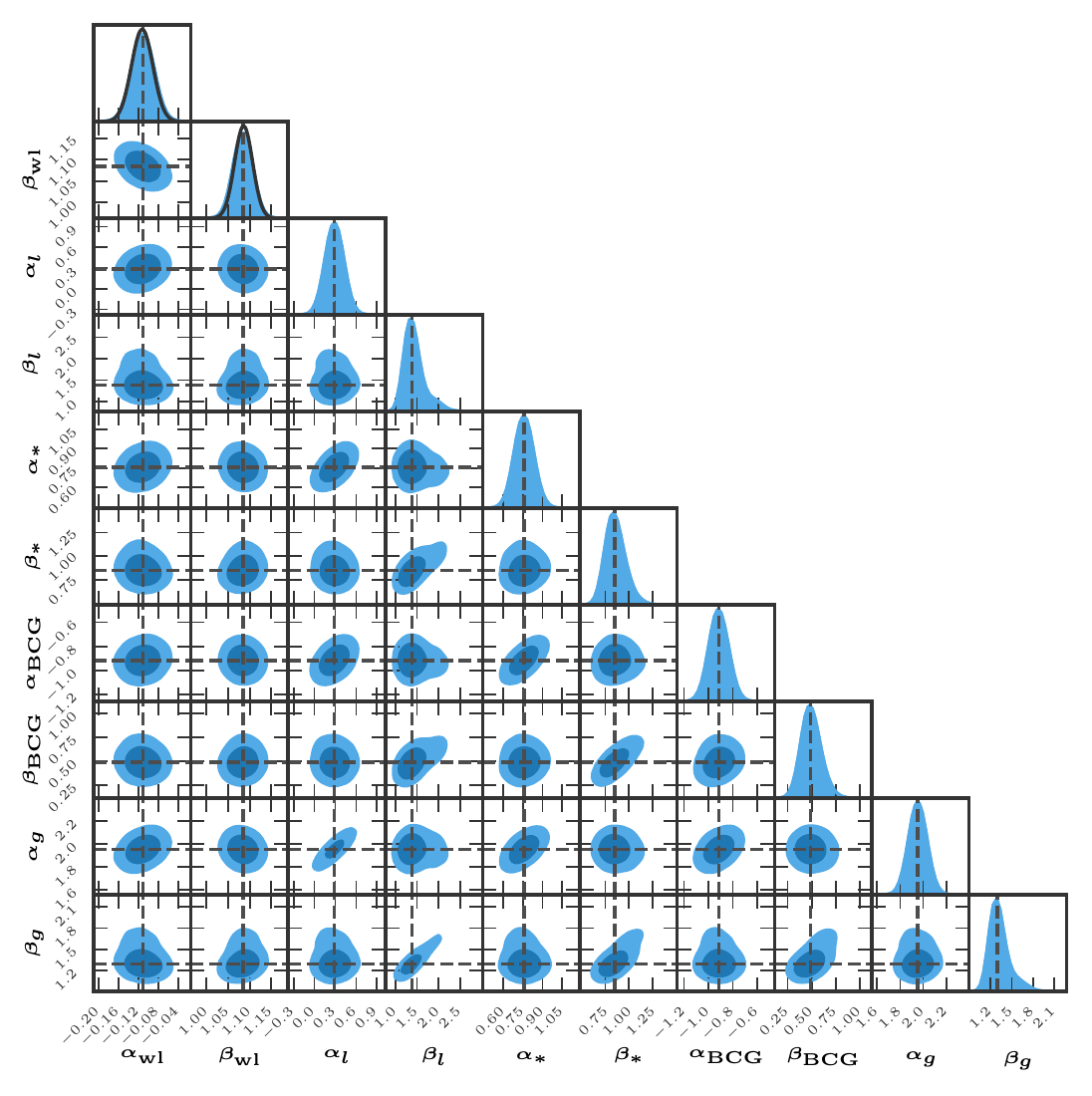}
\end{center}
    \caption{Posterior distributions of the normalization, $\bm{\alpha},$ and slope, $\bm{\beta}$, parameters. Dark and light blue regions in the two dimensional parameter planes denote $1\ \sigma$ and $2\ \sigma$ uncertainty, respectively. Black solid curves are the priors of the weak-lensing mass calibration of $\alpha_{\rm WL}$ and $\beta_{\rm WL}$ (Sec. \ref{subsec:wl_calib}). Black dashed lines are the resulting values.}
    \label{fig:Prob_slope}
\end{figure*}

\begin{figure*}
\begin{center}
    \includegraphics[width=\hsize]{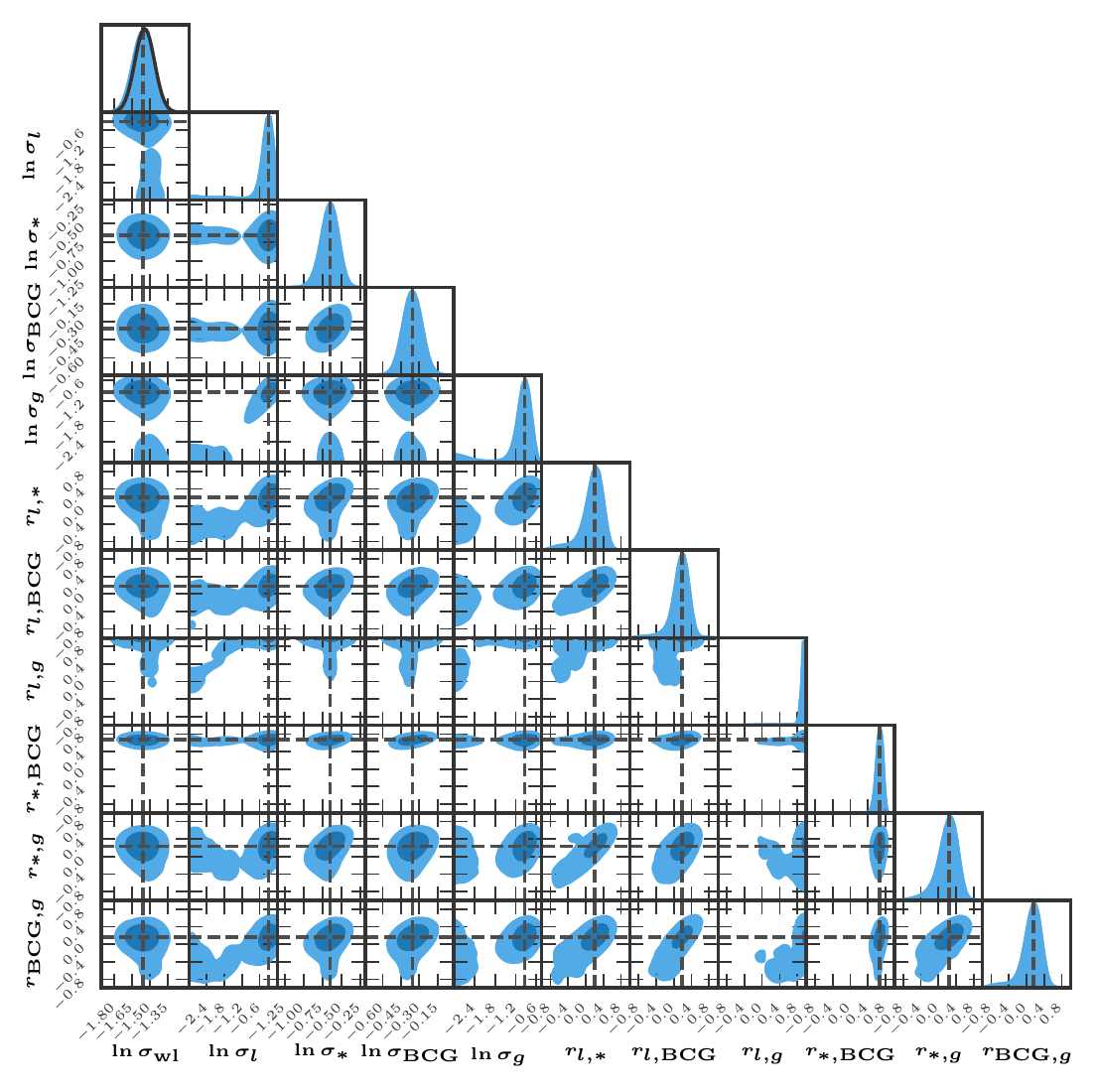}
\end{center}
    \caption{Posterior distributions of the intrinsic covariance of the scaling relations. Dark and light blue regions in the two dimensional parameter planes denote $1\ \sigma$ and $2\ \sigma$ uncertainty, respectively. Black solid curve is the priors of the weak-lensing mass calibration of $\ln \sigma_{\rm WL}$ (Sec. \ref{subsec:wl_calib}). Black dashed lines are the resulting values.}
    \label{fig:Prob_cov}
\end{figure*}

It is worth showing the posterior distributions of our main result (Table \ref{tab:chabrier_4d} and Table \ref{tab:Sigma_int_4d}). Due to the page size limit, we split 
the 25 parameters into two figures for normalization and slopes (Fig. \ref{fig:Prob_slope}) and intrinsic covariance (Fig. \ref{fig:Prob_cov}) for the $r_{*,g}^{\rm err}=r_{{\rm WL},*}^{\rm err}r_{{\rm WL},g}^{\rm err}$ case. Black solid curves represent the trivariate Gaussian distribution prior for the weak-lensing mass calibration (Sec. \ref{subsec:wl_calib}).

\section{Scaling relations without the $E(z)$ correction}\label{app:d}
\setcounter{equation}{0} 
\renewcommand{\theequation}{\Alph{section}.\arabic{equation}}

In some numerical simulations \citep[e.g.][]{2014MNRAS.440.2290M,2014MNRAS.441.1270L,2015MNRAS.452.1982W,2017MNRAS.465.2936M,2017MNRAS.471.1088B,2018MNRAS.478.2618F,2020MNRAS.493.1361F}
and observational papers
\citep[e.g.][]{Lin03,Lin04,2013ApJ...778...14G,2013A&A...555A..66L}, 
they investigated the scaling relations without the $E(z)$ correction. 
It is useful to show the results assuming no redshift evolution. We define the observables independent of redshifts as in eq. \ref{eq:observable_xy}. 
The form of scaling relations is similar to eq. \ref{eq:Yz_Z_Ez} except for redshift-dependent slopes,
\begin{eqnarray}
      &&x=\ln \frac{M_{500}^{\rm WL} }{10^{14}M_\odot}  \\
      &&\bm{y}=\Bigr\{\ln \frac{L_X}{10^{43}\,{\rm erg s^{-1}}}, \ln \frac{M_*}{10^{12}M_\odot},\ln \frac{M_{\rm BCG} }{10^{12}M_\odot}, \ln \frac{M_g }{10^{12}M_\odot}\Bigl\} \nonumber \\
      \nonumber\\
    &&{\bm Y}_Z={\bm \alpha} + {\bm \beta} Z \label{eq:append_alpha_beta}
\end{eqnarray}
Here, we assume that the scaling relations do not evolve with redshift ($\gamma=0$), however, we employ the redshift-dependent mean $\mu_Z(z)$ and standard deviation $\sigma_Z(z)$ of parent population $p(Z,z)$ because the sample selection depends on redshift (Sec. \ref{subsubsec:Pz} and Figure \ref{fig:Pz}).
With this set up, we repeat the Bayesian analysis for scaling relations.
The results are shown in Table \ref{tab:chabrier_4d_noEz} and Table \ref{tab:Sigma_int_4d_noEz}. 

\begin{table}[ht]
    \caption{Resulting regression parameters of the scaling relations between the cluster quantities ($M_{500}^{\rm WL}$, $L_X$, $M_*$, $M_{\rm BCG}$, and $M_g$) and the true mass $M_{500}$ for the 136 XXL clusters without the $E(z)$ correction.
     The normalization, $\alpha$, and the slope, $\beta$, are defined by the linear regressions (eq. \ref{eq:append_alpha_beta}). The intrinsic scatter at a fixed true mass is represented by $\sigma_{\rm int}$. $^\dagger$ : the results using a trivariate Gaussian prior as the WL mass calibration, 
     as described in Sec. \ref{subsec:wl_calib}. The errors denote the $1\sigma$ uncertainty.}
    \begin{center}
    \scalebox{1.15}[1.15]{
    \begin{tabular}{c|rrr}   
              & \multicolumn{1}{c}{$\alpha$}                  &  \multicolumn{1}{c}{$\beta$}                  & \multicolumn{1}{c}{$\sigma_{\rm int}$} \\ \hline
        $M_{500}^{\rm WL}$     &  $-0.11_{-0.02}^{+0.02}$$^\dagger$              & $1.08_{-0.02}^{+0.02}$$^\dagger$                       & $0.21_{-0.02}^{+0.02}$$^\dagger$              \\
        $L_{X}$             & $0.72_{-0.43}^{+0.14}$  & $1.39_{-0.18}^{+0.19}$ & $0.79_{-0.15}^{+0.10}$ \\
        $M_{*}$         &   $0.68_{-0.07}^{+0.15}$   &   $0.76_{-0.08}^{+0.12}$   &  $0.53_{-0.06}^{+0.08}$  \\
        $M_{\rm BCG}$   &  $-1.07_{-0.08}^{+0.18}$   &   $0.39_{-0.08}^{+0.13}$   &   $0.70_{-0.05}^{+0.05}$      \\
        $M_{g}$     &   $2.02_{-0.10}^{+0.08}$   &   $1.28_{-0.09}^{+0.12}$    &    $1.28_{-0.09}^{+0.12}$     \\
    \end{tabular}
    }
    \end{center}
\label{tab:chabrier_4d_noEz}
\end{table}

\begin{table*}[ht]
    \caption{Intrinsic covariance for the 136 XXL clusters ($L_X$, $M_*$ , $M_{\rm BCG}$, and $M_g$) without the $E(z)$ correction. Each column is the same as in Table 2.}
    \begin{center}
    \scalebox{1.2}[1.2]{
    \begin{tabular}{c|cccc}
         & $L_{X}$ & $M_*$ & $M_{\rm BCG}$ & $M_g$  \\ \hline
        $L_{X}$  & $0.79_{-0.15}^{+0.10}$ & $0.08_{-0.09}^{+0.12}$ & $0.10_{-0.08}^{+0.09}$ & $0.30_{-0.10}^{+0.10}$ \\
        $M_*$  & $0.20_{-0.22}^{+0.22}$  & $0.53_{-0.06}^{+0.08}$ & $0.25_{-0.06}^{+0.06}$ & $0.05_{-0.06}^{+0.09}$ \\
        $M_{\rm BCG}$  & $0.18_{-0.14}^{+0.13}$ & $0.68_{-0.10}^{+0.06}$ & $0.70_{-0.05}^{+0.05}$ & $0.04_{-0.04}^{+0.05}$ \\
        $M_g$  & $0.98_{-0.02}^{+0.01}$ & $0.27_{-0.31}^{+0.28}$ & $0.17_{-0.18}^{+0.15}$ & $0.38_{-0.07}^{+0.08}$ \\
    \end{tabular}
    }
    \end{center}
\label{tab:Sigma_int_4d_noEz}
\end{table*}

\end{document}